\documentclass[twocolumn,tighten,twocolappendix]{aastex63}
\pdfoutput=1 
\usepackage[T1]{fontenc}
\usepackage{ae,aecompl}
\usepackage[utf8]{inputenc}
\usepackage[english]{babel}
\usepackage{amsmath,amstext}
\usepackage{apjfonts}
\usepackage[figure,figure*]{hypcap}
\usepackage[hang]{footmisc}
\usepackage{ulem}
\graphicspath{{figures/}}

\newcommand*{\https}[1]{\href{https://#1}{\nolinkurl{#1}}}
\newcommand*{\http}[1]{\href{http://#1}{\nolinkurl{#1}}}
\newcommand*{\fnref}[1]{\textsuperscript{\ref{#1}}}
\newcommand*{\code}[1]{{{#1}}}

\newcommand*{\sneia}{SNe~Ia}

\newcommand*{\opsim}{{OpSim}}
\newcommand*{\gcrcatalogs}{{GCRCatalogs}}

\newcommand{\tagn}{\tau}
\newcommand{\dgv}{d_{\rm gv}}

\definecolor{newgreen}{HTML}{0CA82B}

\shorttitle{The LSST DESC DC2 Simulated Sky Survey}
\shortauthors{LSST~DESC}

\begin{document}
\title{The LSST DESC DC2 Simulated Sky Survey}

\collaboration{1000}{The LSST Dark Energy Science Collaboration (LSST DESC)}
\noaffiliation

\author[0000-0003-1820-8486]{Bela Abolfathi}
\affiliation{Department of Physics and Astronomy, University of California, Irvine, Irvine, CA 92697, USA}

\author[0000-0002-4598-9719]{David Alonso}
\affiliation{Department of Physics, University of Oxford, Denys Wilkinson Building, Keble Road, Oxford OX1 3RH, United Kingdom}

\author{Robert Armstrong}
\affiliation{Lawrence Livermore National Laboratory, Livermore, CA 94550, USA}

\author[0000-0002-5592-023X]{\'{E}ric Aubourg}
\affiliation{Universit\'{e} de Paris, CNRS, CEA, Astroparticule et Cosmologie, F-75013 Paris, France}

\author[0000-0003-2296-7717]{Humna Awan}
\affiliation{Department of Physics and Astronomy, Rutgers, The State University of New Jersey, Piscataway, NJ 08854, USA}
\affiliation{Leinweber Center for Theoretical Physics, Department of Physics, University of Michigan, Ann Arbor, MI 48109, USA}

\author[0000-0002-9162-6003]{Yadu N. Babuji}
\affiliation{Argonne National Laboratory, Lemont, IL 60439, USA}
\affiliation{University of Chicago, Chicago, IL 60637, USA}

\author[0000-0002-8686-8737]{Franz Erik Bauer}
\affiliation{Instituto de Astrof{\'{\i}}sica and Centro de Astroingenier{\'{\i}}a, Facultad de F{\'{i}}sica, Pontificia Universidad Cat{\'{o}}lica de Chile, Casilla 306, Santiago 22, Chile} 
\affiliation{Millennium Institute of Astrophysics (MAS), Nuncio Monse{\~{n}}or S{\'{o}}tero Sanz 100, Providencia, Santiago, Chile} 
\affiliation{Space Science Institute, 4750 Walnut Street, Suite 205, Boulder, Colorado 80301}

\author{Rachel Bean}
\affiliation{Department of Astronomy, Cornell University, Ithaca, NY 14850, USA}

\author[0000-0003-3623-9753]{George Beckett}
\affiliation{EPCC, University of Edinburgh, United Kingdom}

\author[0000-0002-5741-7195]{Rahul Biswas}
\affiliation{The Oskar Klein Centre for Cosmoparticle Physics, Stockholm University, AlbaNova, Stockholm, SE-106 91, Sweden}

\author[0000-0002-1345-1359]{Joanne R. Bogart}
\affiliation{SLAC National Accelerator Laboratory, Menlo Park, CA 94025, USA}
\affiliation{Kavli Institute for Particle Astrophysics and Cosmology, Stanford University, Stanford, CA  94305, USA}

\author[0000-0003-4887-2150]{Dominique Boutigny}
\affiliation{Univ. Grenoble Alpes, Univ. Savoie Mont Blanc, CNRS, LAPP, 74000 Annecy, France}

\author[0000-0002-7370-4805]{Kyle Chard}
\affiliation{Argonne National Laboratory, Lemont, IL 60439, USA}
\affiliation{University of Chicago, Chicago, IL 60637, USA}

\author[0000-0001-5738-8956]{James Chiang}
\affiliation{SLAC National Accelerator Laboratory, Menlo Park, CA 94025, USA}
\affiliation{Kavli Institute for Particle Astrophysics and Cosmology, Stanford University, Stanford, CA  94305, USA}

\author{Chuck F. Claver}
\affiliation{Rubin Observatory Project Office, 950 N. Cherry Ave., Tucson, AZ 85719, USA}

\author[0000-0001-9022-4232]{Johann Cohen-Tanugi}
\affiliation{Univ. Montpellier, CNRS, LUPM, 34095 Montpellier, France}
\affiliation{LPC, IN2P3/CNRS, Université Clermont Auvergne, F-63000 Clermont-Ferrand, France}

\author[0000-0001-6487-1866]{C\'eline Combet}
\affiliation{Univ. Grenoble Alpes, CNRS, LPSC-IN2P3, 38000 Grenoble, France}

\author[0000-0001-5576-8189]{Andrew J. Connolly}
\affiliation{DIRAC Institute and Department of Astronomy, University of Washington, Seattle, WA 98195, USA}

\author{Scott F. Daniel}
\affiliation{DIRAC Institute and Department of Astronomy, University of Washington, Seattle, WA 98195, USA}

\author{Seth W. Digel}
\affiliation{SLAC National Accelerator Laboratory, Menlo Park, CA 94025, USA}
\affiliation{Kavli Institute for Particle Astrophysics and Cosmology, Stanford University, Stanford, CA  94305, USA}

\author[0000-0001-8251-933X]{Alex Drlica-Wagner}
\affiliation{Fermi National Accelerator Laboratory, PO Box 500, Batavia, IL 60510, USA}
\affiliation{University of Chicago, Chicago IL 60637, USA}
\affiliation{Kavli Institute for Cosmological Physics, University of Chicago, Chicago, IL 60637, USA}

\author{Richard Dubois}
\affiliation{SLAC National Accelerator Laboratory, Menlo Park, CA 94025, USA}
\affiliation{Kavli Institute for Particle Astrophysics and Cosmology, Stanford University, Stanford, CA  94305, USA}

\author[0000-0001-6728-1423]{Emmanuel Gangler}
\affiliation{LPC, IN2P3/CNRS, Université Clermont Auvergne, F-63000 Clermont-Ferrand, France}

\author[0000-0003-1530-8713]{Eric Gawiser}
\affiliation{Department of Physics and Astronomy, Rutgers, The State University of New Jersey, Piscataway, NJ 08854, USA}

\author[0000-0001-9649-3871]{Thomas Glanzman}
\affiliation{SLAC National Accelerator Laboratory, Menlo Park, CA 94025, USA}
\affiliation{Kavli Institute for Particle Astrophysics and Cosmology, Stanford University, Stanford, CA  94305, USA}

\author{Phillipe Gris}
\affiliation{LPC, IN2P3/CNRS, Université Clermont Auvergne, F-63000 Clermont-Ferrand, France}

\author[0000-0002-7832-0771]{Salman~Habib}
\affiliation{Argonne National Laboratory, Lemont, IL 60439, USA}

\author[0000-0003-2219-6852]{Andrew P.~Hearin}
\affiliation{Argonne National Laboratory, Lemont, IL 60439, USA}

\author[0000-0003-1468-8232]{Katrin~Heitmann}
\affiliation{Argonne National Laboratory, Lemont, IL 60439, USA}

\author[0000-0001-7203-2552]{Fabio Hernandez}
\affiliation{CNRS, CC-IN2P3, 21 avenue Pierre de Coubertin CS70202, 69627 Villeurbanne cedex, France}

\author[0000-0002-0965-7864]{Ren\'ee~Hlo\v{z}ek}
\affiliation{David A. Dunlap Department of Astronomy and Astrophysics, 50 St. George Street, Toronto ON M5S3H4}
\affiliation{Dunlap Institute for Astronomy and Astrophysics, 50 St. George Street, Toronto ON M5S3H4}

\author[0000-0002-8658-1672]{Joseph Hollowed}
\affiliation{Argonne National Laboratory, Lemont, IL 60439, USA}

\author[0000-0002-6024-466X]{Mustapha Ishak}
\affiliation{Department of Physics, The University of Texas at Dallas, Richardson, TX 75080, USA}

\author[0000-0001-5250-2633]{\v{Z}eljko Ivezi\'{c}}
\affiliation{Department of Astronomy, University of Washington, Box 351580, Seattle, WA 98195, USA}

\author[0000-0002-4179-5175]{Mike Jarvis}
\affiliation{Department of Physics \& Astronomy, University of Pennsylvania, 209 South 33rd Street
Philadelphia, PA 19104-6396, USA}

\author[0000-0001-8738-6011]{Saurabh~W.~Jha}
\affiliation{Department of Physics and Astronomy, Rutgers, The State University of New Jersey, Piscataway, NJ 08854, USA}

\author{Steven M. Kahn}
\affiliation{SLAC National Accelerator Laboratory, Menlo Park, CA 94025, USA}
\affiliation{Kavli Institute for Particle Astrophysics and Cosmology, Stanford University, Stanford, CA  94305, USA}
\affiliation{Rubin Observatory Project Office, 950 N. Cherry Ave., Tucson, AZ 85719, USA}
\affiliation{Department of Physics, Stanford University, Stanford, CA 94305, USA}

\author[0000-0002-6825-5283]{J. Bryce Kalmbach}
\affiliation{DIRAC Institute and Department of Astronomy, University of Washington, Seattle, WA 98195, USA}

\author[0000-0002-4394-6192]{Heather M. Kelly}
\affiliation{SLAC National Accelerator Laboratory, Menlo Park, CA 94025, USA}
\affiliation{Kavli Institute for Particle Astrophysics and Cosmology, Stanford University, Stanford, CA  94305, USA}

\author[0000-0002-2545-1989]{Eve Kovacs}
\affiliation{Argonne National Laboratory, Lemont, IL 60439, USA}

\author[0000-0003-0801-8339]{Danila Korytov}
\affiliation{Argonne National Laboratory, Lemont, IL 60439, USA}
\affiliation{University of Chicago, Chicago, IL 60637, USA}

\author[0000-0002-4410-7868]{K. Simon Krughoff}
\affiliation{Rubin Observatory Project Office, 950 N. Cherry Ave., Tucson, AZ 85719, USA}

\author{Craig S. Lage}
\affiliation{University of California-Davis, Davis, CA 95616, USA}

\author[0000-0001-7956-0542]{Fran\c{c}ois Lanusse}
\affiliation{AIM, CEA, CNRS, Universit\'e Paris-Saclay, Universit\'e Paris Diderot, Sorbonne Paris Cit\'e, F-91191 Gif-sur-Yvette, France}

\author[0000-0001-9592-4676]{Patricia Larsen}
\affiliation{Argonne National Laboratory, Lemont, IL 60439, USA}

\author[0000-0001-7178-8868]{Laurent {Le Guillou}}
\affiliation{Sorbonne Université, CNRS, IN2P3, Laboratoire de Physique Nucléaire et de Hautes Énergies, LPNHE, 75005 Paris, France}

\author[0000-0001-6800-7389]{Nan Li}
\affiliation{School of Physics and Astronomy, University of Nottingham, University
Park, Nottingham, NG7 2RD, United Kingdom}

\author[0000-0002-6758-558X]{Emily Phillips Longley}
\affiliation{Department of Physics, Duke University, Durham NC 27708, USA}

\author[0000-0003-1666-0962]{Robert H. Lupton}
\affiliation{Princeton University, Princeton, NJ, USA}

\author[0000-0003-2271-1527]{Rachel Mandelbaum}\affiliation{McWilliams Center for Cosmology, Department of Physics, Carnegie Mellon University, Pittsburgh, PA 15213, USA}

\author[0000-0002-1200-0820]{Yao-Yuan~Mao}
\altaffiliation{NASA Einstein Fellow}
\affiliation{Department of Physics and Astronomy, Rutgers, The State University of New Jersey, Piscataway, NJ 08854, USA}

\author[0000-0002-0113-5770]{Phil Marshall}
\affiliation{SLAC National Accelerator Laboratory, Menlo Park, CA 94025, USA}
\affiliation{Kavli Institute for Particle Astrophysics and Cosmology, Stanford University, Stanford, CA  94305, USA}

\author[0000-0002-2308-4230]{Joshua E. Meyers}
\affiliation{Lawrence Livermore National Laboratory, Livermore, CA 94550, USA}

\author[0000-0001-8716-6561]{Marc Moniez}
\affiliation{Universit{\'e} Paris-Saclay, CNRS/IN2P3, IJCLab, Orsay, France}

\author{Christopher B. Morrison}
\affiliation{Department of Astronomy, University of Washington, Box 351580, Seattle, WA 98195, USA}

\author[0000-0001-5444-5345]{Andrei Nomerotski}
\affiliation{Brookhaven National Laboratory, Upton, NY 11973, USA}

\author[0000-0002-8718-2235]{Paul O'Connor}
\affiliation{Brookhaven National Laboratory, Upton, NY 11973, USA}

\author[0000-0002-7295-2743]{HyeYun Park}
\affiliation{Brookhaven National Laboratory, Upton, NY 11973, USA}

\author[0000-0002-0692-1092]{Ji Won Park}
\affiliation{SLAC National Accelerator Laboratory, Menlo Park, CA 94025, USA}
\affiliation{Department of Physics, Stanford University, Stanford, CA 94305, USA}
\affiliation{Kavli Institute for Particle Astrophysics and Cosmology, Stanford University, Stanford, CA  94305, USA}

\author[0000-0002-8560-4449]{Julien Peloton}
\affiliation{Universit{\'e} Paris-Saclay, CNRS/IN2P3, IJCLab, Orsay, France}

\author[0000-0002-3988-4881]{Daniel Perrefort}
\affiliation{Department of Physics and Astronomy, University of Pittsburgh, Pittsburgh, PA 15260, USA}
\affiliation{Pittsburgh Particle Physics, Astrophysics and Cosmology Center (PITT PACC), University of Pittsburgh, Pittsburgh, PA 15260, USA}

\author{James Perry}
\affiliation{EPCC, University of Edinburgh, United Kingdom}

\author[0000-0002-1278-109X]{St\'ephane Plaszczynski}
\affiliation{Universit{\'e} Paris-Saclay, CNRS/IN2P3, IJCLab, Orsay, France}

\author[0000-0003-2265-5262]{Adrian~Pope}
\affiliation{Argonne National Laboratory, Lemont, IL 60439, USA}

\author{Andrew Rasmussen}
\affiliation{SLAC National Accelerator Laboratory, Menlo Park, CA 94025, USA}
\affiliation{Kavli Institute for Particle Astrophysics and Cosmology, Stanford University, Stanford, CA  94305, USA}

\author[0000-0002-2234-749X]{Kevin Reil}
\affiliation{SLAC National Accelerator Laboratory, Menlo Park, CA 94025, USA}
\affiliation{Kavli Institute for Particle Astrophysics and Cosmology, Stanford University, Stanford, CA  94305, USA}

\author[0000-0001-5326-3486]{Aaron J. Roodman}
\affiliation{SLAC National Accelerator Laboratory, Menlo Park, CA 94025, USA}
\affiliation{Kavli Institute for Particle Astrophysics and Cosmology, Stanford University, Stanford, CA  94305, USA}

\author[0000-0001-9376-3135]{Eli~S.~Rykoff}
\affiliation{SLAC National Accelerator Laboratory, Menlo Park, CA 94025, USA}
\affiliation{Kavli Institute for Particle Astrophysics and Cosmology, Stanford University, Stanford, CA  94305, USA}

\author[0000-0003-3136-9532]{F. Javier S\'{a}nchez}
\affiliation{Department of Physics and Astronomy, University of California, Irvine, Irvine, CA 92697, USA}
\affiliation{Fermi National Accelerator Laboratory, PO Box 500, Batavia, IL 60510, USA}

\author[0000-0002-5091-0470]{Samuel J. Schmidt}
\affiliation{University of California-Davis, Davis, CA 95616, USA}

\author[0000-0002-4934-5849]{Daniel Scolnic}
\affiliation{Department of Physics, Duke University, Durham NC 27708, USA }

\author[0000-0003-0347-1724]{Christopher W. Stubbs}
\affiliation{Dept. of Physics and Dept. of Astronomy, Harvard University, Cambridge, MA 02138, USA}

\author[0000-0002-9242-8797]{J. Anthony Tyson}
\affiliation{University of California-Davis, Davis, CA 95616, USA}

\author[0000-0002-5631-0142]{Thomas D. Uram}
\affiliation{Argonne National Laboratory, Lemont, IL 60439, USA}

\author[0000-0002-8847-0335]{Antonia Villarreal}
\affiliation{Argonne National Laboratory, Lemont, IL 60439, USA}

\author[0000-0003-2035-2380]{Christopher W. Walter}
\affiliation{Department of Physics, Duke University, Durham NC 27708, USA }

\author[0000-0001-8653-7738]{Matthew P. Wiesner}
\affiliation{Benedictine University, Lisle, IL, 60532, USA}

\author[0000-0001-7113-1233]{W. Michael Wood-Vasey}
\affiliation{Department of Physics and Astronomy, University of Pittsburgh, Pittsburgh, PA 15260, USA}
\affiliation{Pittsburgh Particle Physics, Astrophysics and Cosmology Center (PITT PACC), University of Pittsburgh, Pittsburgh, PA 15260, USA}

\author[0000-0001-9789-9646]{Joe Zuntz}
\affiliation{Institute for Astronomy, University of Edinburgh, Edinburgh EH9 3HJ, United Kingdom}

\begin{abstract}
We describe the simulated sky survey underlying the second data challenge (DC2) carried out in preparation for analysis of the Vera C. Rubin Observatory Legacy Survey of Space and Time (LSST) by the LSST Dark Energy Science Collaboration (LSST DESC). Significant connections across multiple science domains will be a hallmark of LSST; the DC2 program represents a unique modeling effort that stresses this interconnectivity in a way that has not been attempted before. This effort encompasses a full end-to-end approach: starting from a large N-body simulation, through setting up LSST-like observations including realistic cadences, through image simulations, and finally processing with Rubin's LSST Science Pipelines. This last step ensures that we generate data products resembling those to be delivered by the Rubin Observatory as closely as is currently possible. The simulated DC2 sky survey covers six optical bands in a wide-fast-deep (WFD) area of approximately 300~deg$^2$ as well as a deep drilling field (DDF) of approximately 1~deg$^2$. We simulate 5~years of the planned 10-year survey. The DC2 sky survey has multiple purposes. First, the LSST DESC working groups can use the dataset to develop a range of DESC analysis pipelines to prepare for the advent of actual data. Second, it serves as a realistic testbed  for the image processing software under development for LSST by the Rubin Observatory. In particular, simulated data provide a controlled way to investigate certain image-level systematic effects. Finally, the DC2 sky survey enables the exploration of new scientific ideas in both static and time-domain cosmology.
\end{abstract}


\email{papers@lsstdesc.org}


\section{Introduction}
\label{sec:intro}

In the coming decade, several large sky surveys will collect new datasets with the aim of advancing our understanding of fundamental cosmological physics well beyond what is currently possible.  In the language of the Dark Energy Task Force (DETF, \citealt{2006astro.ph..9591A}), Stage IV dark energy surveys such as the Dark Energy Spectroscopic Instrument (DESI) survey\footnote{\https{www.desi.lbl.gov/the-desi-survey}}~\citep{Aghamousa:2016zmz}, the Vera C.~Rubin Observatory Legacy Survey of Space and Time (LSST)\footnote{\https{www.lsst.org}} \citep{2009arXiv0912.0201L,2019ApJ...873..111I}, the Euclid survey\footnote{\https{www.cosmos.esa.int/web/euclid}, \https{www.euclid-ec.org}}~\citep{2011arXiv1110.3193L}, and the Nancy Grace Roman Space Telescope survey\footnote{\https{roman.gsfc.nasa.gov}}~\citep{2015arXiv150303757S,2019arXiv190401174D}
promise to transform our understanding of basic questions such as the cause of the accelerated expansion rate of the Universe.  The LSST Dark Energy Science Collaboration (DESC\footnote{\https{lsstdesc.org}}) was formed in 2012 \citep{Abate:2012za} to prepare for studies of fundamental cosmological physics with the Vera C.~Rubin Observatory LSST. DESC plans an ambitious scientific program including joint analysis of five dark energy probes that are complementary in constraining power within the cosmological parameter space and in handling systematic uncertainties, and together result in Stage IV-level constraints on dark energy \citep{2018arXiv180901669T}.  The challenge faced by the LSST DESC is to build software pipelines to analyze the released LSST data products and unlock the statistical power of the LSST dataset while robustly constraining systematic uncertainties.  Moreover, these pipelines must work at scale on a dataset that is substantially beyond current surveys in size and complexity.

To meet this challenge, the DESC is iteratively developing analysis pipelines based on the current state of the art and then analyzing simulations and precursor data in a series of ``data challenges'' (DCs) that increase in scope and complexity. The first data challenge (DC1) is described in \cite{dc1}; in DC1 a full end-to-end simulation pipeline to generate LSST-like data products was implemented. DC1 covered ten years of data taking in an area of $\approx$40 deg$^2$ and the simulations were carried out in $r$-band only. The input catalog for DC1 was based on the Millennium simulation semi-analytic galaxy catalog~\citep{Springel:2005nw}, which is embedded in the LSST catalog simulation framework, CatSim~\citep{2010SPIE.7738E..1OC, 2014SPIE.9150E..14C}. Image simulations were carried out with imSim (DESC, in preparation), and the resulting dataset was then processed with the Rubin's LSST Data Management
Science Pipelines software stack\footnote{\https{pipelines.lsst.io}} (throughout the paper we refer to it as LSST
Science Pipelines), developed by the Rubin's LSST Data Management (DM) team. The main focus in DC1 was the investigation of systematic effects relevant for large-scale structure measurements (the galaxy catalog within CatSim does not provide shear measurements), as well as the validation and verification of its end-to-end workflow.

In this paper, we describe the second data challenge (DC2) which goes well beyond DC1 in several ways. Working groups within DESC plan to use DC2 for tests of many prototype analysis pipelines that are being developed. A selection of these includes pipelines for measuring weak gravitational lensing correlations, large-scale structure statistics, galaxy cluster abundance and masses based on weak lensing, supernova light curve recovery, and inference of ensemble redshift distributions for samples based on photometric redshifts. To optimize the scientific return of LSST, individual probes cannot be treated in isolation; cross-correlations between them must be properly understood and exploited to sharpen obtainable results as well as to open new avenues of discovery. In order to enable tests across a broad range of science cases, DC2 covers all six optical bands $ugrizy$ that will be observed by the LSST and the area compared to DC1 is increased by a factor of 7.5 to $\approx$300 deg$^2$ to strike a balance between computational cost and analysis value. Another major development compared to DC1 is the integration of a new extragalactic catalog, called cosmoDC2, described in \citet{korytov}. Based on the Outer Rim simulation~\citep{2019ApJS..245...16H}, which has 200 times the volume of the Millennium run~\citep{Springel:2005nw}, cosmoDC2 not only covers a large area to encompass the 300 deg$^2$ required for DC2 but also includes shear measurements and employs an enhanced galaxy modeling approach. A new interface to CatSim was developed, followed by a workflow for the image simulation generation analogous to DC1. The technical implementation of the workflow itself was completely redone to enable scaling to thousands of compute nodes. 

Carrying out an ambitious program such as the one described here requires many careful tests, code optimization and validation, and efficient workflow designs. In order to accomplish this complex set of tasks, we implemented a staged series of activities, following the strategy that would be used for an actual survey: We first executed the equivalent of an engineering run to then advance to a science-grade run. The engineering run, or Run~1, had several stages in which we developed and implemented the full end-to-end pipeline, tested our new approach for generating an extragalactic catalog, investigated two different image simulation tools, PhoSim~\citep{2015ApJS..218...14P} and imSim, processed the simulated images using the LSST Science Pipelines, and created a set of tutorials for the collaboration to enable members to start interacting with the data products. The engineering runs covered a limited area of 25 deg$^2$ out to redshift $z=1$ and were used to identify and eliminate many shortcomings in the overall set-up. Run~2, one of two science-grade runs, covers the full target area of DC2: a 300 deg$^2$ patch out to $z=3$. The other science-grade run, Run~3, covers the Deep Drilling Field (DDF), a 1~deg$^2$ patch within the 300 deg$^2$ that contains additional time-varying objects. In this paper we focus on Runs~2 and~3 but provide information about Run~1 wherever useful.

The paper is organized as follows. First, in \autoref{sec:DC2Universe}, we describe the requirements on the extragalactic components for DC2 as set by the needs of the relevant probes of cosmic acceleration. Next, in \autoref{sec:design} we describe details regarding the DC2 survey design and the observing cadence. The DC2 survey includes a wide-fast-deep (WFD) area as well as a DDF. \autoref{sec:workflow} provides an overview of the end-to-end workflow we have implemented. Detailed descriptions of the different workflow steps are provided in the following sections: starting with the generation of the extragalactic catalog and the input catalogs for the image simulations (\autoref{sec:extra}), to the image simulations themselves in \autoref{sec:image_sims}, to the final image processing in \autoref{sec:processing}. The resulting data products and our data access strategy are detailed in \autoref{sec:data-drp} and \autoref{sec:data-access}. We conclude in \autoref{sec:summary}. Finally, \autoref{sec:calib} provides an overview of the calibration products required to for the data processing with the LSST Science Pipelines, and \autoref{sec:glossary} summarizes the acronyms and the main simulation packages used in the paper.

\section{DC2 Requirements}  
\label{sec:DC2Universe}

The generation of an end-to-end survey simulation, from extragalactic catalogs to processed data products that can be used to test analysis methodology, is a very ambitious undertaking. When designing and planning such a project, several competing considerations must be taken into account. For each component in the simulation we evaluate whether realistic models based on first principles are available and feasible to implement within our available human and computing resources, or whether we must use approximate or empirical models. LSST will enter new observational territory and we have to decide what approximations (if any) we need to predict the unexplored data -- for example, the galaxy populations that will be observed by LSST clearly cannot be predicted from first principles, but rather require approximate modeling approaches and some degree of extrapolation from current observations.  When undertaking detailed image simulations, we may accept approximate models to realize substantial computational efficiencies while still enabling the majority of our expected use cases. Finally, the LSST Science Pipelines are still under very active development.  Therefore, we may exclude certain effects from the simulations if the current version of the LSST Science Pipelines cannot account for them and if they would dominate over smaller effects that are of interest to us.

When DC2 was conceived, the LSST DESC working groups put forward a range of requirements to enable many tests and science investigations that they planned to carry out with DC2. When deciding which features would be truly important for DC2, the interplay between cost and benefit had to be carefully considered, given available time and resources.

In the following we provide an overview of the basic requirements for DC2 in \autoref{sec:basic}, including size, depth and simulated survey duration, followed by a discussion of the science requirements as put forward by the LSST DESC working groups in \autoref{sec:science_req}. We will carefully highlight which science requirements have been met and which remain to be met in future simulation campaigns.

\subsection{Basic Requirements}
\label{sec:basic}

\begin{figure}[t]
\centering
\includegraphics[width=3.5in]{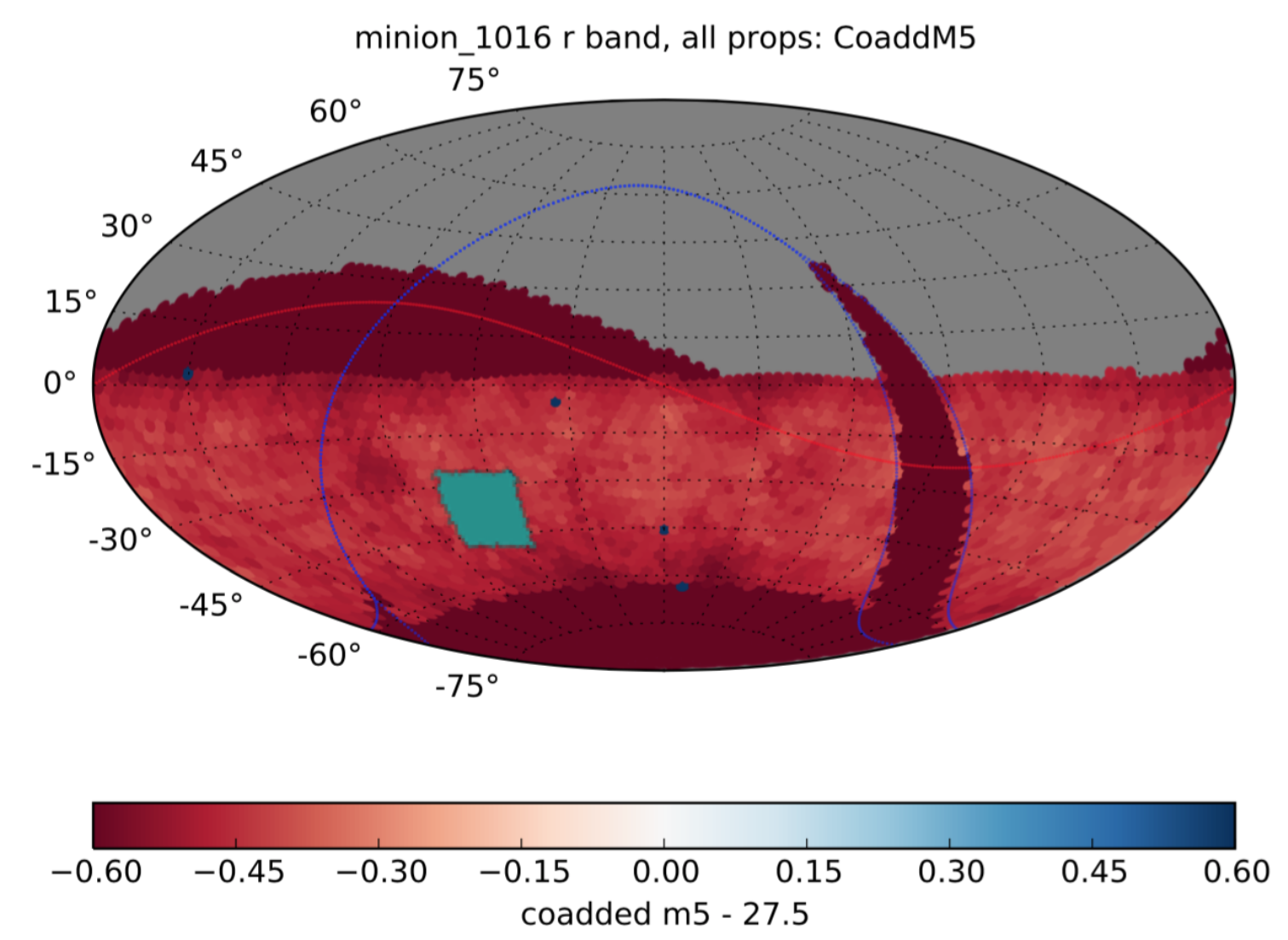}
\caption{Image of the sky along with possible coverage by LSST observations (red, ~\citealt{MAF}) from the \text{minion\_1016} survey simulation shown in Aitoff projection. The blue line marks the Galactic equator and the red line the Ecliptic. More details are provided in \autoref{sec:design}. The green region shows the area on the sky that is covered by DC2 and is simply overlaid on the coadded depth skymap.}
\label{fig:minion1016}
\end{figure}

The DC2 Universe aims to capture a small, representative area of the sky as observed in the LSST. \autoref{fig:minion1016} shows the area that DC2 covers in comparison to an earlier baseline LSST footprint.
Basic specifications for the generation of DC2 concern the size of the simulated area, the number of survey years to be simulated, the bands to be included, and the redshift range to be covered. 

The DC2 area spans 300 deg$^2$. This size provides a good compromise between computational cost for the image simulations and processing and areal size sufficient to derive cosmological constraints for weak lensing and large-scale structure measurements. The area is similar to areas covered by Stage II and early Stage III surveys and therefore has been proven to enable meaningful cosmological investigations. 

When considering the number of years in the survey to simulate, the cost of the image simulations and also the cost for the processing of the data were taken into account. The working groups requested several survey years to investigate improvements of cosmological constraints over time. The difference between 5 and 10 survey years was viewed as having a minor effect on this study; the difference between one and five years of observations, however, in terms of not only depth but also the homogeneity of the dataset, is considerable.      

The number of bands simulated strongly affects the computational requirements for the image simulations. It was decided that the opportunities for science projects with DC2 were greatly enhanced if all bands were included.

Finally, the redshift reach and magnitude limits needed to be set. Here, the biggest challenge is the modeling capability for the extragalactic catalog that underlies DC2. Very faint galaxies, for example, require an extremely high resolution simulation. Our approach for cosmoDC2 is based partly on an empirical modeling strategy and approximations had to be implemented for high-redshift galaxies. An extensive discussion of these challenges and how they were overcome is given in \cite{korytov}.  

\subsection{Science Considerations}
\label{sec:science_req}

\begin{table*}[t]
\centering
\caption{DC2 science opportunities, and simulation requirements to enable those opportunities, for different LSST DESC Working Groups. `I' denotes the image catalog and `EG' the extragalactic catalog. The working group acronyms are: WL -- weak lensing, CL -- clusters, LSS -- large-scale structure, PZ -- photo-$z$, SN -- supernova, SL -- strong lensing. The entry for `cadence' indicates either WFD or DDF for the science cases using image simulations, or N/A for those science cases that primarily rely on the extragalactic catalog. The `Requirement' column also includes a link to the section in this paper or reference to other papers where readers can learn about the feature implementation intended to produce simulations that meet this requirement.} 
\begin{tabular}{ccc p{60mm} p{60mm}}
\toprule
WG   &  Catalog  &   Cadence   &   Study  &   Requirement \\ \hline
WL & I & WFD & Validating weak lensing shear two-point correlation measurements in the presence of realistic image-level systematics and survey masks & Correct galaxy clustering, shear implementation (Sec.~4~\&~5.1, Figs.~7~\&~15 in \citealt{korytov}); at least some image-level systematics (\ref{sec:imsim-features})\\
WL & I & WFD & Investigating the impact of realistic levels of blending 
& Realistic source galaxy populations (Sec.~5.1, Fig.~12 in \citealt{korytov}) and seeing distribution (\ref{sec:imsim-features})\\
WL & I & WFD & Testing the effectiveness of PSF modeling routines, and the impact of residual systematics 
& Reasonably complex (non-parametric and with some spatial correlations) PSF model (\ref{sec:imsim-features})\\
WL & I & WFD & Testing the full end-to-end weak lensing cosmology inference pipeline & Correct clustering, shear implementation (Sec.~4~\&~5.1, Figs.~7~\&~15 in \citealt{korytov}); realistic image simulations (\ref{sec:imsim-features})\\
\hline
CL   & I & WFD & Performance of cluster-finding algorithms & Realistic galaxy colors, spatial distribution (Sec.~5.1--5.3, Figs.~14~\&~15 in \citealt{korytov})\\
CL & I & WFD & Blending and shear estimation biases on cluster mass reconstruction &  Realistic galaxy populations in clusters (Sec.~5.1.2 in \citealt{korytov})\\
CL & EG, I & WFD & Impact of observational systematics on cluster lensing profiles &  At least some sensor effects and other sources of image complexity (\ref{sec:imsim-features})\\
\hline
LSS & EG & N/A & Investigating methods for optimally selecting galaxies for LSS analysis using photometric redshift posteriors &  Realistic photometric redshifts (Sec.~5.1--5.3, Fig.~14 in \citealt{korytov}) \\
LSS & EG & N/A & Investigating methods for optimally selecting galaxies based on colors/magnitudes & Realistic galaxy colors (Sec.~5.1--5.3, Fig.~14 in \citealt{korytov})\\
LSS & EG, I & WFD & Testing LSS analysis pipeline for selected galaxy subsamples & Sufficient area to enable clustering studies (\ref{sec:basic})\\
\hline
PZ   &   EG  &   N/A   &  Testing the impact of incompleteness in spectroscopic training samples on the quality of photometric redshift estimates   &  Realistic galaxy colors, redshift range (Sec.~5.1--5.3, Figs.~13~\&~14 in \citealt{korytov})\\
PZ & EG & N/A & Testing methods for inferring ensemble redshift distributions for photo-$z$-selected samples using cross-correlation analysis & Realistic galaxy colors, spatial distribution, redshift range (Sec.~5, Figs.~12--15 in \citealt{korytov})\\
PZ & I & N/A & Testing the impact of blending on photometric redshift estimates & Realistic galaxy colors, spatial distribution, redshift range (Sec.~5, Figs.~12--15 in \citealt{korytov})\\
\hline
SN & I & WFD/DDF & Estimating the precision and accuracy of photometry of transients  & Realistic modeling of supernovae in the images (\ref{sec:supernovae})\\
SN & I & WFD/DDF & Testing methods to detect and classify transients  & Realistic modeling of supernovae in the images (\ref{sec:supernovae})\\
\hline
SL & I & DDF & Tests of machine learning approach to detect strong lenses & Realistic strong lensing modeling (\ref{sec:lensed_sne}--\ref{sec:lensed_hosts})\\
\hline
WL/LSS & EG & N/A & Evaluating the impact of the ability to model over small-scale theoretical
  uncertainties in the 3x2pt analysis & Realistic galaxy clustering, shear implementation (Sec.~4~\&~5.1, Figs.~7~\&~15 in \citealt{korytov})\\
\hline
\label{tab:projects}
\end{tabular}
\end{table*}

The DC2 Universe has several different components that are important for LSST DESC to enable the science and pipeline tests that will prepare the collaboration for data arrival. First, a representative extragalactic component is needed that covers many features of the actual observed galaxy distribution, including realistic colors, sizes and shapes, and accurate spatial correlations. 
In addition, the DC2 Universe also includes our local neighborhood, e.g., Milky Way stars and galactic reddening. For this we employ a range of observational data provided via the LSST software framework CatSim~\citep{2010SPIE.7738E..1OC, 2014SPIE.9150E..14C}.
Since LSST is a ground-based survey, observing conditions from the ground also need to be modeled
for each 30 second integration, which is  
referred to as a ``visit''. Finally, the telescope and the camera add a number of instrumental effects to the images that have to be either corrected or compensated for. In DC2, the aim is to capture all of these components -- the extragalactic and local environments, observing conditions from the ground
and instrumental and detector artifacts. We defer the detailed discussion of our implementation of effects due to the local environment and observations to \autoref{sec:extra}.

In this section we focus on the science considerations for the DC2 Universe to investigate 
probes of cosmic acceleration relevant to LSST DESC. For a comprehensive review of observational probes of cosmic acceleration, see, e.g., \cite{2013PhR...530...87W}. For an LSST DESC specific discussion, we refer the reader to the LSST DESC Science Requirements Document~\citep{2018arXiv180901669T}.
The LSST DESC working groups that focus on cosmological probes and photometric redshifts 
provided a set of considerations that drove the design of the simulated survey as summarized in \autoref{tab:projects}. These are further discussed in this section. 

\subsection{Weak Lensing}
Weak gravitational lensing is sensitive to the geometry of the Universe and to the growth of structure as a function of time \citep[for a recent review, see][]{2015RPPh...78h6901K}. Cosmological weak lensing analysis involves measuring the minute distortions of faint background galaxies that are induced by the gravitational potential of the matter distribution located between the source galaxies and the observer. Weak lensing is hence sensitive to both dark and luminous matter alike, as it does not rely on luminous tracers of the dark matter density field. However, achieving robust cosmological constraints from weak lensing requires exquisite control of observational and astrophysical systematic effects.  Simulations that are tailored to the survey at hand are one important ingredient in systematics mitigation. For a recent, comprehensive overview of all sources of systematic uncertainty in cosmological weak lensing measurements, see, e.g., \cite{2018ARA&A..56..393M}. The papers presenting the weak lensing catalogs and cosmological weak lensing analyses from ongoing surveys provide examples of how the key systematics are characterized and mitigated in practice \citep[e.g.,][]{2018MNRAS.481.1149Z,2018PhRvD..98d3526A,2018PASJ...70S..25M,2019PASJ...71...43H,2020arXiv200715633A,2020arXiv200701845G}. 

Studying observational systematics such as shear calibration bias and photometric redshift calibration bias requires synthetic catalogs with realistic galaxy shapes, sizes, morphologies and colors. It is important that these quantities scale correctly with redshift, and that galaxy color-dependent clustering is included. Blending of the light from spatially overlapping sources
is one of the most difficult effects to correct for when measuring the shapes and colors of galaxies \citep[e.g.,][]{2018MNRAS.475.4524S}. Hence it is very important that the clustering has flux, size and color distributions that are well matched to real galaxies to ensure that the full challenge of color-dependent blending due to both chance projections and galaxy clustering is present in the simulations. Color gradients in the galaxy spectral energy distribution (SED) and color-dependent galaxy shapes are important to include in order to correctly model the interdependence of measuring shapes in the different photometric bands and to extract the photometric information for a given set of cuts in the catalog \citep{2020ApJ...888...23K}.

The point-spread function (PSF) is another critical systematic effect for weak lensing shear calibration \citep{2008A&A...484...67P}.
The PSF model must include a plausible atmospheric turbulent layer to correctly simulate small-scale spatial variations.
The atmospheric component of the PSF in DC2 uses a frozen flow approximation for six atmospheric turbulent layers with plausible heights and outer scales.
Optical aberrations lead to complex PSF morphology and vary across the field of view (FOV), so this also must be simulated accurately.
For DC2, we used estimates of the variation in aberrations expected for the residuals from the active optics corrections of the Rubin Observatory LSST Camera (hereafter LSSTCam).
Differential chromatic refraction leads to additional chromatic dependence of the PSF and thus is also included~\citep{2015ApJ...807..182M}.
The brighter-fatter effect (see \citealt{2006SPIE.6276E..09D} for an early discussion of the effect) was also identified as a critical confounding factor for PSF determination and
weak lensing shear estimation and is therefore included in the DC2 simulations (see, e.g., \citealt{Gruen_2015} for measurements and modeling approaches of this effect for the Dark Energy Camera and \citealt{Coulton_2018} for the Hyper Suprime-Cam). Finally, accurate simulation of the PSF modeling step requires a realistic stellar catalog in terms of stellar density and SEDs, which is part of DC2.

\subsection{Clusters}

Galaxy clusters, the largest gravitationally bound systems in the Universe, allow us to critically test predictions of structure growth from cosmological models (see, e.g. \citealt{2011ARA&A..49..409A} for an extensive review). Indeed, as identified in the U.S. Department of Energy Cosmic Visions Program \citep{2016arXiv160407626D} and other works, \textit{``The number of massive galaxy clusters could emerge as the most powerful cosmological probe if the masses of the clusters can be accurately measured."} LSST will provide the premier optical dataset for cluster cosmology in the next decade; over 100,000 clusters extending to redshift $z\sim1.2$ are expected to be detected. The DC2 simulations are designed to enable tests of galaxy cluster identification and cosmological analysis.
 
For galaxy clusters, the most critical requirement is that the simulations accurately capture the photometric properties of the cluster galaxy population (e.g., the dominant red sequence, evolving blue fraction, luminosity function and spatial distribution of cluster members). It is also desirable to capture the photometric properties of galaxies as a function of redshift to enable the study of the effects of line-of-sight projections in galaxy clusters. While a large sky area beyond the 300 deg$^2$ presented here will be required for robust statistical characterization of cluster finding and cosmological pipelines, the DC2 image simulations will enable stringent tests of deblending algorithms in dense environments, which will help improve both photometric redshift and shear estimation.

\subsection{Large-Scale Structure}

The Large-Scale Structure (LSS) working group aims to constrain cosmological parameters from the properties of the observed galaxy clustering. The main source of systematic uncertainty for LSS lies in the details of the connection between the galaxy number density and the underlying dark matter density field. Furthermore, unlike weak lensing, LSS is a \emph{local} tracer of the matter distribution, not connected to an integral along a line of sight. The constraining power of LSS is therefore also more sensitive to the quality of photometric redshift estimation
\citep[see, e.g.,][]{2018MNRAS.477.3892C, 2020A&A...640L..14W}.

For this reason, three important astrophysical factors guide the requirements for LSS. The color distribution of the galaxy sample must be realistic, 
with relevant subsamples (e.g., red sequence, blue cloud) having 
number densities in agreement with existing measurements of their luminosity functions, and the clustering properties of these subsamples should also match measured values 
\citep[e.g.,][]{2013MNRAS.432.1961W, 2016MNRAS.455.4122B}. These clustering properties should minimally encompass the large-scale two-point correlation function, but would ideally include the small-scale clustering and higher-order correlations. Although an accurate modeling of the effects of galaxy assembly bias would also be desirable, it is not a priority at this stage.

The DC2 images also need to reproduce some of the most relevant observational systematics for galaxy clustering. These come in the form of artificial modulations in the observed galaxy number density caused by depth variations and observing conditions (e.g., sky brightness, 
seeing, clouds; see \citealt{Awan+2016}). Another important systematic is the spurious contamination from stars classified as galaxies and vice versa. Therefore, the realism of the observed galaxy size, shape and photometry at the image level is also important. Finally, the effect of Galactic dust absorption on galaxy brightness and colors
\citep[e.g.,][]{2017AJ....153...88L}
has to be modeled accurately so that its impact on clustering contamination can be accounted for.

\subsection{Supernovae}

The main aim of the Supernova (SN) working group is the inference of
cosmological parameters using supernovae (SNe) observed during LSST,
in conjunction with other LSST cosmological probes as well as external
datasets. Cosmological inference using
SNe~\citep{1998AJ....116.1009R,1999ApJ...517..565P} proceeds
using the distance-redshift relationship of cosmological models, and
exploits the standardizable candle property
~\citep{1993ApJ...413L.105P,1999ApJ...525..209T} of Type Ia
SNe. LSST is expected to significantly increase the sample of
Type Ia SNe~\citep{2018arXiv180901669T} compared to current surveys
~(e.g., \citealt{2014A&A...568A..22B,2014ApJ...795...44R,2018ApJ...859..101S,2018ApJ...857...51J,2019ApJ...881...19J,2019ApJ...874..106B})
which are already systematics limited. Therefore, an image simulation that provides a truth catalog of the measurable quantities is an
excellent resource for studying potential inaccuracies in quantities
measured by the LSST Science Pipelines. 

The performance of the pipeline in detecting new sources can be characterized by the efficiency and purity of source detections over a range of significance levels, source brightnesses and reference image depths~\citep{2015AJ....150..172K}. Since the performance is usually a function of observing conditions and environmental properties (e.g., the contrast between the transient brightness and the local surface brightness of the galaxy), it must also be studied in diverse conditions. 
Recent time domain surveys have improved their detection performance by using an additional machine learning classifier~\citep{2012PASP..124.1175B,2015AJ....150...82G,2019PASP..131c8002M} that classifies difference image detections as real or bogus. The DC2 data can help in the development and investigation of such algorithms. Forced photometry performed on the difference images in the science pipelines is used to measure the fluxes in  light curves. DC2 also enables the study of bias in such measured fluxes as a function of observational parameters, or truths. In order to use DC2 for such studies, the DC2 cadence (and the distribution of observational properties) and the locations of the SNe (relative to surface brightness) must be representative of realistic data.
DC2 is also useful in the development and investigation of alternative algorithms for building light curves, such as Scene Modeling Photometry~\citep{2006A&A...447...31A,2008AJ....136.2306H,2019ApJ...874..150B}. It additionally allows investigations of optimal stacking procedures for detecting dimmer, higher redshift SNe from multiple daily visits, which will be particularly relevant in the LSST DDFs. Finally, in order to test host association algorithms~\citep{2006ApJ...648..868S,2016AJ....152..154G}, it is essential to have a realistic association of hosts and offsets from the host location~\citep{2020arXiv200809630G}.

\subsection{Strong Lensing}
When a variable background source, such as an active galactic nucleus (AGN) or SN, is strongly lensed by a massive object in the foreground, multiple images are observed and the relative time delays between the images can be measured. Strong-lensing time delays provide direct measurements of absolute distance, independently of early-universe probes such as the cosmic microwave background (CMB) and local probes using the cosmic distance ladder: they primarily constrain the Hubble constant ($H_0$) and, more weakly, other cosmological parameters \citep[see e.g.][for a recent review]{TM16}.

The LSST dataset is projected to contain $\sim$8,000 detectable lensed AGN and $\sim$130 lensed SNe \citep{OM10}. Isolating a pure and complete sample of strong lenses from billions of other observed objects is a major algorithmic and computational challenge for time delay cosmography. 
Developing and testing lens detection algorithms operating on either the catalog or the pixel level leads to a number of time-domain requirements on the DC2 design.  The goal is to enable initial investigations of a catalog-level lens finder that will perform a coarse search for lensed AGN or SNe, before the search can be fine-tuned on the pixel level with more computational resources. This algorithm will be trained on all of the DC2 \text{Object}, \text{Source} and \text{DIASource} tables\footnote{An extensive description of the LSST data products and tables is given in  the Data Products Definition Document, available at this URL: \https{lse-163.lsst.io}}, in order to fully explore the time domain information provided by LSST. For the DC2-trained algorithm to generalize well to the real LSST data, to first order, the light curves of DC2 lensed AGN and SNe must encode the correlated lensing time delays across the multiple images. In addition, the deflector galaxy properties, such as size, shape, mass, and brightness, should agree with the observed population distributions. Lastly, the AGN variability model must be realistic, with the variability parameters following empirical correlations with the physical properties of the AGN, such as black hole mass.  We note that the DC2 lensed AGN and SNe only include the intrinsic variability, not the additional variability caused by microlensing by the stars in the lens galaxy. 
It may be possible to add this effect into the light curves in post-processing. Alternatively, we can model the error on the light curves excluding the effect of microlensing by training a light curve emulator on the DC2 data. Given noiseless light curves with microlensing built in, e.g. according to a separate empirical model, the emulator can then output DC2-like light curves with microlensing included.

The accuracy of time delay measurements directly propagates into the accuracy on $H_0$ inference. The \textit{ugrizy} light curves in the DC2 \text{DIASource} table, with the DM-processed observation noise, should be an improvement on the ones featured in the Time Delay Challenge \citep{liao2015strong}, which only had one filter, assumed perfect deblending, and used a simple, uncorrelated, Gaussian noise model. Without object characterization and deblending algorithms that have been tuned for lensed AGN and SNe, we might expect the automatically generated \text{DIASource} light curves to be blended and sub-optimally measured; if this is the case, the DC2 image data will provide a useful testbed for exploring alternative configurations of the LSST Science Pipelines that can support strong lens light curve extraction. As in the lens finding application, time delay estimation requires a realistic AGN variability model. For $H_0$ recovery tests, the image positions and magnification must be consistent with the time delays.

Massive structures close to the lens line of sight cause weak lensing effects that perturb the time delays. Correcting for these perturbations is an important part of the cosmographic analysis and a potential source of significant systematic error. By embedding lensed AGN and SNe in plausible environments, the DC2 dataset will enable investigations of the characterization of those environments based on the observed \text{object} catalogs. These catalogs should include realistic photometric redshifts, so that this information can be included in the characterization algorithms' inputs.

\subsection{Photometric Redshifts}

Many of the cosmological science cases outlined above require accurate redshifts of either individual galaxies, or well-characterized redshift distributions of ensemble subsets of galaxies (for a recent review on various techniques for obtaining photometric redshifts, see \citealt{2019NatAs...3..212S}).  Rather than precise determinations using spectroscopic observations of emission and absorption lines, photometric redshifts (photo-$z$'s) are estimates of the distance to each galaxy computed using broadband flux information, sensitive to major features such as the Lyman and Balmer/4000{\AA}  breaks passing through the filters.  As the photo-$z$ name implies, these redshift estimates are extremely sensitive to the multiband input photometry, and all modeling and systematic effects that might impact photometric flux measurements and colors in real observations must be modeled in order to evaluate the expected performance of photo-$z$ algorithms for LSST.  A primary concern is the realism of the underlying population of galaxies: the relative abundance of the underlying sub-populations of galaxies is known to evolve with redshift and luminosity, e.g., the fraction of red versus blue galaxies changes dramatically with both cosmic distance and magnitude. In order to match the space of colors expected from observations, a simulation must utilize a realistic set of galaxy SEDs
and apply them to the correctly-evolving relative number densities 
of various galaxy types. Given the small number of available bandpasses, photo-$z$'s are subject to uncertainties and degeneracies where the mapping to colors is not unique; thus, we want the input galaxy population to be as realistic as possible to test that all such degeneracies are captured.

Any systematics that affect the flux determination will impact photo-$z$ estimates for galaxies that will be used in cosmological analyses, e.g. the ``gold" sample of $i<25.3$ galaxies. LSST has to deliver sub-percent accuracy in measured galaxy colors for these samples, largely driven by photo-$z$ requirements. Simulations that have been carried through all the way to simulated images enable tests of multiband photometric measurement algorithms in the presence of realistic observational effects.  Another leading concern is object blending: the tremendous depth of LSST observations over ten years means that LSST will detect billions of galaxies.  Given their finite size, a significant fraction of objects will overlap on the sky, complicating the already challenging problem of estimating multiband fluxes.  Even percent level contamination can lead to biases that exceed targets for LSST photo-$z$ requirements, so blends with even very faint galaxies are important.  The simulations must extend $\sim3$ magnitudes fainter than the galaxies of interest such that low-luminosity blends are properly included (Park et al.,~in prep).  Contamination of the galaxy SED by AGN flux has the potential to skew galaxy colors and bias photo-$z$ estimates.  However, identifying potential contamination through variability over the course of the ten year survey may enable the isolation of such populations, which can be either  excluded from samples or treated with specialized algorithms.

Beyond base photometric redshift algorithms, modern cosmological surveys have developed calibration techniques \citep[e.g.~][]{Newman:2008} that can determine the redshift distribution of ensemble subsets of the data.  Such techniques rely on the shared clustering of samples in space, and thus require samples with realistic position correlations.  As gravitational lensing and magnification change the observed positions and fluxes of objects, simulations must include estimates of the lensing effects if they are to be useful in estimating systematic biases in applying the calibration technique.  Finally, the method is extremely sensitive to exactly how the galaxies populate the underlying dark matter halos and how this galaxy-halo relation evolves with time.  The cosmoDC2 extragalactic catalog contains the necessary complexity and volume of data as described above that is needed to test both the base photo-$z$ algorithms and the redshift calibration methods.  This sample will enable a full end-to-end test of the photometric redshift pipeline for the first time.

\section{DC2 Survey Design and Cadence\label{sec:design}}
In order to simulate realistic visits, we use one output of the LSST Operations Simulator (\opsim), which simulates 10 years of LSST operations and accounts for various factors such as the scheduling of observations, slew and downtime, and site conditions \citep{2016SPIE.9911E..25R, 2016SPIE.9910E..13D}. Specifically for the DC2 runs, we use the cadence output  \text{minion\_1016}\footnote{\https{docushare.lsst.org/docushare/dsweb/View/Collection-4604}}, which contains a realization of five DDFs as well as the nominal WFD area, and uses single 30 sec exposures.

\begin{figure*}[t]
    \center\includegraphics[trim={5 5 5 30}, clip=true, width=0.79\paperwidth]
	{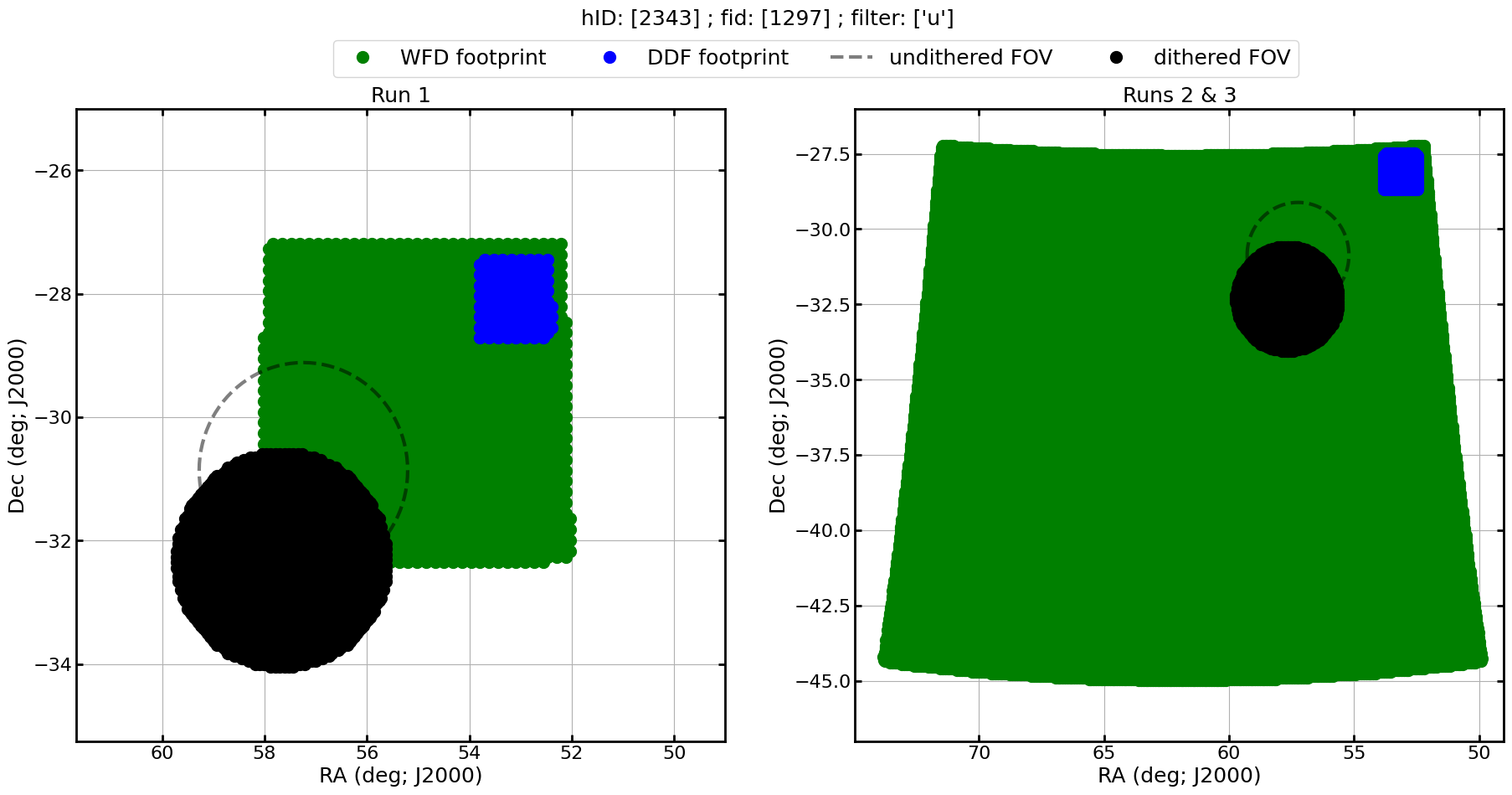}
\caption{DC2 footprints showing the WFD region (green) and the simulated DDF (blue). As an illustration, we show a visit (\opsim\ visit ID 2343 to field 1297 in the $u$-band filter), in both the pre-processed undithered version (dashed) and the implemented dithered one (filled). Left: Run~1 footprint (WFD and DDF region of the engineering run). Right: Run~2 (WFD region for the science-grade run) and Run~3 footprints (DDF region for the science-grade run).}
    \label{fig:footprints}
\end{figure*}

The DDF simulated in DC2 is a square region, with a side length of 68 arcmin, which overlaps with an LSST DDF (the Chandra Deep Field-South; with field ID = 1427 in \opsim); the exact coordinates of the DDF are shown in \autoref{tab: ddf coords}. Situating the DDF in the north-west corner of the DC2 WFD region, we extend the WFD region toward the south-east to cover 25 deg$^2$ in the WFD region for Run~1, while for Run~2, the region is extended to cover 300 deg$^2$, yielding a roughly square region, bounded by great circles\footnote{A great circle is the intersection between a sphere and a plane through the center of the sphere. The WFD region simulated here is bounded on the top and bottom by great circles since we use healpy.query routine to connect the four corners of the region; the healpy routine considers all polygons to be spherical polygons.}, with a side length of $\sim$ 17 degrees; \autoref{tab: wfd coords} lists the coordinates for the corners of the WFD region for Run~1 and Run~2. The south-east extension specifies a region that is typical of the planned LSST WFD survey, avoiding low Galactic latitudes and yielding uniform coverage. We show the WFD and DDF regions for all three runs in \autoref{fig:footprints}.

Before extracting the visits to simulate, we implement dithers, i.e., telescope-pointing offsets, as they significantly improve the depth uniformity of LSST data, as shown in \citet{Awan+2016}. Since dithers are not implemented in the \opsim\ runs, we post-process the \opsim\ output using the LSST Metric Analysis Framework (\text{MAF}; \citealt{MAF}) to produce both translational and rotational dithers -- which is feasible as each \opsim\ output contains realizations of the LSST metadata, including telescope pointing and time and filter of observations. The WFD translational dithers and WFD/DDF rotational dithers were implemented using a \text{MAF} afterburner\footnote{ \https{github.com/humnaawan/sims_operations/blob/master/tools/schema_tools/prep_opsim.py}}, which post-processed the baseline cadence and added the dithered pointing information to the database. Once the DDF translational dither strategy was finalized, we post-processed the afterburner output using a \text{MAF Stacker} to implement it for the DDF visits\footnote{ \https{github.com/LSSTDESC/DC2_visitList/blob/master/DC2visitGen/notebooks/DESC\_Dithers.ipynb}}.

\begin{deluxetable}{ccc}
\centering
\tablecolumns{3}
\tablewidth{0pc}
\tablecaption{Coordinates (J2000) for the simulated DDF.}
\tablehead{ \colhead{ Position} & \colhead{RA (deg)} & \colhead{Dec (deg)} }
\startdata
Center              &   53.125  &   $-$28.100 \\
North-East Corner   &   53.764  &   $-$27.533 \\
North-West Corner   &   52.486  &   $-$27.533 \\
South-East Corner   &   53.771  &   $-$28.667 \\
South-West Corner   &   52.479  &   $-$28.667 \\
\enddata
\tablecomments{ DDF coordinates are the same for Runs 1 and 3. }
\label{tab: ddf coords}
\end{deluxetable}


\begin{table}
\centering
\caption{Coordinates (J2000) for the simulated WFD region.}
\begin{tabular}{*5c}
\toprule
 &  \multicolumn{2}{c}{Run~1} & \multicolumn{2}{c}{Run~2}\\
Position   &   RA (deg)  &   Dec (deg)   &   RA (deg)  &   Dec (deg) \\ \hline
Center              &   55.064  &   $-$29.783   &   61.863   &   $-$35.790 \\
North-East Corner   &   57.870  &   $-$27.250   &   71.460  &   $-$27.250 \\
North-West Corner   &   52.250  &   $-$27.250   &   52.250  &   $-$27.250 \\
South-East Corner   &   58.020  &   $-$32.250   &   73.790  &   $-$44.330 \\
South-West Corner   &   52.110  &   $-$32.250   &   49.920  &   $-$44.330 \\
\hline
\label{tab: wfd coords}
\end{tabular}
\end{table}

For the WFD region, we implement {\em large} translational dithers, i.e., as large as the LSSTCam FOV. Specifically, we use random translational dithers, which have a uniformly random amplitude in the range $[0, 1.75]$~degrees and a uniformly random direction, applied to every visit; this strategy is based on findings in \citet{Awan+2016}.  For the rotational dithers, we use random offsets from the nominal (LSST \opsim\ defined) camera rotation angle between $\pm$ 90 degrees, implemented after every filter change.

For the DDF, we implement small translational dithers, i.e.,
half of the $\sim 7$ arcmin angle subtended by an LSSTCam CCD. This is
sufficient to mitigate chip-scale non-uniformity and is applied to every visit.  We use the same rotational dithering strategy as for WFD: random offsets from the nominal rotation angle between $\pm$ 90 degrees, applied after every filter change.

For both the WFD and DDF regions, we keep the visits at the same cadence as simulated in the baseline and simply extract the visits that fall within our regions of interest. Note that due to the translational dithers, many visits fall only partially in the region of interest.  All of the code for the visit-list generation is in the LSST DESC GitHub repository\footnote{\https{github.com/LSSTDESC/DC2\_visitList}}.

\section{End-to-end Workflow}
\label{sec:workflow}

\begin{figure*}[t]
\centering
\includegraphics[width=7.2in]{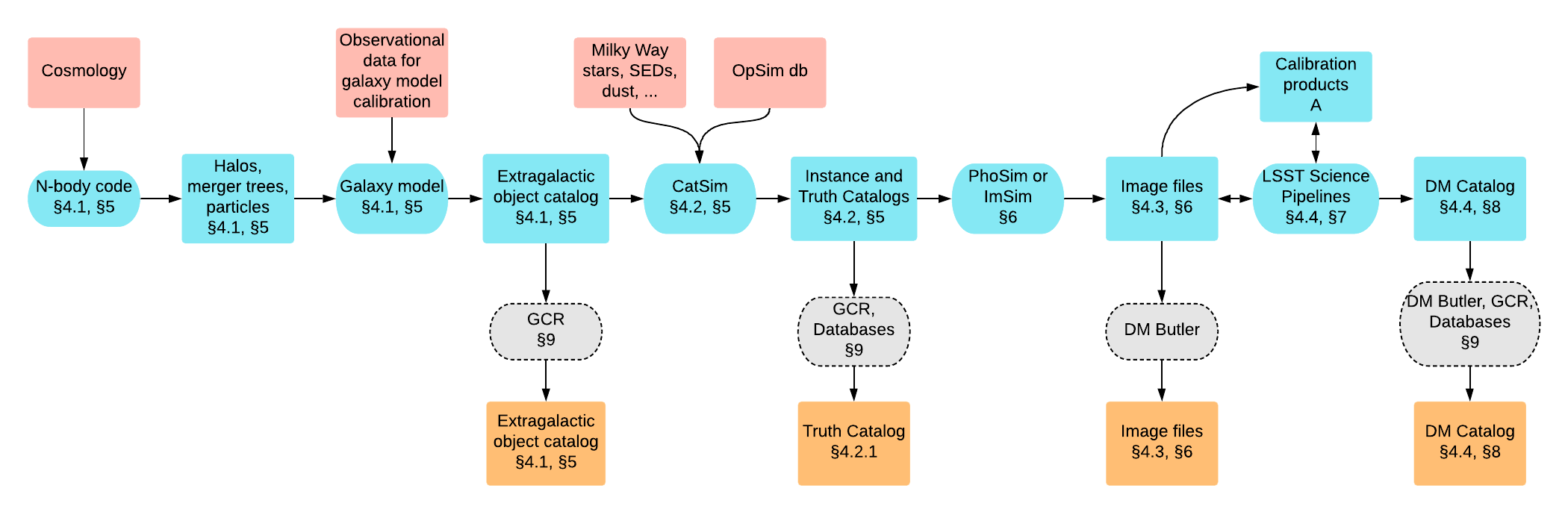}
\caption{Workflow overview: The first row (red) shows external inputs to our workflow, which are both physics and modeling parameters, external datasets, or calibration products. The second row (blue) shows the main workflow and intermediate data products. Overall, the workflow breaks up into four pieces: generation of the extragalactic object catalog, generation of the instance and truth catalogs, generation of the image files, and finally generation of the DM catalog. The last two steps also involve the generation of calibration products which can be derived either from the full image simulations or via specific, additional image simulations. In gray, we show different access tools for interacting with the final data products that are shown in orange. Rectangles show input, output or intermediate data products, while ovals represent codes or access methods. In each box we provide the relevant pointers to the sections that describe the specific part of the workflow.}
\label{fig:workflow}
\end{figure*}

The generation of a simulated dataset that resembles the observational data from the LSST requires a complex workflow that starts with a first-principles structure formation simulation and results in a set of fully processed measurements. \autoref{fig:workflow} provides an overview of the different elements in the workflow as well as data products that are generated at different steps and released to the collaboration. The workflow broadly splits up into four main components: 1) the generation of the extragalactic catalog, 2) the creation of the input catalogs for the image simulations, 3) the image simulations themselves, and 4) the processing of the images with the LSST Science Pipelines. Each of these components results in data products that are used in scientific projects. We discuss the four parts of the workflow briefly in the following. After the broad overview has been provided, we dedicate \autoref{sec:extra} to the extragalactic and input catalog generation, \autoref{sec:image_sims} to the image simulations, \autoref{sec:processing} to the image processing and \autoref{sec:data-drp} and \autoref{sec:data-access} to the data products and access. We provide the relevant section numbers in each box in \autoref{fig:workflow} to enable easy orientation when navigating the paper.

\subsection{The Extragalactic Catalog}
\label{sec:exgal}
The first part of the workflow is based on large cosmological simulations, carried out using major High-Performance Computing (HPC) resources. For Run~1, we generated a small extragalactic catalog covering $\approx$ 25 deg$^2$ out to $z=1$, called protoDC2. protoDC2 is based on a small simulation (``AlphaQ''), carried out with the Hardware/Hybrid Accelerated Cosmology Code (HACC)~\citep{2016NewA...42...49H} on Cooley, a GPU-enhanced cluster hosted at the Argonne Leadership Computing Facility (ALCF). The AlphaQ simulation has the same cosmology and approximately the same mass and force resolution as the main simulation used for DC2,  Runs~2 and~3, but covers a volume 1600$\times$ smaller. This downscaled choice allows for easy handling of the resulting data and many fast iterations to develop and debug the tools needed to create the final catalog. 

The main ``Outer Rim'' simulation was carried out with HACC on Mira, an IBM/BlueGeneQ system that was hosted at the ALCF until the end of 2019. This simulation covers a (4.225Gpc)$^3$ volume and evolved more than one trillion particles, resulting in a particle mass of $m_p=2.6\cdot 10^{9}~{\rm M}_\odot$. Details about the simulation are given in~\cite{2019ApJS..245...16H}. From the simulation, halo and particle lightcones were created and used to generate an extensive extragalactic object catalog called cosmoDC2. A very detailed description of the modeling approach and workflow development is given in~\cite{korytov} and additional information about the validation process will be published in a forthcoming paper. In \autoref{sec:extra} we provide a brief summary of the catalog content most relevant for the DC2 production. Access to cosmoDC2 is provided by the Generic Catalog Reader (GCR) which is described in more detail in \autoref{sec:gcr}.

\subsection{The Instance Catalogs}
\label{sec:image_input}
The second step in the workflow concerns the generation of the input catalogs to the image simulation tools. The image simulator takes as input a series of ``instance catalogs''. An instance catalog is a catalog representing the astrophysical sources whose coordinates are within the footprint of a single field of view at a single time, a concept originating from the image simulation tool PhoSim \citep{2015ApJS..218...14P}.  Producing a separate catalog for each simulated telescope pointing allows us to correctly inject astrophysical variability into the otherwise static cosmological simulation.  This is also the step at which extinction due to Galactic dust and astrometric shifts due to the motion of the Earth are added to each astrophysical source.

This part of the workflow is enabled by the LSST software framework CatSim~\citep{2010SPIE.7738E..1OC, 2014SPIE.9150E..14C}. CatSim provides access to a range of LSST specific data, including position on the LSSTCam focal plane, geocentric apparent position,
luminosity distance, E(B-V) from Milky Way dust, $A_v$ from Milky Way dust, LSST/SDSS (Sloan Digital Sky Survey) magnitudes/fluxes and uncertainty estimates. In addition, during this step, variability is added to the catalog as well as some galaxy features not readily available from the extragalactic catalog. The outputs of the second step in the workflow are 1) a set of instance catalogs, used as input to the image simulations and 2) truth catalogs that can be accessed via \gcrcatalogs{} or the PostgreSQL Database (\autoref{sec:gcr} and \autoref{sec:db}). For DC2, CatSim was optimized to allow the creation of a large number of instance catalogs in a short amount of time. We impose a cut on galaxies from the cosmoDC2 catalog with magnitudes larger than 29 in $r$-band to reduce the catalog sizes, retaining $\sim 42\%$ of the galaxies. Since version 19.0.0 of the LSST Science Pipelines, which we used for image processing, cannot handle proper motion and parallax of stars, we omitted those effects from the instance catalog entries for those objects.

\subsubsection{The Truth Catalogs}
\label{ssec:truth_catalogs}
In order to verify that the inputs to the image simulations, i.e., the instance catalogs, are correct and to assess the fidelity of the output catalogs that are produced by the image processing, we have generated ``truth catalogs'' based on our model of the sky.  These catalogs contain the true values of the {\em measurable} properties of objects as produced by the LSST Science Pipelines software.  As we describe in \autoref{sec:processing}, the image processing outputs comprise catalogs of objects detected and identified in the coadded observations, with measured positions, fluxes, and shape parameters provided for each object.  Catalogs of measured fluxes are also produced for each visit in order to characterize time variability.  Accordingly, our truth catalogs include two tables: a summary truth table that captures the time-averaged properties of objects and a variability truth table that provides for each visit an object's flux with respect to the time-averaged value.  The procedure for assessing the fidelity of the LSST Science Pipelines outputs is then straightforward:  After performing a positional match between truth catalog objects and the LSST catalog objects, the differences between true and measured fluxes and between the true and measured positions can be examined and compared to the expected levels of photometric and astrometric accuracy and precision. 
We introduced our matching procedure in \cite{dc1}. First, a positional query between the true objects and detected objects is carried out. Next, we consider sources in the object catalog as ``matched" if there is a source in the true catalog that is within one magnitude of the measured magnitude in $r$-band (we use $r$-band because it is the deepest). Using this procedure blended objects will still be matched if the deblender performed reasonably well and we eliminate problematic sources that have been shredded. In some cases two or more sources of similar surface brightness are blended and have been detected as just one source. Those will not be considered as matches but we still provide the closest neighbor. The radius for the position matching was chosen to be 1$\arcsec$. This yielded a good compromise between accuracy and speed.

For the verification of the instance catalogs, the procedure is different and somewhat more complicated.  One key difference between the truth table flux values and the information in the instance catalogs is that the truth tables provide the fluxes integrated over each bandpass, including any internal reddening, redshift, Milky Way extinction, and the effects of atmospheric and instrumental throughputs.  These fluxes are the ``true'' values that would be measured for isolated objects with infinite signal-to-noise ratio.  By contrast, the instance catalog entry for an object component provides a tabulated SED, the monochromatic magnitude at 500~nm, the redshift, and internal and Milky Way extinction parameters.   As we describe in \autoref{sec:image_sims}, the image simulation code arrives at the flux for each object by applying each of the ingredients in the instance catalog description in turn.   For PhoSim, this is accomplished by drawing individual photons from the normalized SED and tracing their paths through each element of the simulation.  For imSim/GalSim, the fluxes are computed by direct integration over the observed bandpasses.  For galaxies, another important difference between the summary truth tables and the instance catalogs is that the summary truth tables combine the fluxes from the bulge, disk, and knot components, thereby producing a single entry for each galaxy as a whole, whereas the instance catalogs provide separate entries for each of the three possible galaxy components.   Therefore, to verify an instance catalog, those integrations over bandpasses are computed and the sum over contributions from each galaxy component is made.   Since we have object IDs for the truth and instance catalog entries that allow objects to be matched definitively, positional matching is not needed, and the truth and instance catalog fluxes can be compared directly.  We expect those values to agree to machine precision and verified that this is indeed the case. 

\subsection{The Image Simulations}
The instance catalogs, now containing information about galaxies, stars, the Milky Way, observing conditions and so on, are processed next by image simulation tools. This step delivers simulated pixel data from the LSST focal plane and is described in detail in \autoref{sec:image_sims}. For Run~1, we employed two image simulation tools, PhoSim~\citep{2015ApJS..218...14P} and imSim~(DESC, in preparation). We used the protoDC2 catalog as input for both runs and carried out the PhoSim image simulations (using subversions of v3.7) at the National Energy Research Scientific Computing Center (NERSC) with the SRS Workflow setup~\citep{2009ASPC..411..193F}. The SRS workflow engine was also used for the processing of the image simulations and is briefly described in~\autoref{ssec:image_processing_computing}.
Run~1 with imSim (v0.2.0-alpha) was carried out on Theta at the ALCF, a Cray XC40 with Intel Knights Landing (KNL) processors. We used a Python-based script to define the overall workload, manage submission and monitoring of jobs, and to validate output images, in an iterative fashion, on up to 2,000 nodes; a great majority of these runs were completed over a long weekend. This run was done after the PhoSim run, using the instance catalogs that had been generated already.

For the final DC2 image data, we divided the simulations into two separate runs, Run~2 and Run~3.  Run~2 includes all of the \text{minion\_1016} visits and covers the entire 300 deg$^2$, but since these data would be used primarily by the static dark energy probes (weak lensing, large-scale structure, clusters) that do not rely on the analysis of time-varying objects, we omitted the AGN at the centers of galaxies, although we do include ordinary SNe and variable stars, as well as non-varying stars, as these latter objects are needed for performing the astrometric and photometric calibration for the image processing.   By contrast, Run~3 is designed specifically for the time domain probes, SNe and strong lensing cosmography. It just covers the DDF region and includes the time-varying objects, i.e., the ordinary AGN and SNe, the strongly lensed AGN and SNe, as well as the strongly lensed hosts for those objects.  Since the non-varying sky for the DDF regions comprises the same static scenes that were produced in Run~2, in order to save computing resources, we rendered the additional strongly lensed and time-varying objects on top of the Run~2 images, doing so before applying the electronic readout so that the instrumental effects would be simulated consistently. In the following we will use the shorthand Run~2/3 whenever the full set of DC2 image simulations is discussed.

Due to limited resources, for the Run~2/3 simulations, we employed only imSim (Run~2 simulations were carried out with imSim v0.6.2 and Run~3 with v1.0.0). The decision to use imSim for the main run was made after carefully evaluating the results from the engineering runs with PhoSim and imSim, including the results from a range of validation tests, and performance when comparing the two codes using settings that met the validation criteria.  The conclusion was that the setup chosen for Run~2/3 would permit the production of simulations that would enable the science goals outlined in \autoref{sec:DC2Universe} within the available human and computing resources.

A new workflow setup was developed based on Parsl~\citep{parsl} to allow scaling up the simulation campaigns to thousands of nodes. The workflow implementation is described in detail in \autoref{sec:parsl}. Run~2 was carried out on Cori, a KNL architecture-based system at NERSC and on grid resources in the UK and France; Run~3 was carried out on Theta. Overall, Run~2/3 generated just under 100TB of simulated image data.

\subsection{The Image Processing}

Finally, in the fourth step, the data is processed with the LSST Science Pipelines. For all three runs, the processing was carried out at the Centre De Calcul -- Institut National de Physique Nucl\'eaire et de Physique des Particules du CNRS (CC-IN2P3) using an SRS-based workflow setup. Since Run~2 consists predominately of WFD observations, the processing of the data is limited to the analysis of the coadded images. The results from coadd processing comprise the catalog outputs needed by the weak lensing, large-scale structure, and clusters dark energy probes. By contrast, for the Run~3 data, we will focus on the difference image analysis (DIA) processing\footnote{See \https{project.lsst.org/meetings/lsst2019/content/difference-image-analysis-dia-parallel-workshop} for a discussion of DIA processing of Rubin Observatory data.} since the purpose of Run~3 is the simulation of the time varying objects that are needed by the supernova and strong lensing probes.

We could, in principle, perform the coadd and DIA processing for both Run~2 and Run~3, but Run~2 lacks those time varying objects, thereby making its DIA processing of limited value, and the much greater numbers of visits for the DDF (roughly 2 orders of magnitude more visits than for non-DDF regions) make it too computationally costly to justify the full coadd processing of the data in the vicinity of the DDF for either Run~2 or Run~3. Specifically, for the Run~2 data, we exclude from the coadd processing tracts and patches\footnote{``Tracts'' and ``patches'' are regions of the sky defined for the image processing pipeline.  They are described in \autoref{sec:processing}.} that enclose regions with $> 4000$\,seconds exposure time in the $i$-band.  We note that the DIA processing is the subject of ongoing work and will be presented in a future paper.

During the image processing, intermediate and final data products were generated at the scale of 1PB.
Approximately 80\% of those data products consist of calibrated single-visit exposures, versions of those exposures that have been resampled onto a common pixelization on the sky, and coadded images in each band, which were generated from the resampled visit-level frames.
The final object catalogs added up to less than 2.5TB of data. 
The image processing for all three runs is described in detail in \autoref{sec:processing}.

In the following we provide an extensive description of the modeling approaches, codes and workflows used in each of the four key steps of survey generation as well as of the resulting data products.

\section{Modeling the DC2 Universe}
\label{sec:extra}

In this section we describe the first and the second step of our end-to-end workflow, i.e., the generation of the extragalactic catalog, cosmoDC2, and the additional components of the Universe that are included in the instance catalogs. We divide the description into three parts -- the static components of the DC2 Universe, the variable components, and the local DC2 Universe. All three parts are combined via the CatSim framework to create the input to the image simulations.

\subsection{The Static DC2 Universe}

The cosmoDC2 extragalactic catalog \citep{korytov} is based on the gravity-only Outer Rim simulation~\citep{2019ApJS..245...16H}, which evolved more than a trillion particles in a ($4.225$~Gpc)$^3$ volume. CosmoDC2 covers $440$~deg$^2$ of sky area to a redshift of $z=3$ and is complete to a magnitude depth of 28 in the LSST $r$-band. The sky area of cosmoDC2, which is delivered in HEALPix\footnote{\https{http://healpix.sourceforge.net}} (Hierarchical Equal Area iso-Latitude Pixelization) format~\citep{Gorski_2005},  was chosen so that the predefined image area would be covered. Hence it is slightly larger than DC2 in order to account for edge effects. The catalog also contains many fainter galaxies to magnitude depths of $\approx 33$ for use in weak lensing and blending studies. Faint galaxies with $r$-band magnitudes $> 29$ are removed from the image simulations to reduce the number of objects that need to be rendered (see \autoref{sec:image_input}).

The catalog is produced by means of a new two-step hybrid method that combines empirical modeling with the results from semi-analytic model (SAM) simulations. In this approach, the Outer Rim halo lightcone is populated with galaxies according to an empirical model that has been tuned to observational data such that the distributions of a restricted set of fundamental galaxy properties consisting of positions, stellar masses and star-formation rates are in good agreement with a variety of observations \citep{korytov, behroozi_etal18}. Additional modeling for the distributions of LSST $r$-band rest-frame magnitude and $g-r$ and $r-i$ colors is also done in this step. These properties are not sufficient for performing the image simulations, most notably since observer-frame magnitudes have not yet been specified. In order to provide a full complement of galaxy properties, we invoke the second step in the hybrid approach. The galaxies from the empirical model are matched with those from the SAM by performing a KDE-Tree match on the rest-frame magnitude and colors. These matched SAM galaxies now provide all the required galaxy properties including morphology, SEDs, and broadband colors.  The properties are self-consistent, incorporate the highly nonlinear relationships that are built into the SAM and capture some of the complexity inherent in the real Universe. In our hybrid method, the SAM galaxies function as a galaxy library from which to draw a suitable subset of galaxies that match observations and have the complex ensemble of properties that are required by DC2. The method is predicated on the assumption that the properties that have been tuned in the empirical model are sufficiently correlated with the other properties obtained from the SAM library to ensure that the latter will also be realistically distributed.

The empirical model used for cosmoDC2 is based on UniverseMachine \citep{behroozi_etal18}, augmented with additional rest-frame magnitude and color modeling. This model is applied to Outer Rim halos to populate its halo lightcone with galaxies. Then, as described above, these galaxies are matched to those that have been obtained by running the Galacticus SAM \citep{benson_2010b} on the small companion simulation, AlphaQ, which was also used for protoDC2 (see \autoref{sec:exgal}).  The Galacticus model provides the remaining LSST rest-frame magnitudes not obtained from the empirical model, as well as LSST observer-frame magnitudes,\footnote{Total throughputs for six LSST filters were obtained from \https{github.com/lsst/throughputs/releases/tag/1.4}} 
SDSS rest-frame and observer-frame magnitudes, and coarse-grained SEDs obtained from a set of top-hat filters spanning the wavelength range from 100 nm to 2 $\mu$m. The magnitudes are available separately for disk and bulge components, and with and without host-galaxy extinction corrections and emission-line corrections.  Galaxy shapes, orientations and sizes are obtained by additional empirical modeling based on properties obtained from the matched SAM galaxies (e.g., the bulge-to-total ratio). The shapes and sizes are provided separately for the disk and bulge components and the light profiles are assumed to be $n=1$ Sersic and $n=4$ Sersic for the disk and bulge components, respectively. Additional information on the implementation of the modeling of SEDs and morphologies is given in \cite{korytov}.

The cosmoDC2 catalog also contains host halo information derived from the Outer Rim halo catalog and weak lensing distortions and deflections calculated from the Outer Rim particle lightcone catalog. Weak lensing quantities are derived from the particle lightcone by projecting particles onto a series of mass sheets and performing a full ray-tracing calculation to produce weak lensing maps. The redshift shells have a median width of approximately 114~Mpc. The ray-tracing calculation involves following photon paths backward in time from an observer grid to the source planes, with deflections based on the surface density of particles at each mass sheet. The shear and convergence values for each galaxy are obtained from the source maps by first shifting the galaxy to its observed position and then interpolating the source map to the observed position. Many more details are provided in \cite{korytov}.

The catalog is delivered as a set of
HEALPix pixel files with resolution parameter Nside=32, split into redshift ranges $0<z<1$, $1<z<2$, and $2<z<3$, covering the image simulation area.
The reason for this  pixelization is to ensure that the memory footprint for generating the instance catalogs does not exceed the available memory for running the DC2 image simulation pipeline. In the worst-case scenario, 4 pixels could, in principle, meet in the FOV. Since the file for one Nside=8 HEALPix pixel barely fits into the available memory, the choice of Nside=32 offers a generous safety margin.

The realism of the extragalactic catalog has been assessed by applying a series of validation tests  using the automated DESCQA validation framework~\citep{2018ApJS..234...36M}. This procedure invokes a set of tests and uses a series of validation criteria developed by the LSST DESC. The tests and criteria are designed to enable checks that the distributions of various galaxy properties are sufficiently realistic to enable the science goals for DC2. For example, several tests compare the distributions of the number density of galaxies as a function of magnitude, redshift and color with selected sets of  observational data. Meeting the validation criteria associated with these tests is important to guarantee that cosmoDC2 will be useful for assessing the performance of algorithms including photometric redshift calibration and galaxy deblending. Other validation tests, called readiness tests, check the distributions of basic galaxy properties to make sure that selected summary statistics have sensible values and that the galaxy properties do not contain egregious outliers that would be problematic for the image simulation code. Once the catalog passes the validation and readiness tests, it is released for the next step of generating the instance catalogs. More details and validation results will be given in a forthcoming paper.

\subsubsection{Random Walk Model for Galaxy Morphologies}
\label{sec:knots}

As described above, the extragalactic catalog represents each galaxy as a combination of bulge and disk S\'ersic profiles. However, such parametric light profiles may prove too simplistic to thoroughly test crucial elements of the pipeline, including deblending and shape measurement. In an effort to address these concerns, at the level of the instance catalog generation (described in \autoref{sec:image_input}) we increase the complexity of the galaxy models by adding a \textit{random walk} component \citep{2008MNRAS.383..113Z, 2015JCAP...01..024Z, 2017ApJ...841...24S} to the bulge and disk representation. This new component comprises a number of point sources of equal flux, with the same SED as the disk component, and with positions drawn from a Gaussian distribution matching the size and ellipticity of the disk. Finally, the flux allocated to the random walk is subtracted from the disk component to preserve the original flux. These point sources can be thought of as a simple model to emulate knots of star formation in the disks and produce non-trivial light profiles.

The only free parameters with this approach are $N$, the number of point sources, and $f_r$, the ratio of total disk flux allocated to the random walk component. To build a model for these parameters as a function of galaxy properties, we perform three-component fits (bulge, disk, and random walk) on observational data from the HST/ACS COSMOS survey \citep{2007ApJS..172...38S,2007ApJS..172....1S}. This dataset \citep{Mandelbaum2012}, originally compiled for the GREAT3 challenge \citep{2014ApJS..212....5M}, is composed of postage stamps of individual galaxies, along with the parameters of a bulge+disk parametric fits. Starting from these parametric models, we allow for an extra set of point sources to improve the fit to the actual postage stamps. This procedure yields for each COSMOS galaxy the $(N, f_r)$ parameters of a three-component fit, which we can relate to other galaxy parameters such as the size and flux of both bulge and disk components. Using a simple Mixture Density Network \citep{Bishop1994}, we build from this dataset a probabilistic model for the joint distribution of $(N, f_r)$ given galaxy parameters provided in the extragalactic catalog. In addition, to ensure that large, well-resolved galaxies do not exhibit non-physical isolated point sources, we impose a size-dependent bound on $f_r$ so that most of the flux remains in the disk for larger galaxies. This model is then used to conditionally sample plausible random walk parameters for each DC2 galaxy.

\begin{figure}
    \centering
    \includegraphics[width=\columnwidth]{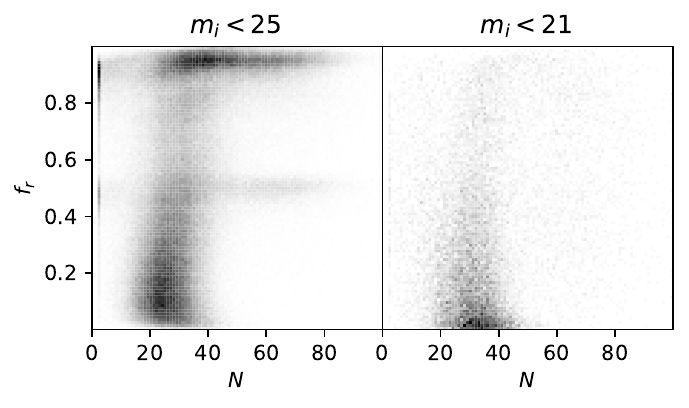}
    \caption{Joint distribution of number of point sources $N$ and random walk-to-total disk flux ratio $f_r$ of the bulge + disk + random walk model, for different cuts in $i$-band magnitude.}
    \label{fig:knots}
\end{figure}

\autoref{fig:knots} illustrates how the model yields distinct distributions of $(N, f_r)$ for different galaxy populations (in this case, two different cuts in $i$-band magnitude $m_i <25$ and $m_i < 21$). One feature to note is that a significant fraction of fainter galaxies are found to have $f_r$ close to 1, which corresponds to allocating all of the original disk flux to the random walk component. This indicates that for these typically smaller galaxies, a bulge + random walk model provides a better fit to COSMOS galaxies than a bulge+disk model. For brighter galaxies, however, the random walk component remains subdominant, as point sources are inefficient at modeling an extended disk. The higher concentration observed at low $N$ in the $m_i < 25$ plot is simply caused by the hard positivity constraint $N>0$ on the number of point sources. 

\subsection{The Time Domain DC2 Universe}
\label{sec:variable}

Next we describe our modeling approaches for the variable DC2 Universe. SNe were inserted in both the WFD and DDF region; all other components -- strongly lensed SNe, AGN, strongly lensed AGN, and strongly lensed galaxies -- are only sprinkled into the DDF region. In the following we provide details about the modeling approach and the implementation.

\subsubsection{Supernovae}
\label{sec:supernovae}
Type Ia SNe (\sneia) were inserted in the redshift range $z<1.4$ with a redshift dependent volumetric rate of $r_v(z) = 2.5\times 10^{-5} (1 + z)^{1.5} \rm{Mpc}^{-3} \rm{yr}^{-1}$ which is compatible with the observed rate (e.g. ~\citealt{2010ApJ...713.1026D}) over the WFD region, and in the DDF region at about twice the observed rate\footnote{The factor of two was chosen arbitrarily. The goal was to provide a bigger sample, but not a sample so large that it could interfere with static science processing in a way that would be unrealistic in the real universe.}. The time evolution of the SNe brightness is described by a family of time series of spectra which are slightly modified versions of the SALT2 model~\citep{2007A&A...466...11G}. The modifications, which replaced small negative values of the spectra interpolated in wavelengths by zero, were necessary as image processing software uses the spectra as a probability density to sample from, and thus requires them to be positive semi-definite. Additional deviations from the SALT2 spectral surfaces, which describe the diversity of SNe not captured in the SALT2 spectral templates, were not added to these spectra. The properties of individual \sneia\ are then completely determined by the SN parameters of the SALT2 model, namely, $\{z, t_{0}, x_{0}, x_{1}, c \},$ and the description of the simulation inputs is completed by describing the prescription of assignment of these parameters to the SNe, and their relation to the environment, as described by the properties of the host galaxy.

The rate determines the number of \sneia\ in any redshift bin (chosen to be of width $0.05$) in the DC2 survey region. To assign them further properties, we first decided on their environment.  All \sneia\ at  $z> 1.0$ were chosen to be hostless and do not trace the large-scale structure. Even at $z < 1.0$, $10 \%$  of the SNe~Ia were randomly selected to be hostless. This choice was made to provide a control sample free from the potential problems of image subtraction with a host galaxy, while the remaining $90 \%$ of the \sneia\ were matched to cosmoDC2 host galaxies in the specified redshift bin, through a prescription described below. The redshifts of the SNe were thus assigned to be a sample from the cosmological volume (for the hostless redshifts) or the specific redshift of the host galaxy, even though the procedure respects the distribution of the redshifts according to cosmological volume to about a bin width of $\Delta z = 0.05.$ While it is known that \sneia\ of higher stretch and redder colors tend to explode more frequently in more massive, high-metallicity galaxies, such correlations were ignored and these parameters were drawn from a global distribution of $x_1$ and $c$ that was normal and centered around $0$ with a standard deviation of $1.0$ and $0.1$ respectively. An intrinsic dispersion was added in the form of an absolute magnitude distribution in the rest-frame Bessell $B$ band taken to be a normal distribution centered on $-19.3,$ with a standard deviation of $0.15$ mag. This corresponds to a reasonably realistic sample of \sneia\ with correct amounts of cosmological dimming applied for the redshifts of host galaxies. Lensing magnification is not applied to the \sneia.

\begin{figure}[t]
\centering
\includegraphics[width=0.95\columnwidth]{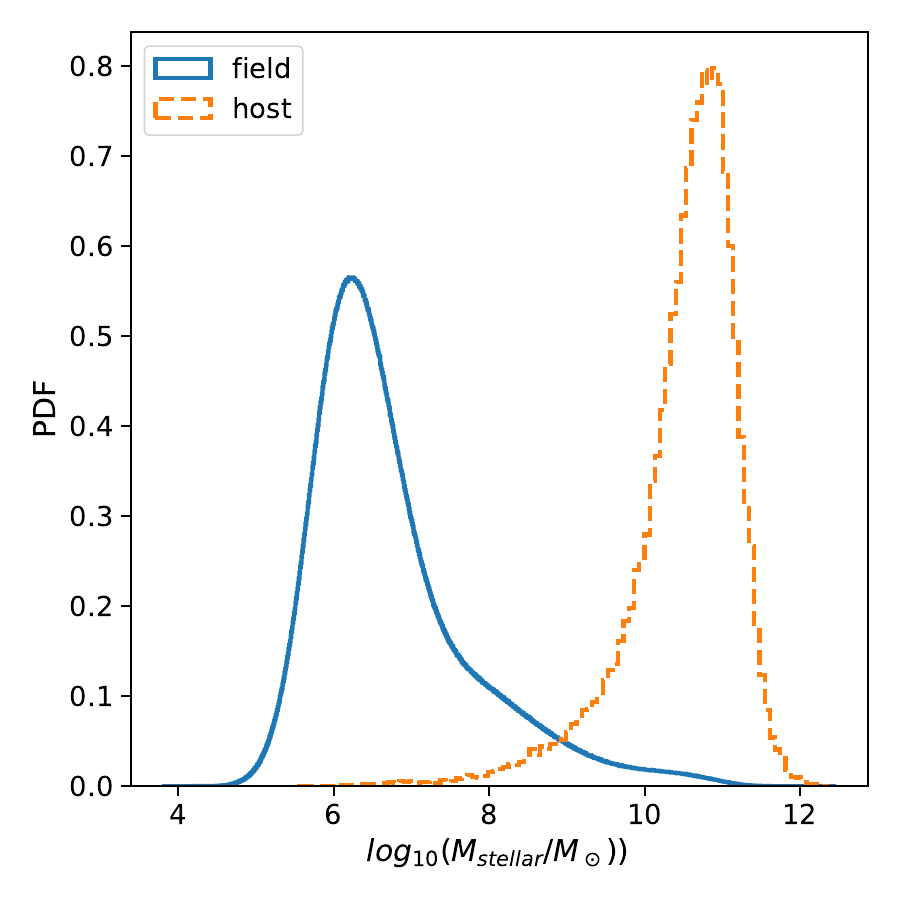}
\caption{Stellar mass distribution of \sneia\ host galaxies (orange) in the redshift range $z < 1.0$ and galaxies in the cosmoDC2 catalog (blue) showing that the SN host distribution peaks at a mass scale far greater than those of the most abundant galaxies in cosmoDC2.}
\label{fig:snhost_mass}
\end{figure}

\begin{figure}[t]
\centering
\includegraphics[width=0.95\columnwidth]{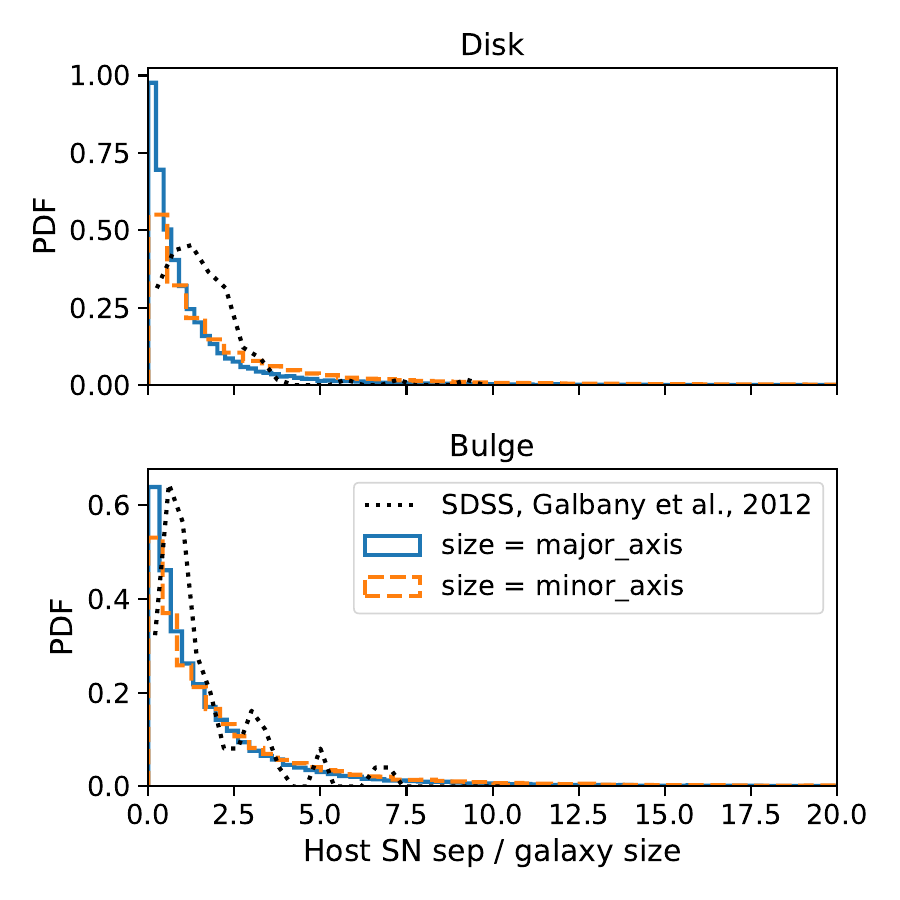}
\caption{Distribution of distances of SN from host relative to host galaxy size in the redshift range $z < 1.$ We consider different measures of the galaxy size by using the major and minor axes of the component used to host the supernova. The supernovae hosted by the disk are shown in the top panel, while the supernovae hosted in the bulge are shown in the bottom panel. For a comparison to observations from SDSS, we also show an approximate distribution (dotted lines) estimated by eye from a histogram of \sneia\ distances from their host galaxies following ~\citealt{2012ApJ...755..125G}. While the SDSS result includes selection effects, particularly at the core, there is good agreement between these results.}
\label{fig:sn_distances_2}
\end{figure}

\begin{figure}[t]
\centering
\includegraphics[width=0.9\columnwidth]{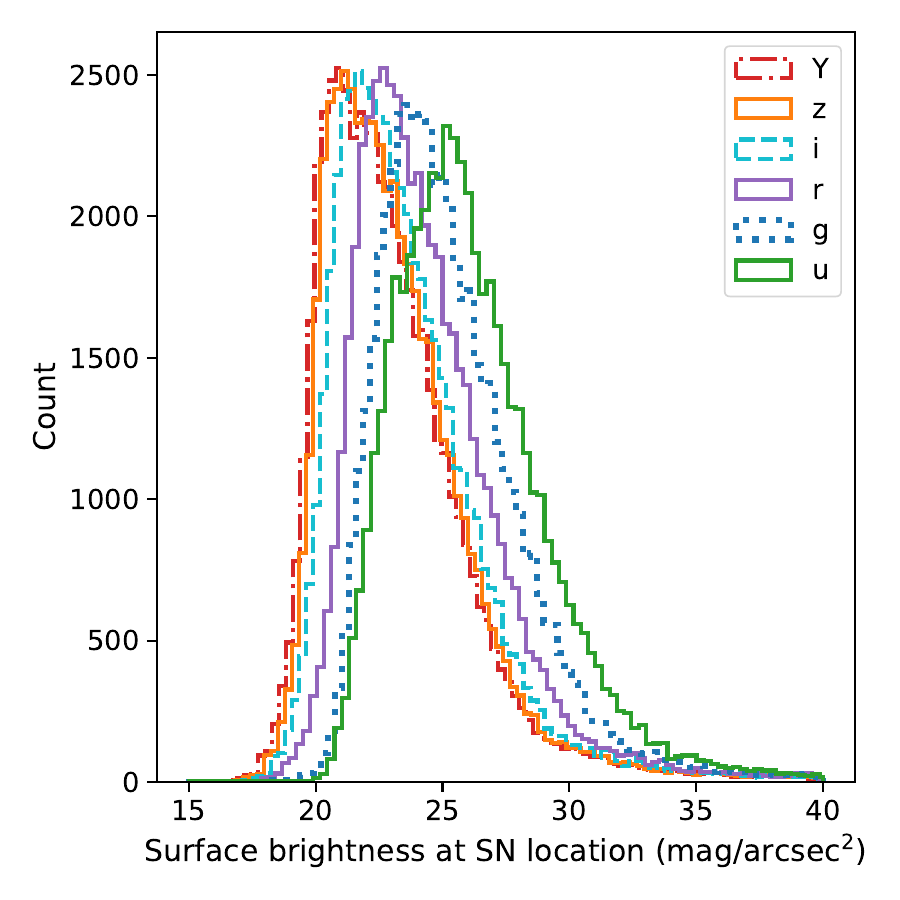}
\caption{Distribution of the surface brightness of the host galaxy at the location of the inserted supernovae.}
\label{fig:surface_brightness}
\end{figure}

The probability of occurrence of \sneia\ has been observed to be roughly proportional to the stellar mass of the host galaxy, while the stellar mass and other properties of the host galaxy such as its star formation rate are also correlated to the abundance of \sneia\ per unit stellar mass. Ignoring the latter, we chose the host galaxies in the redshift bin, such that the probability of occupation of a galaxy is proportional to its stellar mass. As there are many more host galaxies of lower stellar mass, and the probability of occurrence of very high mass galaxies is exponentially lower, this leads to a stellar mass distribution of the host galaxies shown in \autoref{fig:snhost_mass}. Finally, an important aspect of the planned analysis is related to the position of the SNe with respect to the host galaxy, and the surface brightness of the galaxy in the pixels around the position. Hence, for the SNe that are hosted by the galaxy, we developed a prescription so that its position traces the light in the galaxy. Since cosmoDC2 galaxies have bulges, disks, or both, we first assigned the SN to one of the components randomly, assuming that the probability of a component hosting it was proportional to its stellar mass. Then, finally, we assigned a position by sampling the surface brightness profile of the hosting component (which in the cosmoDC2 catalog was by definition sersic and only had a sersic index of 1 or 4)\footnote{\https{github.com/rbiswas4/SNPop}}. This prescription results in angular distance of the hosted \sneia\ from the galaxy center described in \autoref{fig:sn_distances_2}. The figure shows the probability density function of the distance to the SNe from the centers of their host galaxies in DC2, relative to the size of the host galaxy over the entire redshift range ($z < 1$). In normalizing these distances by host galaxy size, we use of both the semi-major and semi-minor axes as estimates of the galaxy size and get very similar results. To compare with past studies of such distances of \sneia\ from galaxies on the basis of real observations (e.g. \citealt{2012ApJ...755..125G}), we estimate the probability density function by eye from Figure 2 of \citealt{2012ApJ...755..125G} and overplot on our results. In doing so, we have have identified  `spiral` and `elliptical` hosts in \citealt{2012ApJ...755..125G} with disks and bulges in DC2 simulations, respectively. We also ignore that their normalization used a slightly different estimate of galaxy size.  It should be noted that while our results apply to all simulated supernovae with hosts, their results apply to detected and spectroscopically  identified \sneia\ that passed selection criteria for a light curve analysis sample and therefore include selection effects. In particular, this is likely to miss supernovae near the galaxy cores, partly because these regions of the galaxy have higher surface brightness which is a problem for difference imaging algorithms, but more importantly, because detections near the core were not followed up as they were likely to be AGN or tidal disruption events. This is a likely explanation for the difference near the core. In general, the simulated distances show good agreement with the observations from SDSS.   
It also leads to the distribution of surface brightness at the location of the SNe (as predicted by the truth catalog) shown in~\autoref{fig:surface_brightness}.

As these simulations will largely be used for studies of the image processing pipeline, it is important to ensure that there is sufficient diversity in properties of the SNe for such image processing needs. Some of the properties of SNe that are of interest are the host brightness (or its proxy as stellar mass), the (angular) distances to the hosts and the surface brightness of the host galaxy at the location of the SN.
The code for generating these properties is publicly available\footnote{\https{github.com/LSSTDESC/SN_image_catalog_validation}}.

\subsubsection{Strongly Lensed Supernovae} \label{sec:lensed_sne}

Strongly lensed \sneia~are added (``sprinkled'') into the DDF region with the Strong Lensing Sprinkler (\text{SLSprinkler}) code\footnote{\https{github.com/lsstdesc/slsprinkler}}. The lensing systems come from the \citet{2019ApJS..243....6G} catalog. \text{SLSprinkler} only adds components to the simulations and does not remove any cosmoDC2 galaxies. To do this, the code selects large, elliptical cosmoDC2 galaxies as potential foreground deflector galaxies. \text{SLSprinkler} then matches these galaxies to the lensing systems from the \citet{2019ApJS..243....6G} catalog by selecting systems where the deflector galaxy matches the velocity dispersion and redshift of the candidate galaxy, to better than 0.03 within 0.03 dex for each property. The cosmoDC2 catalog does not provide the velocity dispersion, however. In order to obtain these values, we use the Fundamental Plane (FP) relation of \cite{hyde+bernardi2009} with its $r$-band apparent magnitude and half-light radius. This matching defines a set of potential lens galaxies, from which \text{SLSprinkler} randomly selects a set of the cosmoDC2 galaxies with at least one matching system so that we end up with 1,129 lensed SNe systems in the DDF region.

For each deflector galaxy, \text{SLSprinkler} then randomly selects one of the lensing systems that matched to that galaxy and uses the new geometry of the lens to update image positions, time delays, and magnifications from the original \citet{2019ApJS..243....6G} catalog. To compute the new lensing observables, \text{SLSprinkler} uses the software package lenstronomy\footnote{\https{github.com/sibirrer/lenstronomy}} \citep{lenstronomy}. \text{SLSprinkler} also assigns host galaxies to the lensing systems by matching cosmoDC2 galaxies from outside the DDF to the redshift and half-light radius of the lens catalog entries, to a matching tolerance of 0.05 in dex for each property.

Finally, at each visit, \text{SLSprinkler} queries the SEDs for each of the various time-delayed images of the SNIa. If the SED for an image has non-zero flux at 500 nm at the epoch of the visit, then it is added to the instance catalog for that visit. At all epochs, the foreground deflector galaxy and images of the SNIa host galaxy are added to the catalog.

\subsubsection{Active Galactic Nuclei}
The time variability of AGN in the DDF region is modeled by assuming that the light curves in each of the LSST bandpasses can be represented as damped random walks added to quiescent magnitudes based on synthetic photometry using the template AGN spectra from \citet{2001AJ....122..549V}.  This model was originally fit to 9,000 spectroscopically verified quasars from SDSS by \citet{2010ApJ...721.1014M}.  It is described by a characteristic timescale and a structure function in each of the bands.  Equation~(7) and Table~1 in \citet{2010ApJ...721.1014M} give a phenomenological fit to these parameters based on black hole mass, redshift, and the absolute $i$-band magnitude of the AGN. The black hole mass and Eddington ratio are provided by the cosmoDC2 extragalactic catalog (see Section 5.4.2 of \citealt{korytov} for the implementation details in cosmoDC2). We use Figure~15 from \citet{2010ApJ...721.1014M} to construct a linear mapping from black hole mass and Eddington ratio to absolute $i$-band magnitude, allowing us to use Equation~(7) from \citet{2010ApJ...721.1014M} to assign the variability parameters to the AGN.  We impose a cut that only black holes with masses greater than $10^7$~M$_{\odot}$ and apparent $i$-band magnitudes less than 30 exhibit AGN activity.  Using this information, we populate a database of AGN variability parameters, which are added to the cosmoDC2 catalog before passing data into the image simulator.

The procedure described above adds an AGN to every sufficiently bright galaxy with a sufficiently massive black hole at its center, and results in a universe with too many bright AGN.
In order to obtain realistic number densities comparable to what is observed by SDSS for bright quasi-stellar objects \citep{Richards_et_al}, it is necessary to impose a duty cycle on the AGN. We developed an effective--duty-cycle model which assigns a probability that each AGN is ``on'' or ``off''  based on the stellar mass, $g-r$ color, $r-i$ color and redshift of its host galaxy. The model parameters were tuned by matching the number of active AGN to observational data.
Once the duty-cycle model is imposed on the simulated AGN, we obtain AGN number densities as a function of the magnitude of the host galaxy plus mean AGN contribution in both $g$- and $i$-band that broadly agree with the distributions shown in \citet{Richards_et_al}. 

We model the fraction of galaxies whose AGN is ``on" via the fraction $\tagn,$ defined to have a dependence on both redshift, $z,$ as well as on rest-frame colors $(g-r)$ and $(r-i).$ Galaxies at the reddest end of the two-dimensional color-color space are assigned a constant value of $\tagn(\dgv, z)=\tagn^{\rm rs}=10^{-4}$ at all redshifts; galaxies at the bluest end are assigned $\tagn(\dgv, z)=\tagn^{\rm ms}_{\rm lo-z}=0.025$ at low redshift, with $\tagn(\dgv, z)=\tagn^{\rm ms}_{\rm hi-z}=0.1$ at high redshift, where both the redshift- and color-dependent behavior is regulated by a sigmoid function:

\begin{equation}
\label{eq:duty_cycle_model}
\tagn(\dgv, z)=\tagn^{\rm rs} + \left[\tagn^{\rm rs}-\tagn^{\rm ms}(z)\right]/\left[1 + \exp\left(-k_{\dgv}\cdot\dgv\right)\right],
\end{equation}
with
\begin{equation}
\tagn^{\rm ms}(z)=\tagn^{\rm ms}_{\rm lo-z} + \left(\tagn^{\rm ms}_{\rm hi-z}-\tagn^{\rm ms}_{\rm lo-z}\right)/\left[1 + \exp\left(-k_z\cdot(z-z_0)\right)\right],
\end{equation}
where $k_z=k_{\dgv}=4$ and $z_0=0.5.$

The quantity $\dgv$ is defined in terms of an ``eigencolor", $c_{\rm eigen},$ that is a linear combination of $(g-r)_{\rm rest}$ and $(r-i)_{\rm rest}:$

$$c_{\rm eigen}\equiv0.93\cdot\left(g-r\right)_{\rm rest} + 0.37\cdot\left(r-i\right)_{\rm rest}.$$ This choice is motivated in part by recent results characterizing the dependence of optical broadband color upon star formation history, which indicate that this linear combination of $(g-r)$ and $(r-i)$ captures most of the information available in these bands about the galaxy's underlying star formation history \citep{jcm_aph20}. 
We compute the value of $c_{\rm eigen}$ for every galaxy,  and define $\dgv$ as the difference between $c_{\rm eigen}$ and the stellar mass-dependent location of the green valley in eigencolor-space:
$$c_{\rm gv}(M_{\star}) \equiv 0.8 + 0.1\cdot\left(\log_{10}M_{\star}-10\right).$$
For every galaxy, the distance from the green valley is then given by $\dgv\equiv c_{\rm eigen}-c_{\rm gv},$ which is the quantity used in \autoref{eq:duty_cycle_model} to define $\tagn(\dgv, z),$ the probability that the galaxy hosts an AGN.

\begin{figure}[t]
\centering
\includegraphics[width=0.95\columnwidth]{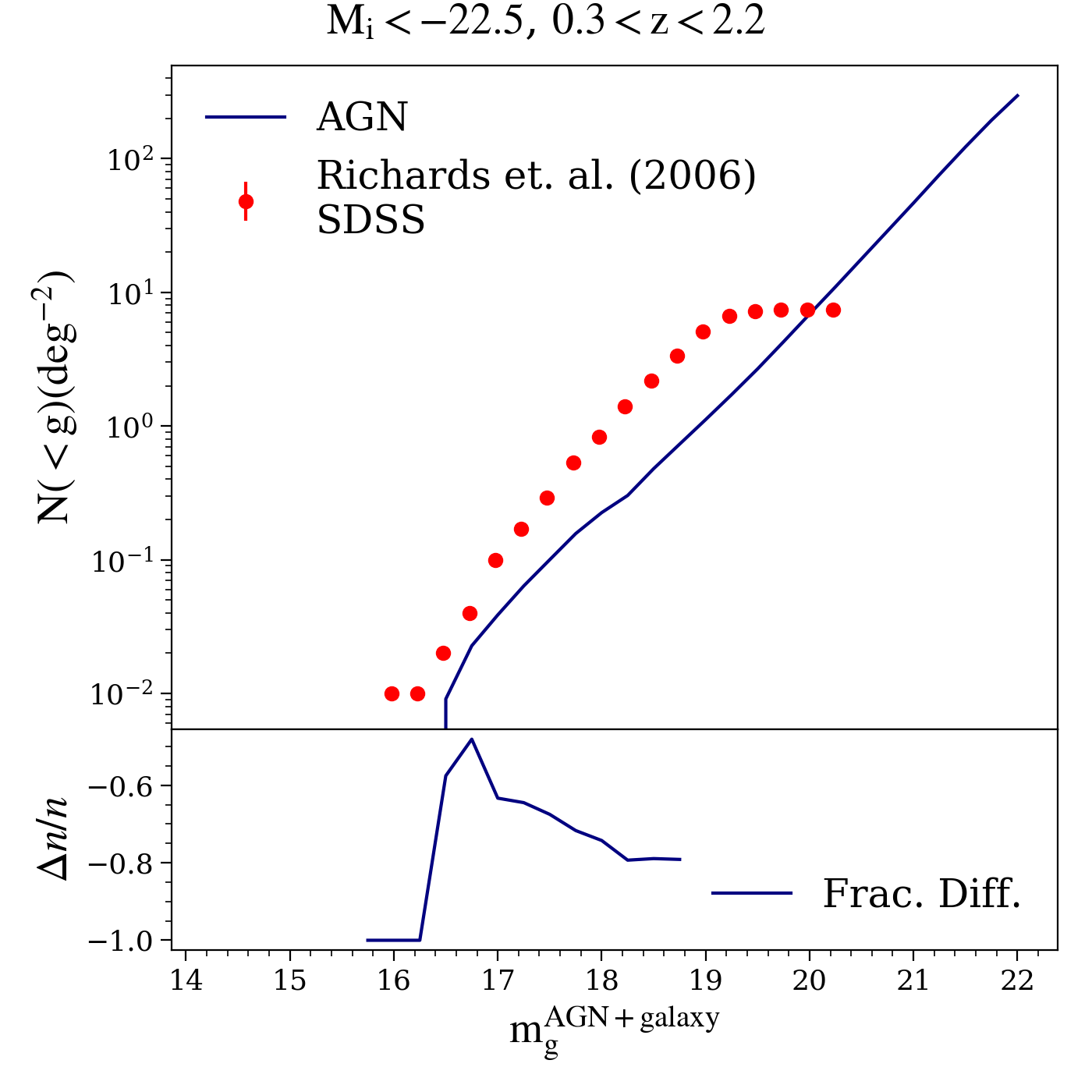}
\includegraphics[width=0.95\columnwidth]{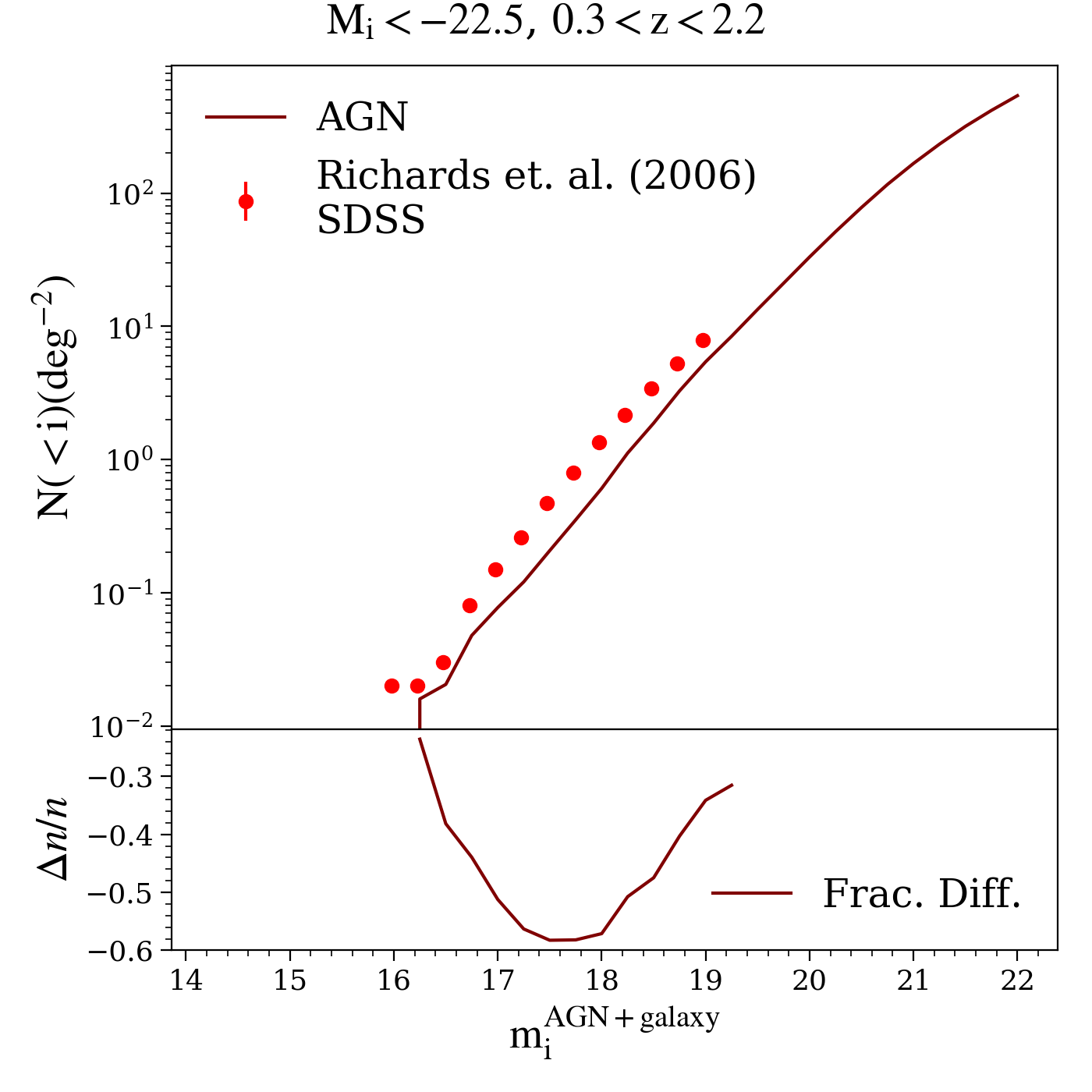}
\caption{Cumulative AGN number densities as a function of $g$-band magnitudes (upper plot) and $i$-band magnitudes (lower plot) compared to data from \citet{Richards_et_al}. The bottom panels for each plot show the fractional difference between the SDSS and DC2 data (normalized by the SDSS data). The SDSS $g$-band data (upper plot) is not complete for $m_{i}^{\rm{AGN + galaxy}} \gtrsim 19$, so we do not show the fractional difference above this point.
}
\label{fig:agn_dndmag_i}
\end{figure}

In~\autoref{fig:agn_dndmag_i} we show the cumulative number densities for the active simulated AGN, as selected by the duty-cycle model, as a function of the observed $g$- and $i$-band magnitudes. In order to generate this figure, we imposed selection cuts on the simulated AGN to match those used to select the observed SDSS data. The magnitudes shown are obtained from the sum of the fluxes of the galaxy and the average magnitude of the AGN component, with no cut imposed on the fraction of the flux coming from the AGN component. Whilst the agreement is not perfect, the complexities of tuning and improving the duty-cycle model preclude the use of a more detailed approach. The present level of fidelity of the simulation allows us to use the results to explore techniques for data analysis of strongly lensed AGN and associated time variability.

\subsubsection{Strongly Lensed AGN} \label{sec:lensed_agn}

Strongly lensed AGN are inserted into the instance catalogs in the DDF region using the LSST DESC \text{SLSprinkler} code, which we also use for the lensed SNe (see~\autoref{sec:supernovae}). The deflector galaxies are assumed to follow the Singular Isothermal Ellipsoid (SIE) mass profile in an external shear field. The \cite{OM10} (OM10) catalog provides the base mock catalog of strongly lensed AGN describing plausible distributions in deflector and source properties. Given an OM10 system, \text{SLSprinkler} matches the OM10 SIE mass to the elliptical S\'{e}rsic stellar profile of a cosmoDC2 galaxy based on two criteria: the deflector redshift and the velocity dispersion of the cosmoDC2 galaxy as predicted by the FP relation of \cite{hyde+bernardi2009} with the cosmoDC2 $r$-band apparent magnitude and half-light radius. The external convergence was assumed to be zero at the position of the source and the external shear was taken from the OM10 catalog, regardless of the matched DC2 environment. Each OM10 AGN is also assigned a cosmoDC2 galaxy by matching to a cosmoDC2 galaxy outside the DDF that hosts an AGN with similar $i$-band magnitudes and redshift (to a matching tolerance of 0.05 in dex for both properties). The final number of lensed AGN systems (where the AGN, its host galaxy, and the deflector galaxy comprise one system) matched this way was 1,056.

Because the \text{SLSprinkler} matching introduces slight offsets from the original OM10 lens geometry and redshifts, some lensing observables in the OM10 catalog must be updated. We thus recompute the lensed image positions, time delays, and magnifications, and populate the truth catalogs for the lensed host galaxies and lensed AGN with this information. The relevant code uses lenstronomy and lives within \text{SLSprinkler}. To enable lens environment investigations, the truth catalogs additionally include the external convergence and shear in the OM10 catalog as well as the ray-traced weak lensing shear and convergence at the cosmoDC2 source positions. We use the same variability model for the lensed AGN as for the unlensed AGN, and each lensed AGN image follows the same variability appropriately offset in time by the lensing time delay.

\subsubsection{Strongly Lensed Galaxies} \label{sec:lensed_hosts}
\begin{figure}[t]
\centering
\includegraphics[width=\columnwidth]{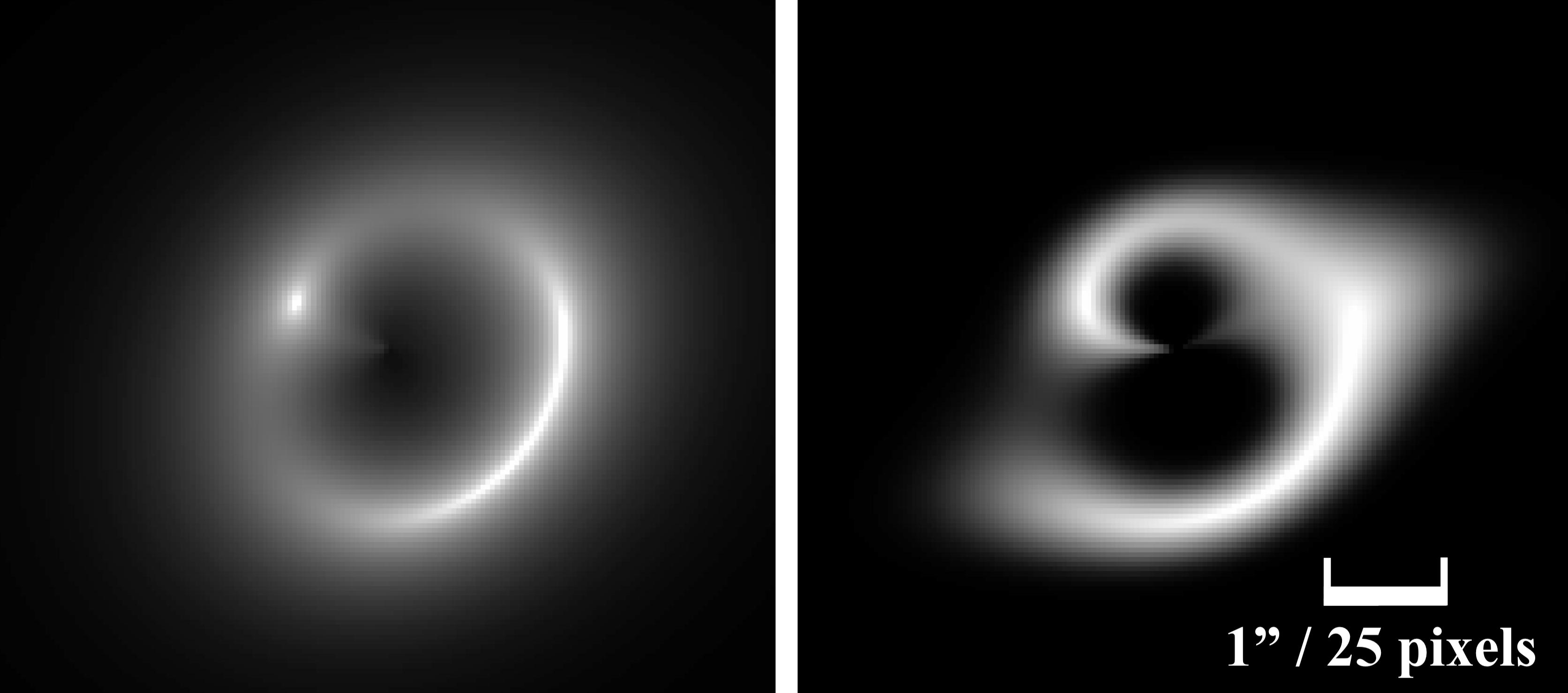}
\caption{High-resolution (250x250 pixels, 0.04"/pixel) postage stamps of the lensed AGN host galaxy bulge (left) and disk (right) created with the \text{SLSprinkler} code and used as input for imSim.
}
\label{fig:lensed_host_postage_stamp}
\end{figure}

As mentioned in \autoref{sec:lensed_sne} and \autoref{sec:lensed_agn} we include strongly lensed AGN and SNe in the DDF region along with the host galaxies for the corresponding point sources. However, unlike the lensed AGN and SNe, the host galaxies for these objects are extended sources that, in their lensed form, require special treatment to be added to image simulations. Recall that cosmoDC2 models galaxies as a Sersic bulge plus disk. \text{SLSprinkler} thus takes in the geometry and relevant lensing parameters of the lensing systems and generates two postage stamps for each host galaxy -- one for the lensed bulge component and the other for the lensed disk component. The imSim software can then accept the postage stamps as input and render the lensing systems in the images. \autoref{fig:lensed_host_postage_stamp} shows the two postage stamps corresponding to the lensed bulge and lensed disk for an example host galaxy.

\subsection{The Local DC2 Universe: The Milky Way}

Stars are drawn from the Galfast model of \citet{2008ApJ...673..864J}, which is based on the densities and colors of sources in the SDSS. Stellar catalogs were modeled to $r>27$ by extrapolating the luminosity functions derived from the SDSS data to fainter luminosities. Comparisons of the predicted stellar number counts at $r=27$ with those from the Besan\c{c}on model \citep{robin2003} show agreement within 80\% at all Galactic latitudes, $l$, and within 20\% at $l<20\deg$. Metallicities, proper motions, and parallaxes are assigned from the Fe/H and kinematic models of  \cite{bond}. SEDs are fit to the simulated SDSS photometry using \cite{kurucz} for main-sequence stars and giants, \cite{Bergeron} for white dwarfs, and a combination of spectral models and SDSS spectra for the M, L, and T dwarfs \citep[e.g.][]{cushing05,bochanski07,burrows06,kowalski10}. These SEDs are used to derive the LSST photometric magnitudes for each star. Galactic reddening is simulated using the three-dimensional Galactic model from \cite{2005AJ....130..659A}, which is renormalized to match the values in the \cite{schlegel} dust maps at a fiducial distance of 100kpc. This ensures that the stellar and extragalactic source catalogs share a common extinction model. All sources, their photometry, and associated SEDs  are stored within a Microsoft SQLServer database.\footnote{Access to the database is available upon request.} 

Variability is incorporated within the stellar catalogs with approximately 10\% of the stellar sources expected to be variable at a level detectable by LSST \citep{2014ApJ...796...53R}. We adopt two separate mechanisms for assigning variability. For sources with well-defined periodic variability (e.g., RR Lyrae and Cepheids) a light curve is generated from an amplitude, period, and a phase. For non-periodic variables  (e.g., cataclysmic variables, flaring M-dwarfs, and microlensing events) or transient sources  the period of the light curve is set to $>10$~years such that the sources will not repeat within the timespan of the simulated LSST observations. For stars that have not been assigned a definitive variability class, we model variability by assigning light curves taken from the Kepler Q17 data release \citep{2016ksci.rept....3T} where we represent each light curve by 51 Fourier components (derived from an iterative Lomb-Scargle algorithm).  Frequencies, amplitudes, and phase offsets associated with these Fourier components are used to reconstitute the light curves as needed.  Kepler light curves are assigned to the stars in our simulated Milky Way by taking the quiescent properties of the Kepler sources from the Kepler Input Catalog \citep{2011AJ....142..112B} and associating each simulated star with its nearest Kepler neighbor in ($g$-$r$), $r$ color-magnitude space. For queries that contain time constraints, the magnitude of the source is modified based on the properties of the light curve (the current implementation only allows for monochromatic variations in the fluxes).

\section{Image Simulations}
\label{sec:image_sims}

In this section we describe the third step in our end-to-end workflow, the image simulations that underlie DC2. In particular, we describe the image simulation tool, the workflow implementation to carry out the image simulations, and our quality assurance approach for the images. As discussed in some detail in \autoref{sec:workflow}, for the engineering run, Run~1, we employed two image simulation tools, PhoSim and imSim. Due to resource limitations for Runs~2 and~3, the science-grade runs, we only use imSim. In the following, we focus therefore on the imSim approach, including workflow set-ups and validation. 

\subsection{imSim Introduction}

The imSim software package produces pixel data for simulated observations of the Rubin Observatory.  The raw image data produced by imSim resemble the actual data that will be obtained from LSSTCam~\citep{2018SPIE10705E..0DR}: The LSSTCam focal plane is populated by 189 science CCDs, 8 guider CCDs, and 8 wavefront sensors, for a total of 3.2 gigapixels.  Each science CCD has approximately 16 million pixels, divided among 16 imaging segments.  Since the raw image files produced by imSim have the same data format as will the real LSSTCam data, they can therefore be processed by the LSST Science Pipelines~\citep{2017ASPC..512..279J}.
imSim takes as input a catalog of astronomical sources as described in \autoref{sec:image_input} along with information about how their light is modified on the way to Earth including lensing and extinction information.

To produce the simulated images, imSim calls the GalSim software library~\citep{2015A&C....10..121R} for astronomical object rendering and is run in the LSST Science Pipelines and LSST Simulation Framework software environment~\citep{2014SPIE.9150E..14C}.  The LSST software libraries provide the 
telescope and hardware-specific information necessary to simulate the exposure, such as pixel coordinates on the focal plane, telescope filter characteristics, and the brightness of the sky.  Using that description of LSSTCam, imSim produces output files that simulate the pixel data after readout.

A key motivation for building imSim is to provide the capability to implement new models developed by DESC members via an open-source paradigm. This capability, together with the very flexible GalSim base, allows imSim to supply multiple models for the same effects.  As a result, varying levels of fidelity and speed can be selected for features depending on the particular need.  Some of the available models for PSFs and detector effects are simple approximations, while others include full physics treatments such as a physics-based silicon model and ray-traced stochastic atmospheres.  The execution speed of the code depends sensitively on the chosen fidelity for certain aspects of the simulation processes (e.g., fully realistic galaxy morphology vs.\  parametric models; ray-tracing through stochastic atmospheres vs.\  using parametric PSF models;  simple pixel response vs.\ the full silicon model for the LSST sensors). Additionally, more accurate, slower features in the code can be used to tune faster parametric ones.  Detailed descriptions of and references to the algorithms implemented in imSim can be found in \autoref{sec:imsim-features} and in a forthcoming paper.

\subsubsection{imSim Features}
\label{sec:imsim-features}

For the DC2 simulations, imSim used several features that are available from the GalSim package and from the Rubin Observatory software environment.  GalSim provided the physics-based silicon sensor model, the multi-layer atmospheric ray-tracing implementation as described in~\cite{2011PASP..123..596J, 2015ApJS..218...14P}, and several options for rendering galaxies, including parameterized Sersic descriptions, postage stamps of real or simulated images (\autoref{sec:lensed_hosts}), and the ability to add knots of star formation to make galaxy morphologies more realistic (\autoref{sec:knots}). imSim uses the LSST Science Pipelines~~\citep{2017ASPC..512..279J} and LSST Simulation Framework software~\citep{2014SPIE.9150E..14C} and information contained therein about the Rubin focal plane, CCD geometry, and electronics readout to simulate the final raw pixel data, including effects such as cosmic rays, and bleed trails.

For each visit that is simulated, conditions are taken from the \opsim\ database and encoded into the instance catalogs with CatSim.  These include, e.g., sky brightness and conditions of atmospheric seeing. These inputs are then used by imSim to calculate the sky background level and the source-SED specific atmospheric PSF to use for rendering objects.
To generate background photons, imSim takes the \opsim\ sky brightness as input to the LSST sky model, which is the ESO sky model with a twilight component added~\citep{2016SPIE.9910E..1AY}. 
Galactic SEDs for the rendered objects are derived from the Galacticus model as described in \autoref{sec:extra}. In order to have a fully interpolatable SED while employing a reasonably sized stored SED library, we use as a base the library of templates of SEDs used by the Rubin project that are based on~\citep{2003MNRAS.344.1000B}.  For each galaxy produced by Galacticus we find the full SED closest matched in spectrum to that of Galacticus in the library.  Then, for each bandpass, we re-weight the full SED template so that its observed flux matches that expected using the Galacticus model.  In this way, we reproduce the colors and fluxes of the Galaticus model.

To run efficiently when rendering large numbers of objects in DC2, imSim is configured to use alternative algorithms when an object is either extremely dim or extremely bright.  Very dim objects unlikely to be detected as distinct sources in the measurement process are rendered with a simplified SED and silicon model. Additionally, the very bright stars, for which the numbers of realized photons make ray tracing prohibitively expensive, are rendered using a simplified surface brightness profile via GalSim’s Fast Fourier Transform technique. These stars will saturate the central pixels of their rendered images, as well as produce bleed trails for the very brightest stars, and therefore cannot be used for astrometry, photometry, or PSF estimation. Accordingly, they will be masked in the image processing and so will not be used for PSF estimation.  Thus, any slight differences in the realized PSF between the ray tracing and FFT implementations cannot bias any aspects of the object measurement pipelines.

Several physical processes contribute to the PSF in real astronomical observations.  For LSST observations, these include effects from the transmission of light through the atmosphere and telescope optics and from electrostatic effects on the accumulated charge in the sensors.  For these simulations, we used physically motivated models where possible in order to capture the relevant emergent effects in the resulting images, but we also include a final ad hoc adjustment to the overall PSF sizes in order to match those that are expected for real Rubin Observatory data.

Following \citet{2011PASP..123..596J} and \citet{2015ApJS..218...14P}, for each simulated exposure we generate a series of atmospheric phase screens with different altitudes and weights to realize the atmospheric PSF. The altitudes and weights are centered around those used in~\citet{2002JOSAA..19.1803E}, though we introduce further exposure-to-exposure variation by randomly perturbing the weights $\sim$~10\% in each exposure. We also move the ground layer altitude from 0~m to 200~m to decrease the correlation length of generated atmospheric PSFs, such that the patterns of PSF ellipticity better match those seen in other surveys, e.g., carried out with the Canada-France-Hawaii Telescope, CFHT~\citep{10.1111/j.1365-2966.2011.20312.x}.  For each layer, a wind speed is uniformly drawn between 0 and 20 ms$^{-1}$ and a wind direction is isotropically selected.

The phases in each screen are a realization of a Gaussian random field with a Von~Karman power spectrum, parameterized by the Fried parameter $r_0$ which sets the turbulence amplitude and an outer scale above which phase power asymptotes.  In each exposure, the outer scale, which is common across layers, is drawn from a log-normal distribution with mean 25-m and width $\sim 15$-m.  The overall effective Fried parameter is computed such that the realized atmospheric PSF full width at half maximum (FWHM) is consistent with the target FWHM from \opsim\ for the given exposure; separate Fried parameters for each layer are then determined from the layer weights described above. When generating the phases and using them to ray-trace the paths of individual photons, we follow \citet{2015ApJS..218...14P} and truncate the power spectrum above a value $k_{crit}$. The effects of phase fluctuations below this scale are accurately modeled by deflecting the angles of incoming photons in proportion to the local gradients of the generated screens. In contrast, the PSF effects of phase fluctuations above this scale are better modeled as a convolution by a second term, dubbed the “second kick” by \citet{2015ApJS..218...14P}, which is expressible as a numerical integral and implemented in GalSim explicitly for this purpose.

For DC2, we used the results of a simulation of the LSST active optics system (AOS) which implements a parameterized model of the LSST optics response run in closed-loop mode~\citep{2012SPIE.8444E..4PC, 2014SPIE.9150E..0HA, 2016SPIE.9911E..18A,2018SPIE10705E..0PX, 2019ApJ...873...98P}.  In this mode, aberrations due to distortions of the system from gravity and thermal factors are corrected with a set of actuators that manipulate the optics system. The corrections are determined by measuring excursions from ideal focusing at the edge of the focal plane in the LSST wavefront sensors~\citep{2018SPIE10705E..0DR}. This simulation produced a sensitivity matrix derived from the Zemax suite\footnote{\https{www.zemax.com}} which maps the degrees of freedom of the actuator system to changes of the wavefronts as seen at the focal plane.

These wavefront distortions are parameterized as a set of Zernike polynomials.  The LSST systems engineering team utilized a custom program to simulate the closed-loop system, computing the full optical response with PhoSim \citep{2012SPIE.8444E..4PC, 2014SPIE.9150E..0HA, 2016SPIE.9911E..18A,2018SPIE10705E..0PX, 2019ApJ...873...98P}. This study yielded a set of fully simulated cycles of moving the LSST actuators to keep the system in focus.  The output included the amount each actuator was required to move to keep the system in lock.  Even after the AOS is employed, small residual deviations from the perfect system remain.  These deviations were used to choose a set of random inputs to the Zemax sensitivity matrix which added appropriate optical aberrations to the incoming wavefront before the photons impinged on the focal plane.
However, even with the inclusion of those deviations, the resulting PSFs had ellipticities that were significantly smaller than the science requirements for the Rubin Observatory\footnote{The requirements for the PSF ellipticity distributions are presented in \https{www.lsst.org/scientists/publications/science-requirements-document}}. It is likely that this underestimate was due to the limited numbers of runs of the closed loop simulation described above, resulting in an under-sampling of the parameter space we used as inputs to imSim. In order to have more realistic PSF shapes, we found that an additional factor of 3 applied to the misalignment coefficients of the Zemax sensitivity matrix produced ellipticity distributions that are more in agreement with the expectations for those distributions as expressed in those science requirements.

To capture the components of the optical system that contribute to the overall size of the PSF but which are not yet included in the imSim model, an additional convolution with a Gaussian with FWHM 0.4 arcsec was included during the run at all wavelengths. This accounts for dome seeing and for artifacts on the optics from polishing errors and similar optics defects that have a range of spatial frequencies.  With this addition, the resulting PSFs have FWHM sizes that match the predictions from \opsim.

For detailed simulation of sensor response, GalSim contains a fast interpolation code built from the output of a detailed model of the LSST sensors that includes electrodynamic calculations of charge redistribution in the silicon~\citep{2019arXiv191109038L}.  Careful treatment of the physics in the sensor, including static or dynamically generated lateral fields, is necessary in order to properly simulate the charge distribution in the LSST CCDs.  One example is the so-called brighter-fatter effect.  The brighter-fatter effect is an observed property of thick CCDs: as the intensity of the light source increases (becomes brighter), the PSF becomes wider (fatter)~\citep{2006SPIE.6276E..09D, 2014JInst...9C3048A,  2015A&A...575A..41G,  2014SPIE.9150E..17R}.  This has potentially serious consequences as the PSFs that are used to measure the properties of galaxies are typically derived from much brighter stars.  A full stand-alone Poisson solver~\citep{2019arXiv191109038L}, along with a fast interpolator to apply the electrostatic solutions inside of GalSim, was developed to implement this effect.  In addition to the brighter-fatter effect, other features seen in the LSST CCDs such as tree rings~\citep{2015JInst..10C5027B} are also implemented. The parameters of the model for tree rings are based on both lab-bench data~\citep{2017JInst..12C5015P} and doping profiles of the LSST sensors, which were determined by performing physical measurements~\citep{2019arXiv191109577L}.

After the distribution of the deposited charge in the CCDs is computed, the effects of the electronics readout are simulated. These include system gain, crosstalk, read noise, bias levels, dark current,
and charge transfer inefficiency. The parameters describing these effects are based on laboratory measurements of LSST sensors and are retrieved from the LSST software package that will contain the final as-built specifications of LSSTCam. After the readout effects have been simulated, imSim produces FITS files that conform to LSST file output specifications.

\begin{figure*}
   \centering
   \includegraphics[width=0.45\textwidth]{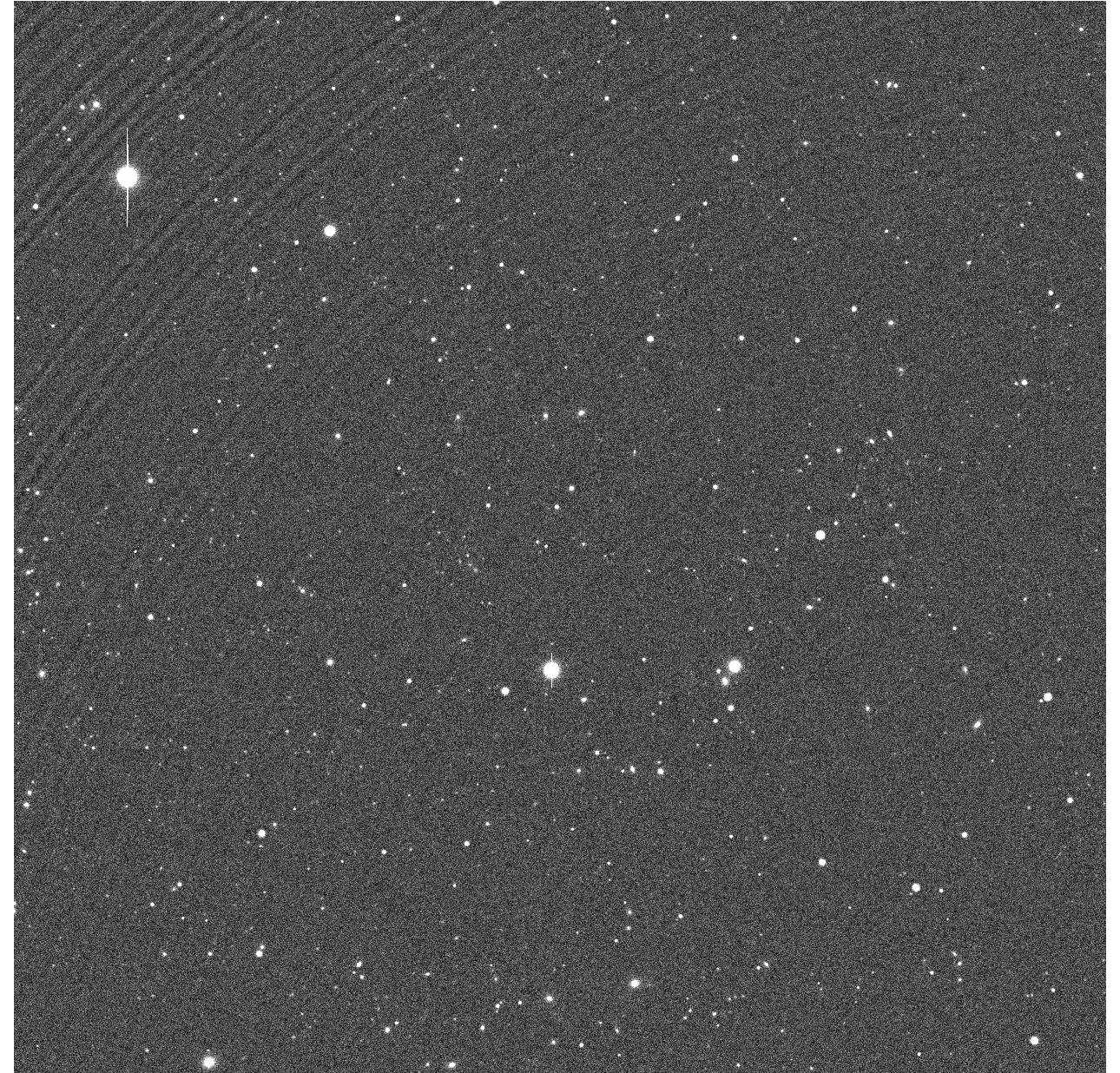}
   \includegraphics[width=0.45\textwidth]{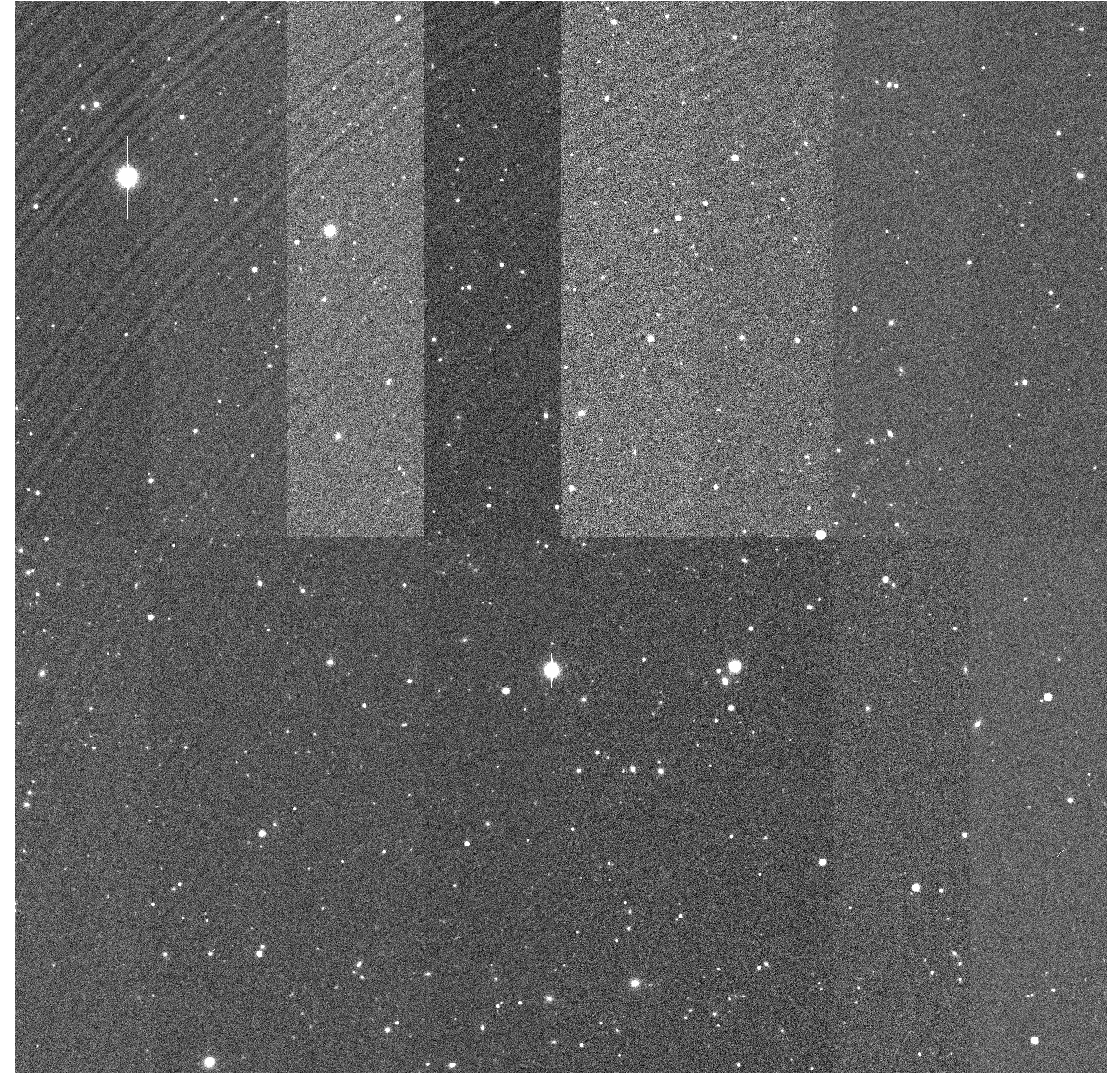}
   \caption{Left: An ``e-image'' of a typical DC2 region in $i$-band for a single CCD on the LSSTCam focal plane. Both simulated tree ring sensor effects (see text)  
   and pixel saturation bleed trails are clearly visible in the image.  Right: A mosaic of the data after simulating the electronics readout.  The different amplifier regions are evident because each region has a separate mapping of pixel to grayscale values.}
   \label{fig:eimage_and_raw_image}
\end{figure*}

In \autoref{fig:eimage_and_raw_image}, we show pixel data from a single CCD in the LSSTCam focal plane for a typical $i$-band observation.  The image on the left is an ``e-image'', a true image of the collected electrons after the exposure.  Various sensor effects such as tree rings, the brighter-fatter effect, and saturation trails can be seen.  The image on the right is a mosaic of the pixel data in each of the 16 amplifier regions per CCD after simulating the electronics readout.

\subsubsection{imSim Architecture}

The data flow of imSim proceeds as follows. Upon reading in the instance catalogs, the list of objects is trimmed to those whose light will intersect with the sensors requested for simulation in the visit.  Each sensor is then simulated individually, with the processing for each sensor carried out in parallel, spread over the CPU cores available to each job. For each sensor, the objects are drawn serially. The photons for each object are traced through a multi-layer atmosphere shared by all sensors in a visit. The objects are all simulated with a 30 second exposure, resulting in a single spatially varying atmospheric PSF across the focal plane for all of the objects.  Other atmospheric effects such as differential chromatic refraction are accounted for here as well.

Due to the current lack of a truly ray-traced optical system, several important effects are missing in the imSim output.  For example, we do not include vignetting, which removes up to $\sim$70\% of simulated photons near the edge of the outermost sensors (and $\sim 2\%$ of all simulated photons) and also changes the optical PSF in this region.  Missing effects like ghosting, stray light, and diffraction spikes will introduce artifacts that will either be identified and masked or effectively become part of the sky background.  While we expect the impact of these effects to be primarily a decrease in simulated survey depth, it is possible that they will also affect the performance of some measurement algorithms.  Other current approximations, to which we found that our results are insensitive, include employing the same gain and quantum efficiencies across amplifiers and sensors, and holding the sky brightness constant across each sensor (intra-CCD sky variations were found to be less than the Poisson noise of the sky background).  Finally, there are currently no clouds in our simulation, nor time-dependent sky glow.

The imSim code is available on Github\footnote{\https{github.com/LSSTDESC/imSim}}.  A more complete description of imSim will be presented in a future publication.

\subsection{imSim Simulations: Parsl Workflow}
\label{sec:parsl}
In order to create a portable workflow that could be used across the various computing resources available to LSST DESC, we choose to take advantage of the Parsl parallel scripting library for Python~\citep{parsl} in combination with the Singularity \citep{singularity} (for ALCF resources) and Shifter (for NERSC resources) containerization software.
New underlying approaches were developed in order to scale the workflow to utilizing all available nodes on Cori and Theta. Additional load balancing was included in order to avoid having worker processes idling for large periods of time given tasks with significantly different runtimes (on the order of several hours variance). This allows us to take advantage of HPC systems with a tool (imSim) that does not inherently lend itself well to cross-node communication.

The workflow can be best understood as having several key steps -- work identification, work scheduling, and compute on available resources. First, for each visit, we identify which detectors on the camera will require an image to be simulated; this may not include all 189 CCDs due to some fraction of detectors lying outside of the simulation area. This is primarily a pre-processing step to avoid expending compute time on CCDs that do not contain images. Upon completion of this pre-processing step, the Parsl workflow has been designed to operate entirely within a compute node, as opposed to many similar workflow engines which require additional software running on service nodes.

At run time, each compute node receives tasks based on an optimization scheme that addresses the nature of the imSim multi-process schema. The core limitation is the significant differences in available memory on the various computing resources employed. In particular, the Cori system at NERSC has compute nodes containing 68 Xeon Phi Knights Landing (KNL) cores with a total of 98 GB of RAM available, while the Theta system at ALCF has compute nodes with 64 KNL cores and a total of 196 GB of RAM. For a given visit run with imSim, there are two main contributions to the memory requirements. The first is shared memory for atmospheric screens and read-in of instance catalogs, which have a large per-node cost. The second is per-sensor-visit memory that can have large spikes during FFT operations used to speed up drawing very bright objects. The latter is a per-thread cost, which leads to it being a significant limitation on nodes with limited memory per core. Due to this limitation, we simulate 33 sensors per node on Cori. In contrast, the larger available memory on Theta allows us to simulate 64 sensors per node, corresponding to the number of available threads before hyper-threading. 

For each visit on a given compute node, a container is initialized to run imSim for that specific visit and the associated set of detectors. As such, a visit with many detectors may be spread over multiple nodes trivially {\em or} a single node may contain many visits with a few sensors each. If all containers assigned to a given compute node finish during the allotted time, Parsl will assign a new set of visits to that node, helping to avoid wasted compute time. Check-pointing at the imSim level allows for the objects that are completed to then be carried over to a new run, simplifying scheduling of compute resources.

This approach has proven to be highly scalable on both Theta and Cori. Early iterations of DC2 were run on up to 4000 nodes on Theta, effectively making use of the entire machine. Later iterations have been scaled up to 2000 nodes on Cori. While in practice we could scale to higher node counts, we find this to be a practical limit for purposes of pushing jobs through the queue in a reasonable time frame.

\subsection{imSim Simulations: The Grid}

In order to take advantage of the grid computing resources in the UK and France, imSim has been deployed on the CVMFS\footnote{\https{cernvm.cern.ch/portal/filesystem}} (CernVM File System) and scripts have been developed to manage the submission of large numbers of imSim jobs to the grid\footnote{\https{github.com/LSSTDESC/DC2_grid_scripts/releases/tag/Run2.2i}}. These scripts utilize the DIRAC\footnote{\http{diracgrid.org}} middleware to submit jobs, monitor their status, automatically resubmit failed jobs, and manage the associated data.

This approach scales up to large numbers of jobs, with typically 50,000-100,000 total jobs in the system at a time, and 1,000-2,000 running simultaneously. Each job processes four sensors, so 48 jobs are required to simulate the 189 sensors of a single visit. Up to 15,000 jobs can run to completion each day, equivalent to around 300 visits per day. Due to resource limitations at some grid sites it is necessary to allocate eight CPU cores per job to ensure sufficient memory, though only four cores are actually used. imSim is executed from the same Singularity container image used for DC2 at other HPC sites. The same configuration files are also used to ensure consistency with data generated elsewhere.

Input data (instance catalogs) and output data (FITS images and checkpoint files) are stored on grid storage elements. They are registered in the DIRAC File Catalog, which allows DIRAC to automatically download the required input files for each job from a suitable storage node, and to upload and register the output data when the job completes. Instance catalogs are transferred from NERSC to the UK via GridFTP and then replicated to multiple grid storage elements to spread the considerable load placed on the network by several hundred imSim jobs fetching their input data simultaneously. Output data is uploaded directly from the jobs to CC-IN2P3 where they are further processed.

\subsection{Image Validation}

Most of the quantitative checks of the simulated images are performed indirectly by comparing the astrometric and photometric measurements from the LSST Science Pipelines against the input truth catalog values.  This is described in \autoref{ssec:DM_product_validation}. We perform basic checks of the raw image data, i.e., comparing the simulated sky background levels against the \opsim\ predictions and checking that the power spectrum of the sky background has the expected properties. In addition, we visually inspect some of the raw data to check cosmic-ray rates and morphologies, bleed trails, and general image quality.

\section{Image Processing}
\label{sec:processing}

For the image processing of the DC2 data, we use the LSST Science Pipelines to produce the same data products as will be provided by Rubin Observatory during survey operations.
We have here focused on producing the static science products: coadded images and associated image catalogs. For the static science cases -- weak lensing, clusters, large-scale structure -- the image data is processed with the Data Release Production (DRP) pipeline.  This pipeline generates catalogs of object properties derived from the observations associated with a given yearly Data Release interval. It is beyond the scope of this paper to discuss the DRP pipelines in detail, but we here describe some of their basic functionality in order to provide context for the data products that will be covered in the next section.  More detailed descriptions of the LSST Science Pipelines code can be found in \cite{2018PASJ...70S...5B} and \cite{2018arXiv181203248B}.

\subsection{The DRP Pipeline}
\label{ssec:DRP_pipeline}
\begin{figure}
    \centering
    \includegraphics[width=0.35\textwidth]{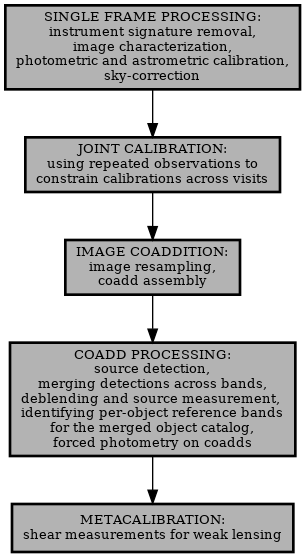}
    \caption{High-level flowchart showing the main processing steps in the Data Release Production pipeline used for DC2.}
    \label{fig:DRP_flowchart}
\end{figure}

Image processing with the DRP pipeline involves four major steps -- single-frame processing, joint calibration, image coaddition, and coadd processing \citep{2018PASJ...70S...5B}.  In~\autoref{fig:DRP_flowchart}, we show a high-level flowchart of the major processing steps with the key substeps listed.  To support our weak lensing studies, we have included a final ``metacalibration'' step for shear measurement which is not a part of the standard DRP pipeline.  In the actual execution of the DRP pipeline, we have omitted the joint calibration step, which uses repeated obversations of the same sources to constrain the photometric and astrometric calibrations across visits.  We have found that because of the lack of throughput variation across the focal plane for the DC2 simulations, that step does not improve those calibrations.

For single-frame processing, individual visits are processed on a per-CCD basis.  The first part of this step is instrument signature removal (ISR) that consists of bias subtraction, crosstalk correction, non-linearity correction, flat-fielding, brighter-fatter correction, and masking of bad and saturated pixels.  A detailed description of the calibration products that were used for ISR is given in \autoref{sec:calib}.  ISR is followed by an image characterization step that performs background estimation and subtraction, PSF modeling, cosmic-ray detection and repair, source detection, source
deblending, and source measurement.  For the sources, various measurement algorithms are applied including centroiding, aperture photometry, PSF photometry, CModel photometry and shape fitting.
The image catalogs are compared to a reference catalog to generate photometric and astrometric calibrations for the images and associated catalogs.
For DC2, the photometric and astrometric calibrations are based on a simulated reference catalog.  We use the CatSim inputs to the instance catalogs as the basis for a simulated reference catalog of stars and galaxies.  For actual Rubin Observatory operations, the reference catalog will likely be based on the GAIA catalog~\citep{2018A&A...616A...1G}.

The resulting calibrated images are known formally as ``Processed Visit Images'' (PVI) and informally as ``calibrated exposures'' (\code{calexps}).  To accompany the calibrated images, a calibrated source catalog is generated for each CCD in each visit containing those measurements.  
As described in \autoref{ssec:DM_product_validation}, we use the contents of the source catalogs as part of our visit-level validation. 

The calibrated exposures are then used to generate coadded images.  The background models that are stored along with the calibrated exposures have been modified by the ``sky-correction'' step in the single-frame processing. This step uses an empirical background model that extends over the entire focal plane and a model of the mean response of the instrument to the sky in each filter (which is flat in the DC2 simulations) to control the extent to which extended features are included in the background model.  In the absence of the sky-correction step, the over-subtraction of bright object ``halos'' affecting nearby objects is present in the PVIs. Inclusion of the sky-correction step mitigates this over-subtraction. See~\autoref{fig:sky_correction_effect} for an example of this effect.

Coadded images in each band are generated from single visit images that are resampled onto a common pixel grid on the sky.  This grid is defined in terms of ``tracts'' and ``patches'', where each tract is composed of 7$\times$7 patches, and each patch is 4100$\times$4100 pixels with a pixel scale of 0.2 arcsec.  A patch is roughly the same size as a CCD, a square 13.7 arcminutes on a side. Tracts are squares measuring 1.6 degrees on a side. Patches overlap by 100 pixels along each edge so that objects lying on the edge of one patch are typically fully contained on the neighboring overlapping patch.  Similarly, tracts overlap their neighbors by $(1/60)^\circ$.  A distinguishing feature of a tract is that all of its patches have a single ``World Coordinate System'' (WCS).  When producing the coadded images we remove variable sources and artifacts using resampled PSF-matched images to produce a static image of the sky.  As described in \cite{2019PASJ...71..114A}, each image is resampled, PSF-matched and stacked into a 2-sigma-clipped mean coadd which serves as a model of the static scene.  Then a difference image is created for each image with respect to this model to identify regions associated with transient detections that only appear in a small number of epochs.  With these regions identified, the final coadded image is created as a weighted mean stack of images where the transient detections are ignored.  The PSF at any point in the coadded image is computed by taking a weighted sum of the individual visit PSFs that have been resampled and weighted in the same way as the coadds.  Regions that have clipped areas will not have the correct PSF, and these are flagged for individual objects.

As described in \cite{2018PASJ...70S...5B}, the coadd processing consists of five main steps: 1) above-threshold detection in each band, 2) merging the detections across bands, 3) deblending the merged detections to generate ``objects'' and measuring object properties in each band, 4) identifying a reference band for each object and merging the per-band catalogs into a single object catalog to use for forced photometry, and 5) performing forced measurements in each band using the reference band positions and shapes.\footnote{The standard DRP pipeline \citep{2018PASJ...70S...5B} also includes forced photometry of the objects in individual visits.  Since those measurements are not used for static dark energy science, we have omitted them from our processing.}  This last step produces a catalog of independent per-band object measurements containing the key object data provided to the science working group analysis pipelines. As noted, in addition to the standard LSST Science Pipelines measurements, we perform DESC-specific processing, i.e., the application of the ``metacalibration'' shear inference algorithm for use in weak lensing measurements \citep{2017ApJ...841...24S,2017arXiv170202600H}.

\begin{figure*}
    \centering
    \includegraphics[width=0.9\textwidth]{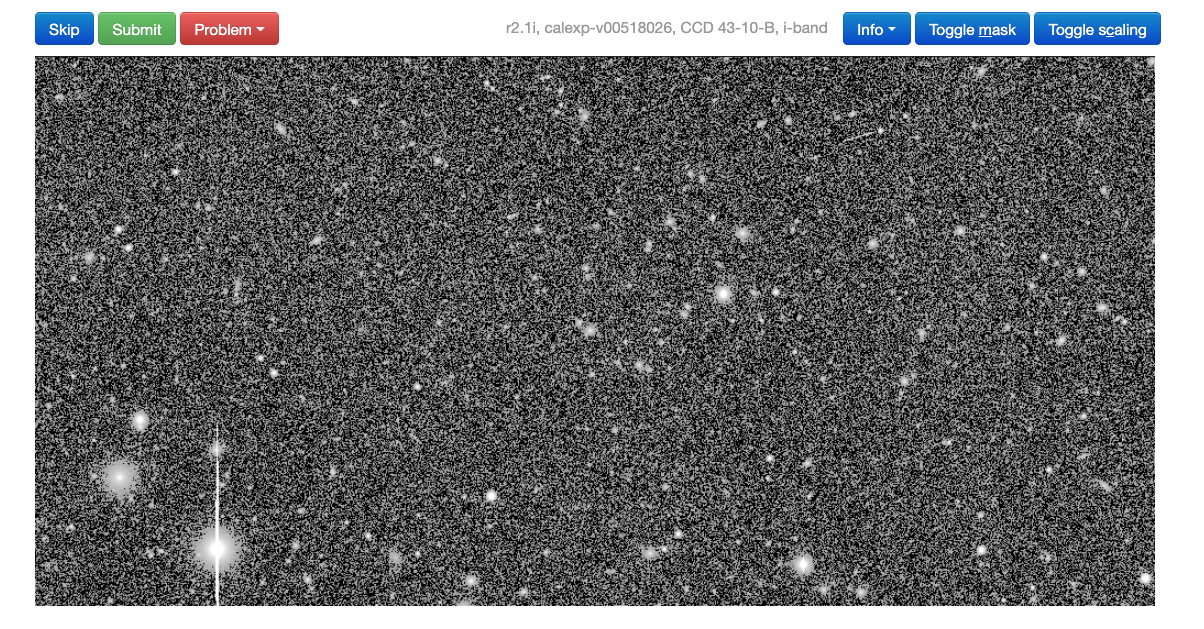}
    \caption{Interface of the exposure checker used for image validation. The checker includes the ability to show the mask plane, change the scale, show the full focal plane, report problems, and show the problems reported by other users, among others. The original images are rebinned (each pixel corresponds to 4 pixels in the original exposure) for storage purposes, but the original FITS files are accessible via this interface for more careful inspection. For more details we refer the reader to~\citet{2016A&C....16...99M}.}
    \label{fig:DC2_exp_checker}
\end{figure*}

\begin{figure*}
    \centering
    \includegraphics[width=0.45\textwidth]{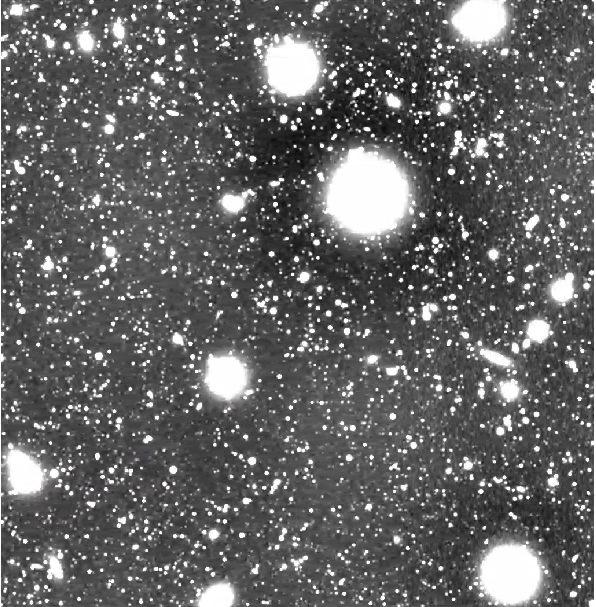}
    \includegraphics[width=0.45\textwidth]{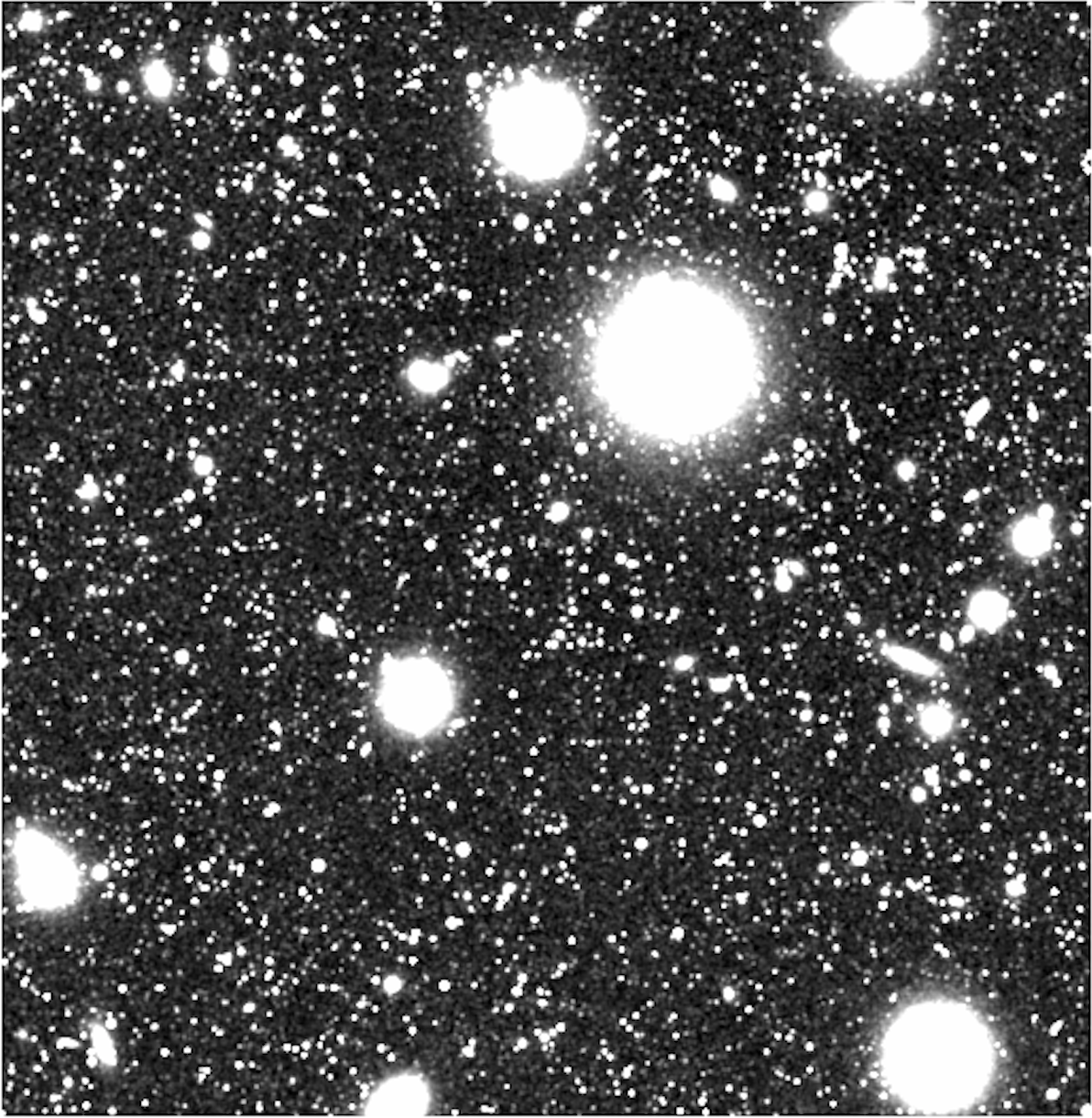}
    \caption{Left: Example of coadded $r$-band image (tract: 3263, patch: 0,3) with background over-subtraction near bright sources. Right: Same coadd after applying the new sky-correction algorithm. The prominent dark halos close around the bright sources are no longer present.}
    \label{fig:sky_correction_effect}
\end{figure*}

\subsection{Workflow Management}
\label{ssec:image_processing_computing}

The various LSST Science Pipelines steps are all implemented as programs that are executed at the shell command line.  The LSST~DM team provides ``driver'' scripts that chain together more finely grained command line tasks that are commonly run together in sequence, and these driver scripts provide some level of parallelization in order to take advantage of multi-core and multi-threaded computing environments. In order to process the large volume of DC2 data, we use the ``SRS workflow engine'' that was originally developed at SLAC to handle the processing of data from the Fermi Gamma-ray Space Telescope \citep{2009ASPC..411..193F}.  The SRS workflow engine was designed to orchestrate very complicated data processing pipelines using a standard interface that can handle job submissions to any underlying batch job handling software. The system integrates batch job control, processing complex directed acyclic graphs (DAGs), and extensive real-time monitoring, control, and bookkeeping. We have implemented the DRP and DIA image processing pipelines using the SRS workflow engine such that these pipelines can be executed at either CC-IN2P3 or NERSC, which have very different batch job handling systems.

Although NERSC is the central computing facility for DESC, the LSST Science Pipelines processing of the image data for DC2 is performed primarily at the IN2P3 Computing Center (CC-IN2P3).  CC-IN2P3 is a High Throughput Computing Center located in Villeurbanne, France and serves the French High Energy Physics, Astroparticles and Dark Energy community through a large farm of computing servers orchestrated by the Grid Engine batch system\footnote{\https{www.univa.com/products}} coupled to a multi-petabyte parallel file system (GPFS\footnote{\https{en.wikipedia.org/wiki/IBM\_Spectrum\_Scale}}).  As noted in \autoref{sec:image_sims}, the raw image data is produced at NERSC, ALCF, and on the Grid.  Once the data for a particular visit is generated, the raw image files are transferred to CC-IN2P3; and after image processing, the LSST Science Pipelines output catalogs are repackaged to follow the LSST Data Products Definition Document (DPDD)~\footnote{\https{ls.st/dpdd}} specification for what will be provided by an annual LSST data release. The calibrated images, coadded images, and catalogs are transferred back to NERSC where they are stored and served to the DESC science groups, as described in \autoref{sec:data-drp} and \autoref{sec:data-access}.

For making the LSST Science Pipelines software available at CC-IN2P3 (both for the login and batch farm nodes), at NERSC and at the individual scientist's personal computer, we used CVMFS~\citep{1742-6596-331-4-042003} to distribute the software and ensure that the exact same releases were available and used by all the sites participating in the survey.\footnote{More details on this are available at \https{sw.lsst.eu}}

Data exchange between CC-IN2P3 and NERSC was handled by an experimental, secure HTTP-based transfer system. The secure HTTP protocol (technically, HTTP over TLS) guarantees confidentiality and integrity of the data transported. By keeping simultaneous network connections between the data transfer nodes on both sites open, we were able to handle multi-terabyte data transfer campaigns involving millions of relatively small files and reaching throughput of up to 1.2 GB/s for disk-to-disk transfers sustained over periods of up to 20 hours. The round-trip time between those two sites is about 150ms.

\subsection{Image Processing Quality Assurance}
\label{ssec:DM_product_validation}

The initial image quality assurance effort is performed using a web-based ``exposure checker'' which enables ``crowd-sourced'' visual inspection of a sub-sample of the $> 10$ million calibrated images. The exposure checker is a customized version of the DES exposure checker~\citep{2016A&C....16...99M} for use with LSST data. A screenshot of the exposure checker can be seen in \autoref{fig:DC2_exp_checker}. The ``crowd'' comprised DESC members, who collectively have a broad range of expertise in examining astronomical images. In total, more than 9,000 images were inspected by $\sim 40$ DESC members.

Among the problems that were identified in an initial scan of images were over-subtraction near bright objects (31\% of the reported problems) and problems with cosmic-ray masking (11\%) or identification (12\%).
The left panel of \autoref{fig:sky_correction_effect} shows an example of sky background over-subtraction which can result in mis-estimation of source fluxes in those regions.
This particular effect was introduced during image processing. Working with the LSST~DM team, we corrected these effects. The right panel in \autoref{fig:sky_correction_effect} shows the processed images after applying these fixes.

\begin{figure}[t]
    \centering
    \includegraphics[width=0.9\columnwidth]{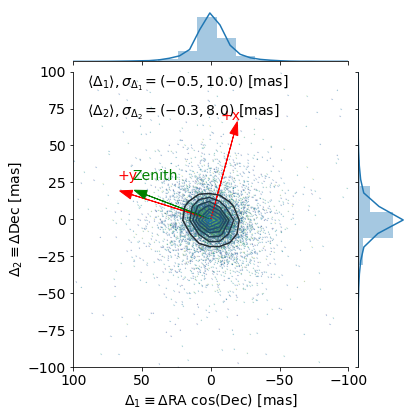}
    \caption{Example of absolute astrometric accuracy test for a single $r$-band visit. We compute the difference between the input and measured positions in the right ascension (corrected by $\cos{(\rm{Dec})}$) and declination axes. The contours are the 2D kernel density estimation (KDE), and they are made to look for possible asymmetries in the distribution. We also show the 1D projection in each axis as well as the corresponding KDE. We show the orientation of the camera axes (labeled as $+x, +y$) and the parallactic angle (PA) to check for the impact of differential chromatic refraction (DCR). We also show the values for the mean, $\langle \Delta_{i} \rangle$, and a "Gaussianized`` standard deviation, $\sigma_{\Delta_{i}}$, for each axis. We check that these values are within a certain range given the observing conditions.}
    \label{fig:example_dm_validation}
\end{figure}

\begin{figure*}
\center\includegraphics[width=7.2in]{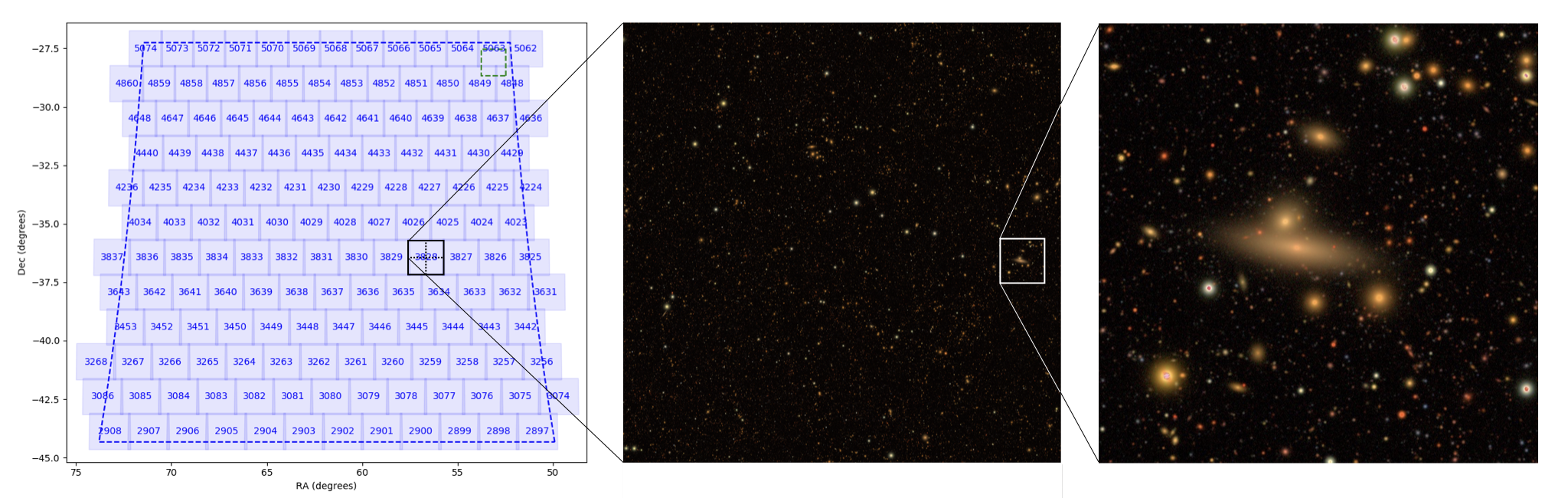}
\caption{Illustration of the detailed image simulations available in DC2. The left image shows the tracts in the DC2 area. The middle panel shows the upper quadrant of tract 3828 in $gri$. The right panel shows a further zoom-in to the image simulations.}
\label{fig:DC2zoom}
\end{figure*}

The second layer in the verification procedure takes place after the catalog production. The LSST Science Pipelines produce catalogs at the single-visit level and coadd level. We test that the catalogs pass the requirements described in both the LSST Science Requirement Document~\citep{LPM-17} and the DESC Science Requirements Document~\citep{2018arXiv180901669T}, since they establish basic requirements that ensure that the data quality is adequate to address the different science goals considered by the LSST and DESC. These tests include checks for astrometric and photometric repeatability and accuracy, requirements on PSF ellipticity, correlations, and residuals, minimum depth amongst others.  As an example, we show  an absolute astrometric accuracy test in~\autoref{fig:example_dm_validation}. Details about the validation procedure will be reported in a forthcoming paper.

\section{Data Release Data Products}
\label{sec:data-drp}

The version of the LSST Science Pipelines that we used for image processing (19.0.0) does not include a step for generating the final DRP data products following the LSST Science Data Model. Therefore we developed our own interim processing step to generate Data Release-like data products, so that DESC members can familiarize themselves with datasets that are very similar to those defined in LSST System Engineering Data Products Definition Document (DPDD; LSE-163\footnote{\https{lse-163.lsst.io}}).
In this section we describe our Data Release data processing pipeline and selected validation tests on the Data Release data products. 

The DC2 LSST Science Pipelines first create single-frame \code{calexps} and then coadd the single frames into \code{deepCoadd} images\footnote{using \code{singleFrameDriver.py} and \code{multiBandDriver.py} respectively}. On \code{calexps} or \code{deepCoadd} images, the Science Pipelines run detection and create merged lists of objects, and then do forced-position and forced-shape photometry on each of the individual per-band coadds and per-epoch images. The result of this processing is a catalog output file for each image (for \code{calexps}) or for each tract and patch (for \code{deepCoadd}). \autoref{fig:DC2zoom} illustrates the tracts covered by the DC2 simulation and includes a panel showing one quarter of one of the tracts, and a second panel showing a small sub-region of the central panel. For an in-depth discussion of the LSST Science Pipelines and terminology used here we refer the reader to \cite{2018PASJ...70S...5B} and \cite{2018arXiv181203248B}.

These individual catalog files need to be collated, merged, and translated to provide the database tables for end users.  The merging of these files is handled with a set of scripts\footnote{\https{github.com/LSSTDESC/DC2-production/tree/master/scripts}}.
Running these scripts produces sets of files, organized by visit for individual exposure-based files or by tract for coadd based products.  The Object tables and Source tables are stored in Apache Parquet format to better support modern data access paradigms, including Apache Spark, and to match the choice of the Rubin Observatory Project to standardize on Apache Parquet for post-processing analysis.

While the resulting Apache Parquet files can be accessed directly, we suggest that end users use the GCR interface described in \autoref{sec:gcr}. In the GCR interface we specify translations (including relabeling, transformation, and combinations) between the column names used in the LSST Science Pipelines and the DPDD.  For example, for the DC2 Object table, we relabel \code{base\_ClassificationExtendedness\_value} as  \code{extendedness}, transform \code{base\_psfFlux\_instFlux} calibrated to zeropoint of 27~mag AB to \code{psFlux} in nanoJansky, and combine \code{base\_SdssShape\_psf\_\{xx,xy,yy\}} in pixels into \code{psf\_fwhm} in arcsec).  We also provide convenience columns, such as \code{tract}, \code{patch}, and \code{r\_mag}, that are not formally part of the data products definitions.  We include  \code{tract} and \code{patch} because it is relatively difficult for a user to recreate these in general, and there are many debugging and validation tasks that benefit from comparison against these indexes to the skymap.  We include magnitudes and magnitude uncertainties,\footnote{They are set to NaN (not a number) for entries with ill-defined magnitude (e.g., due to negative fluxes from the forced photometry).} as they are commonly used quantities that most users would expect from the catalog.  Providing these quantities as part of the catalog leads to more standard code and usage across DESC.

The planned set of data releases from the LSST survey start with a ``DR1'' based on the first 6 months of data, then ``DR2'' based on the first 12 months of data, followed by additional data yearly data releases.  Thus the processing of the first five years of data will be ``DR6''.
For example, \code{dc2\_object\_run2.2i\_dr6} is the Object Table based on the first five years of data.

\subsection{Selected Validation Results }
\label{sec:drp-validation}

\begin{figure}[b]
    \centering
    \includegraphics[width=1.0\columnwidth]{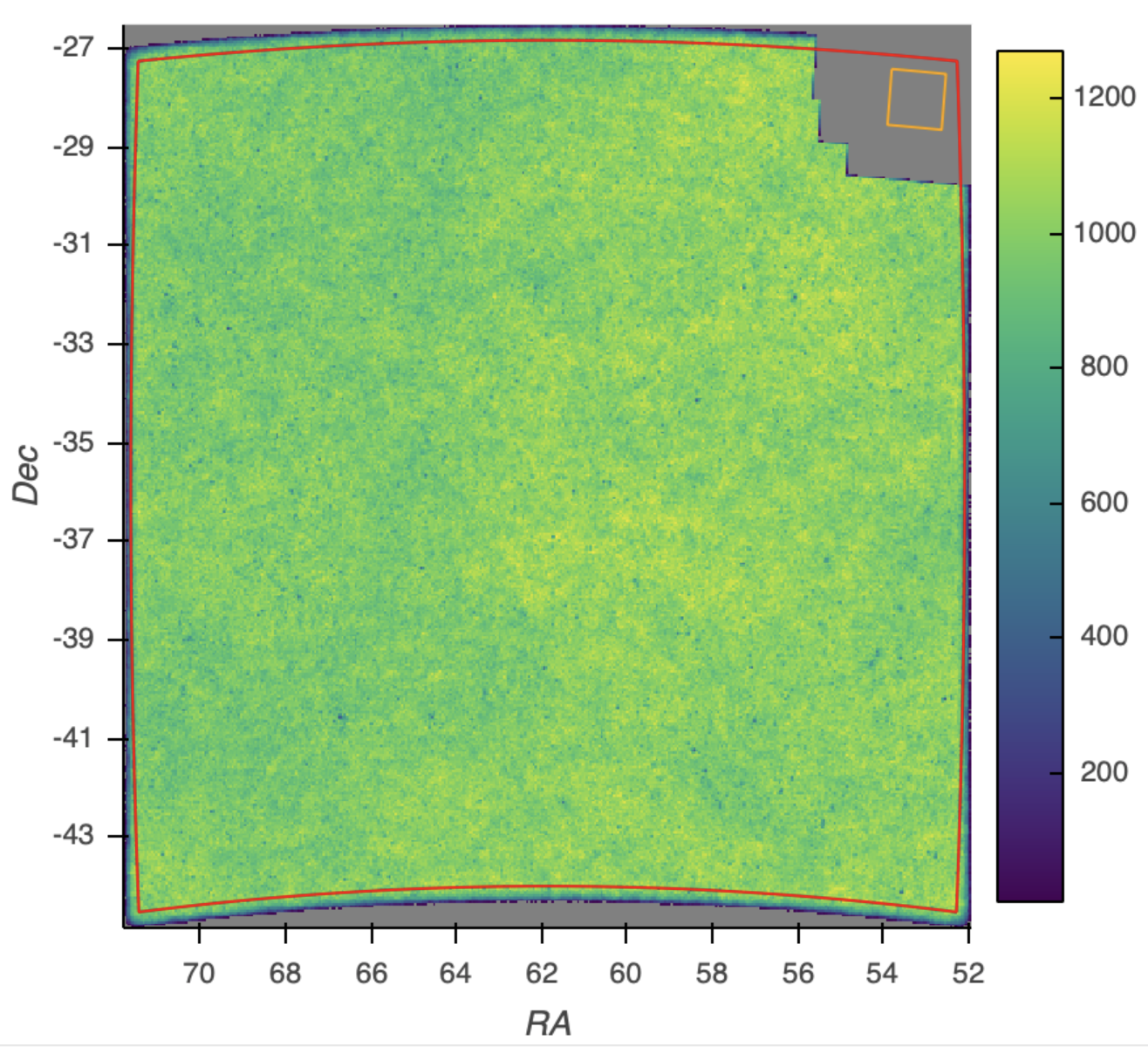}
    \caption{DC2 RA, Dec object density / 40~sq.~arcmin for the DC2 wide field.  The WFD region (red outline) traces out the outline of the DC2 region, which is shown here in Mollweide (equal-area) projection centered on the center of the DC2 region.  The missing corner in the upper right is our small-scale DDF region (orange box).  The sharp edges of the missing range are the three tracts that we have not processed due to the influence of the large number of simulated visits in the DDF region.  The coadd processing is done in tracts that are defined along lines of Declination (which are curved in this Mollweide projection) and have equal areas.}
    \label{fig:ra_dec}
\end{figure}

\autoref{fig:ra_dec} shows the RA, Dec density of sources in our processed coadded images for the wide-field DC2 observations.  The structure visible in that figure represents true variation in galaxy density from our large-scale structure.

Next we show three examples of tests that are important across the cosmological probes targeted by DC2 and which have been carried out on the final data product, the DPDD catalogs. Just as for cosmoDC2, we used the DESCQA framework to run the validation tests. By using the same validation framework, we can readily include a comparison to the cosmoDC2 extragalactic catalog and to the observational data. 
For both the cosmoDC2 catalog and the final DC2 object catalogs, detailed validation papers are in preparation.

\begin{figure}[b]
    \centering
    \includegraphics[width=1.0\columnwidth]{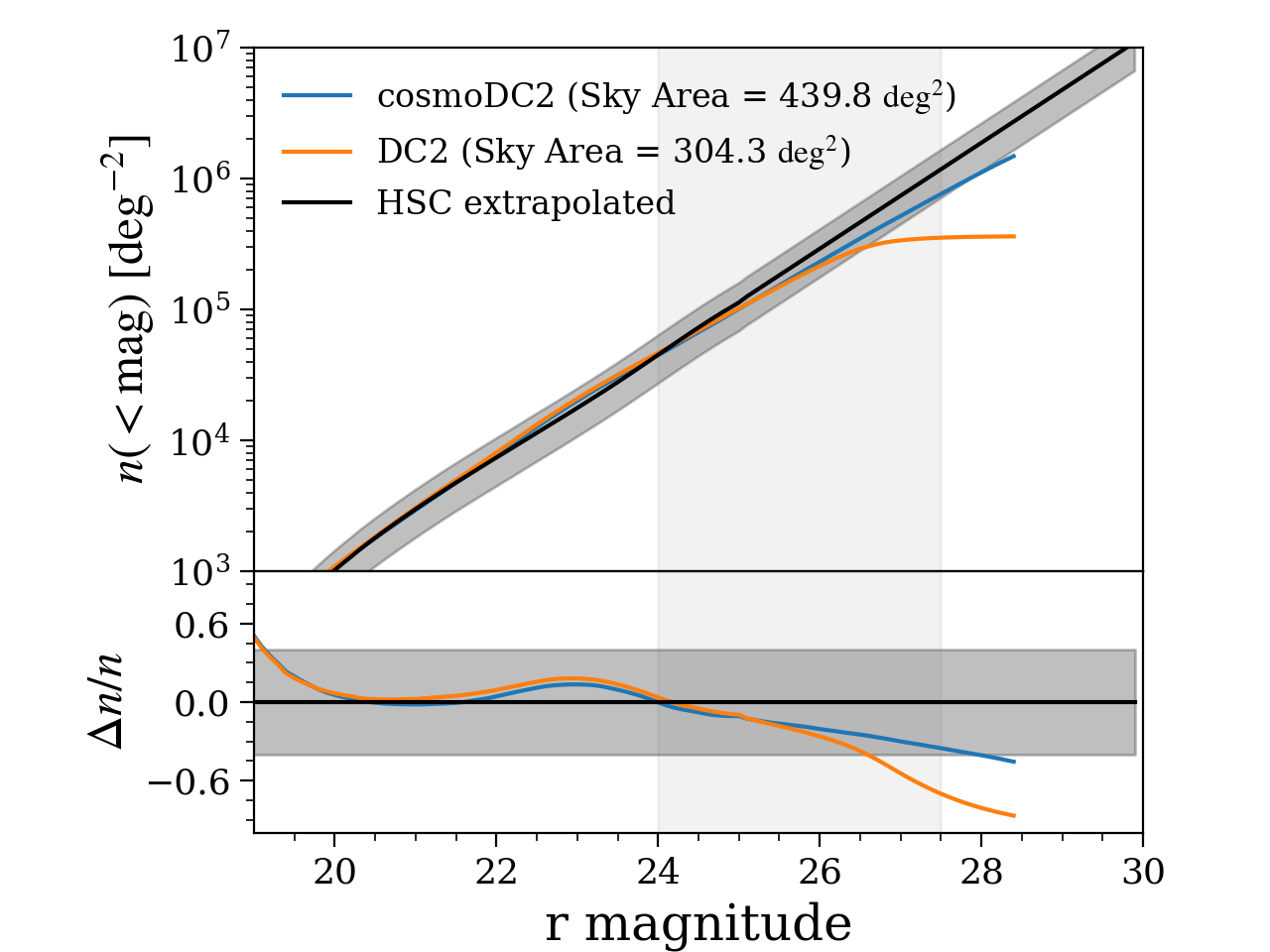}
    \caption{Comparison of cumulative galaxy number density as a function of $r$-band magnitude
for the DC2  object  catalog  from the 5-year (orange) coadded  catalog  with  the  extragalactic  catalog,
cosmoDC2 (blue), and HSC extrapolated measurements (black). The light-gray shaded region indicates the magnitude range over which validation criteria for the cosmoDC2 catalog were provided by the DESC working groups (see text for details).}
    \label{fig:dndmag}
\end{figure}

The first validation test shown in \autoref{fig:dndmag}, examines the cumulative galaxy number density as a function of $r$-band magnitude. The light gray vertical band in the upper panel shows the validation region identified by the DESC working groups for cosmoDC2 ($24 < r < 27.5$). The design specification for cosmoDC2 required that the galaxy number density in this region lie within $\pm$40\% of the HSC data (dark gray horizontal band in the lower panel). Since HSC data does not cover the full range of magnitudes relevant for LSST, the validation data was extrapolated out to a magnitude of 30 in $r$-band. As was shown in \cite{korytov}, the extragalactic catalog (depicted here as the blue line) met the requirement. As mentioned above, for the instance catalogs we impose a cut on galaxies with magnitudes larger than 29 in $r$-band to reduce the catalog size. The orange line shows the results for five years of DC2 data. As expected, this falls below the blue line for cosmoDC2 at magnitudes $r>26$ due to blending effects and the limited depth of the DC2 data. 

\begin{figure}
    \centering
    \includegraphics[width=1.0\columnwidth]{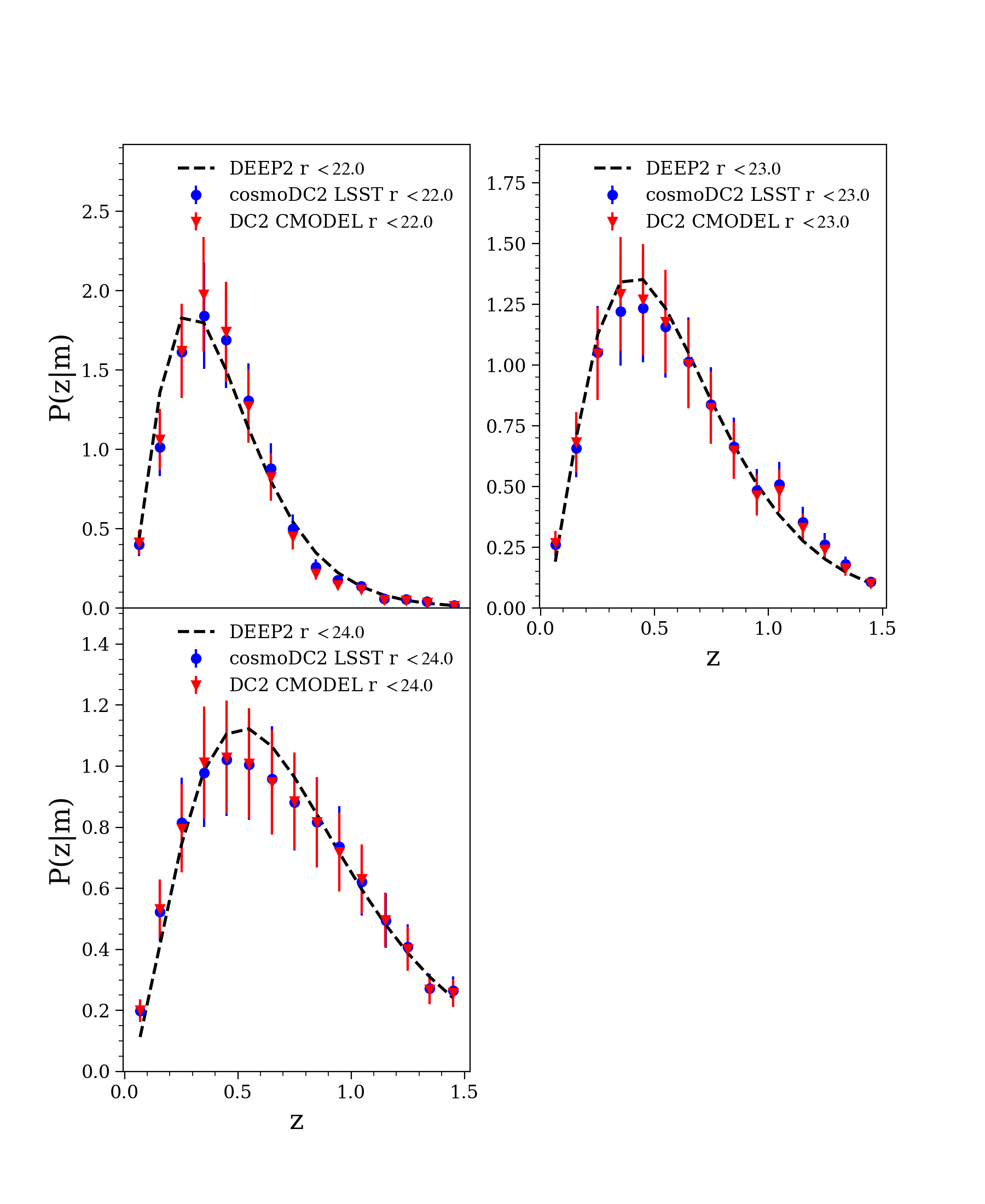}
    \caption{Comparison of the matched redshift distribution of galaxies in the 5-year coadded catalog (red) with cosmoDC2
(blue) and DEEP2 data (black line).}
    \label{fig:Nz_r}
\end{figure}

For our next example, shown in \autoref{fig:Nz_r}, we present the redshift distribution of the galaxies in our 5-year coadded DC2 object catalog (red triangles) compared to the input cosmoDC2 catalog (blue points) and measurements from DEEP2 (black dashed line) for samples of galaxies with different $r$-band magnitude cuts. The agreement is well within the error bars. The object-catalog results presented here were computed using the matching procedure described in \autoref{ssec:truth_catalogs} between galaxies in the object catalog and galaxies in the cosmoDC2 input catalog. That is, we match the sources in the DC2 object catalog with their closest match in magnitude within 1 arcsec radius in cosmoDC2. Then each source in the DC2 object catalog is associated with the true redshift of its corresponding match in cosmoDC2. This matching procedure, as opposed to a simple positional matching, helps to eliminate possible ambiguities in the presence of blending, and eliminates potential artifacts and poorly determined objects from the sample. The performance of this matching procedure was deemed sufficient for the majority of the applications foreseen by DESC. More complex matching schemes (e.g, tying each photon in each pixel to its original source) can further break degeneracies while matching but introduce an additional complexity (e.g., data overhead). Such complex schemes are outside the focus of this work and are left to future studies. This test can therefore be seen as a validation of the obtained galaxy (CMODEL) magnitudes and angular positions relative to the input catalog, as well as serving as a placeholder for validation of upcoming photometric redshift measurements.   

\begin{figure}
    \centering
    \includegraphics[width=1.0\columnwidth]{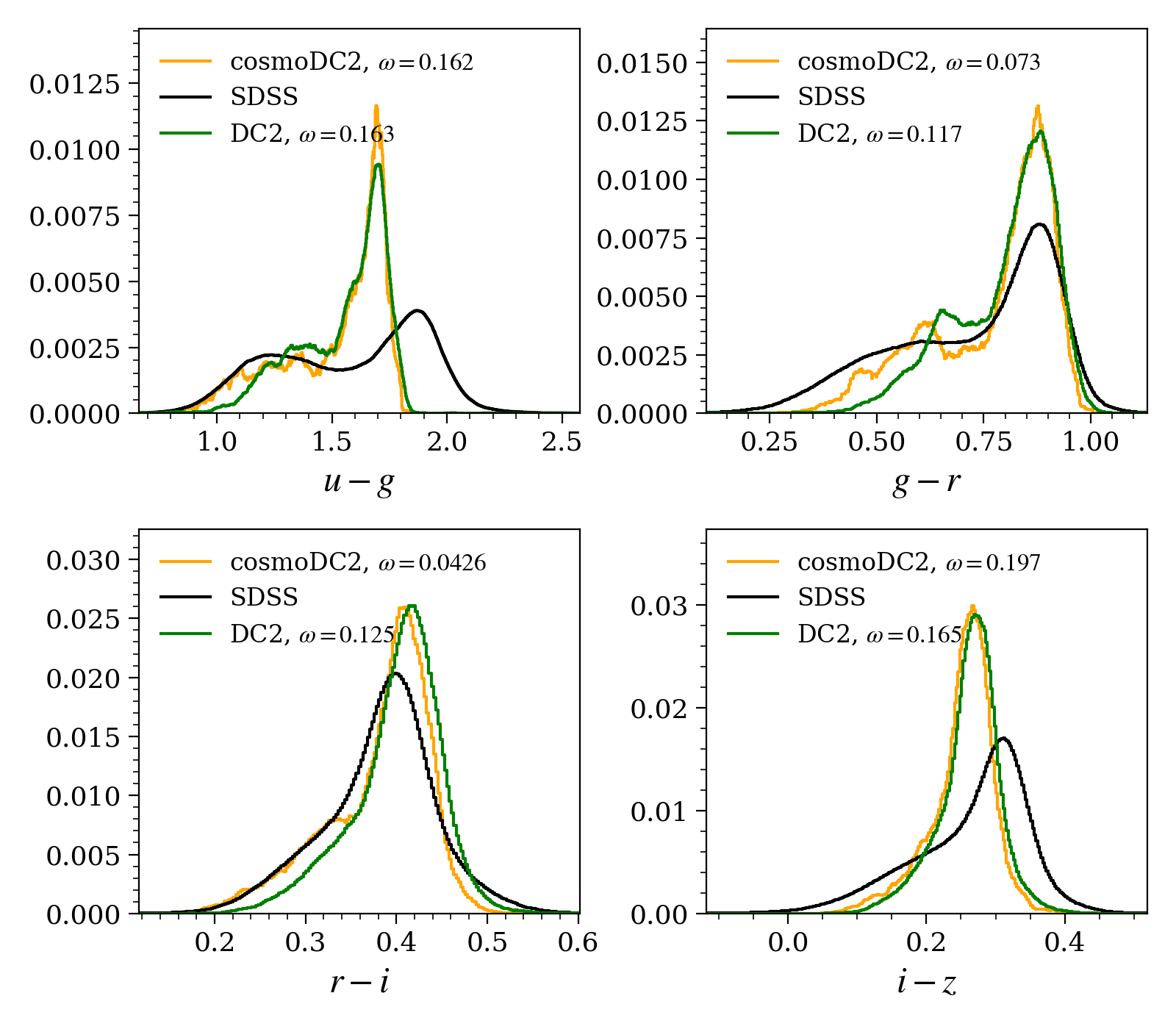}
    \caption{Comparison of color distributions for the 5-year coadded DC2 object catalog with SDSS data. Plotted are smoothed histograms of the normalized bin counts for data from the DC2 object catalog (green), the cosmoDC2 input catalog (orange) and the SDSS validation data (black). The $w$ statistic, included in the plot legends, is described in~\citet{descqa}  and provides a measure of the difference between the distributions.}
    \label{fig:color_SDSS}
\end{figure}

Finally, we show an example of a color test in \autoref{fig:color_SDSS}. Here we show a comparison of the 5-year coadded DC2 object catalog (orange) in comparison to SDSS data for different bands. The DC2 data encompass galaxies that have a unique match to the cosmoDC2 catalog in angular position, and whose true redshifts fall within a range of $0.05<z<0.1$. While the agreement of this color distribution with SDSS data is not perfect, it is very similar to that of the cosmoDC2 catalog, which was deemed sufficient to fulfill the needs for the DESC working groups as detailed in~\citet{korytov}. 

These three tests give a brief impression about the overall quality of the final catalogs. Many more probe specific tests have been carried out, including cluster finding and correlation functions tests that will be shown in forthcoming papers.

\section{Data Access and DESC Science Platform}
\label{sec:data-access}

Part of the DC2 effort is to explore data access methods that are easy-to-use and persistent so that collaboration members can more easily engage in DC2. 
We also want to explore access methods that can be integrated with analysis pipelines. 
The large number of datasets in various formats that DC2 produces serve as an excellent playground for this exploration. 
In this section we describe the DESC Science Platform, our approach to provide uniform and persistent access to multiple DC2 datasets, and our exploration of several different data access methods. 

\subsection{DESC Science Platform}

We chose Jupyter\footnote{\https{jupyter.org}} as our primary interactive science platform for several reasons.
First, it resembles the Notebook aspect of the LSST Science Platform Design (LDM-542\footnote{\https{ldm-542.lsst.io}}), allowing DESC members to adapt early to the use of Jupyter Notebook and JupyterLab\footnote{\https{jupyterlab.readthedocs.io}}.
Second, thanks to its interactive nature, Jupyter is an ideal platform for data exploration, validation, and down-stream pipeline development. It serves the needs of the DC2 users well, and also enables us to develop pedagogical tutorials with Jupyter notebooks.
Third, NERSC provides a native JupyterHub service for its users\footnote{\https{docs.nersc.gov/connect/jupyter}}, and since most of the DC2 datasets are accessible at NERSC, using NERSC JupyterHub enables direct access to those datasets. We also have established a Jupyter service at CC-IN2P3.  

To provide a uniform experience for all DESC members when they use Jupyter at NERSC or CC-IN2P3, we develop and maintain custom Jupyter kernels that include commonly-used Python packages, DESC-specific packages, and the LSST Science Pipelines. DESC members can use these DESC-custom Jupyter kernels at NERSC or CC-IN2P3, without the need of maintaining their own Python environments. The DESC custom kernels are centrally maintained\footnote{\https{github.com/LSSTDESC/desc-python}} and are built into Docker containers making it easy to share the same environment across sites. This also aids the collaborative development of analysis notebooks, as DESC members will use the same environment for development.

To promote the DESC Science Platform within the DESC, we have developed a set of tutorial notebooks\footnote{\https{github.com/LSSTDESC/DC2-analysis}\label{tutorials_repo}} which provide a starting point for DESC members to learn about and to access various DC2 data products. Because of the DESC custom kernels, only a minimal setup is required to run these tutorial notebooks on NERSC JupyterHub. We also encourage DESC members to contribute their own notebooks to benefit other collaboration members and the broader astronomical community (while the DESC Science Platform itself is currently accessible for DESC members only, the tutorial notebooks are publicly available).

\subsection{Generic Catalog Reader and Catalog Registry}
\label{sec:gcr}

The various DC2 data products, ranging from the extragalactic catalogs to the Data Release-like data products, are stored in different formats on disk when initially produced. 
As the DC2 effort progresses, many data products continue to receive updates and the data models are also evolving. At the same time, we wish to shorten as much as possible the turnaround time between the generation of new data products and their availability to end users. The need of a persistent data access method during the rapid development and data production period therefore poses an interesting challenge.

A custom Python package, \gcrcatalogs{}\footnote{\https{github.com/LSSTDESC/gcr-catalogs}},
originally developed as part of the DESCQA framework \citep{2018ApJS..234...36M},
serves well as a catalog registry and provides a unified access interface to data products that are stored in different formats.
In particular, \gcrcatalogs{} enables end users to access data products in their original formats and eliminates the need for having to write custom data ingestion code or to hard-code file paths in their analysis code.
In addition, the \gcrcatalogs{} package is versioned and installed in the kernels that we provide on the DESC Science Platform to ensure that all users access a common set of data products.

Each data product is registered in \gcrcatalogs{} by means of a plain-text YAML file, that records the file paths or directories, some basic metadata, and a corresponding \textit{reader} that can read in the raw data. A \textit{reader} is a Python class, also part of the \gcrcatalogs{} package, that implements the ingestion of the raw data files. These reader classes are subclasses of a Generic Catalog Reader (GCR) abstract parent class\footnote{\https{github.com/yymao/generic-catalog-reader}}, which provides convenient methods to rename or alias columns, to define derived columns, to filter data, and to iterate over chunks of data. In essence, the GCR abstract class provides the basic functionalities that a database language commonly provides, but without actually using databases to shorten the turnaround time.

While \gcrcatalogs{} is currently the main data access method for DESC members, we note that this approach is not meant to provide a long-term solution. \gcrcatalogs{} is particularly useful for data exploration and validation, and shortens the time needed to deliver a persistent data access method for end users during the development stage. 
In the long term, we expect that data models and storage formats will converge such that stable schemas in databases or flat files will replace much of the translation feature that \gcrcatalogs{} is currently being used for.
Below we describe some other data access methods that we have explored. 

\subsection{PostgreSQL Database Access}
\label{sec:db}

The Qserv database \citep{LDM-135_4.0}, currently under development, is designed to handle the massive data volume to be produced by the LSST.  In order to gain experience with ingest into and user access from an SQL (Structured Query Language) database, a package developed for ingesting Hyper Suprime-Cam (HSC) data \citep{2018PASJ...70S...8A,2019PASJ...71..114A}, in particular object catalogs, into PostgreSQL has been adapted for use with DESC data challenge catalog outputs.  New features in the revised package DC2-PostgreSQL\footnote{\https{github.com/LSSTDESC/DC2-PostgreSQL}}  include  greater flexibility and generality to accommodate differences between HSC and various generations of DC1 and DC2 data, support for forced source as well as object catalogs, and modifications to allow concurrent ingest jobs.

Data (FITS files produced by the LSST Science Pipelines) is ingested into a PostgreSQL database at NERSC; tables arising from each DC2 run are grouped together in a unique PostgreSQL schema. A separate table keeps track by schema name of DC2 runs ingested and their provenance. The largest object catalog produced with this method, about 40 million objects, took about three hours to ingest and a similar amount of time to index. Users typically query the database from Python. Some knowledge of SQL is necessary, but the complications of joins are avoided in most common queries by the definition of DPDD-like views for object catalogs and forced source catalogs. Jupyter notebook tutorials\fnref{tutorials_repo} have been provided to demonstrate access patterns for the database.

For the Run 2 dataset, a different method of ingest is employed. Instead of LSST Science Pipelines output, the DRP parquet files are the starting point. This avoids duplicate calculations, is significantly faster, and results in a single object table with no need for views.

A database representation (as opposed to a column-primary format like parquet) is especially suitable for datasets having a very large number of objects and few attributes, such as the forced source dataset described in the Data Products Definition Document LSE-163, previously cited. Efficient access and filtering will be required for studies involving light curves. However the forced source catalog is not self-contained; it must be matched with the object catalog and a visit table in most cases. In order to gain experience with similar strategies, summary and variable truth catalogs  as well as the \text{minion\_1016} observation database have also been ingested into PostgreSQL.

The DC2-PostgreSQL package, like the HSC original, takes advantage of PostgreSQL support for geometric types to index data by location, speeding up searches by location dramatically. Other standard queries to the database are noticeably faster than comparable \added{single-threaded} searches of the same information from file-based repositories.  While the PostgreSQL database is not meant to scale up to the thousands of users that Qserv is designed to support, it should be adequate for DC2 catalog outputs.

\subsection{Data Mining and Processing at Scale: Apache Spark}

Cluster-computing frameworks such as Apache Spark\footnote{\https{spark.apache.org}} \citep{spark180560} are a compelling and complementary approach to current tools to process simulated or real data products at the volumes that will be collected by the Rubin Observatory. The intrinsic fault-tolerance of Apache Spark, its scalability with respect to data size and its high level Python API (pyspark) make it a robust tool for many applications in astrophysics and cosmology \citep{Plaszczynski2018}.

Most of the data products available from the LSST Science Pipelines (object catalogs and source tables) are generated in Apache Parquet format which is natively supported by Apache Spark.  We also developed a Spark connector to read and distribute data generated in FITS format \citep{peloton2018fits}, allowing access to previously-processed catalogs stored as FITS files. A central object in Apache Spark is the DataFrame, a distributed collection of data organized into named columns. The structure of a DataFrame is extremely close to that of astronomical catalogs where each object corresponds to a row whose properties may be stored as components of the DataFrame columns, hence Spark tools are suitable for processing the data as-is. In addition to the Spark standard tools, such as data distribution, filtering, aggregation, the \text{pyspark} API provides an easy interface and allows the use of any Python third-party libraries.

In order to ease the use of Apache Spark for DESC users, we develop and maintain a Jupyter kernel\footnote{\https{github.com/astrolabsoftware/spark-kernel-nersc}} that includes commonly-used Python packages, DESC-specific packages, and pyspark at NERSC. Users will typically prototype their applications or perform data mining in the NERSC JupyterHub, and then launch production jobs at scale using the regular Apache Spark installation on the Cori machine using the same code. Note that despite the fact that Apache Spark is traditionally deployed and used on the cloud, our results show that its use on current HPC systems gives good performance. Apache Spark enables interactive and deep exploration of the DC2 datasets ($\sim$200 million objects, or $\sim$50~GB of end-user catalogs).

\section{Summary and Outlook}
\label{sec:summary}

In this paper we introduced the process for the generation of the second data challenge (DC2) carried out by the LSST DESC. DC2 is a full end-to-end simulation campaign that provides several years of simulated observed data for a $\sim$300~deg$^2$ area of the planned LSST WFD survey and 1~deg$^2$ of a DDF area. The dataset is based on the Outer Rim simulation from which a galaxy catalog, cosmoDC2, is derived. After adding stars, AGN, SNe, and strong lenses by employing CatSim and the SLSprinkler code, the image simulation tool imSim is used to generate LSST-like images that are then processed with the LSST Science Pipelines. The final catalog resembles as closely as possible the expected data products that will be delivered by Rubin Observatory to all data rights holders, given current tools and resources. 

The area, depth, and observational realism of DC2 allow it to be used not just for individual analyses in isolation, but also for a wider class of investigations that employ cross-correlations to bring in new information and better understanding and control of systematic effects. As a synthetic sky survey, DC2 provides several important opportunities for the DESC community. First, it allows the scientists to become familiar with the LSST data formats and data products and process them with the LSST DESC analysis pipelines under development. Second, the correctness of the analysis pipelines can be carefully tested since the underlying truth is known and, since the data products are based on simulations, no unexpected observational systematics can cause problems. Third, the image simulation output enables DESC to exercise the LSST Science Pipelines on a controlled dataset. 
Finally, the DC2 dataset provides a testbed for collaborative analysis projects across different working groups.

In the future, several improvements to the end-to-end simulation pipeline will be implemented. The extragalactic catalog generation is currently being extended to incorporate several updates. These include improving the realism of the spatial distribution of cluster satellite galaxies to follow an ellipsoidal Navarro-Frenk-White profile~\citep{1996ApJ...462..563N} that is aligned with the large-scale tidal field and improving the color model to eliminate the halo-mass dependence in the colors of red-sequence galaxies. The area of cosmoDC2 will be increased considerably, to cover 5000~deg$^2$ in the forthcoming SkySim5000 extension. The resolution of the weak lensing maps will be improved by a factor of two in angular scale. Work is underway to include intrinsic alignments as an add-on to the extragalactic catalog, so as to enable tests of mitigation schemes for intrinsic alignments, one of the key sources of astrophysical uncertainty for LSST weak lensing measurements \citep[e.g.,][]{2016MNRAS.456..207K}. As Rubin Observatory commissioning nears, imSim is being made ever more realistic and efficient. The eventual aim is to use imSim to provide simulations that aid the investigation and estimation of systematic errors. Planned features include true ray tracing for the optical system, improvements to the sky model including varying clouds and time-dependent sky glow, fringing effects in the LSST sensors and improvements to chromatic PSF response.  All of this will be coupled with careful tuning and incorporation of parameters as measured on LSSTCam, which is being integrated at SLAC. After these improvements have been incorporated, we are planning to carry out smaller, targeted simulations. 

The LSST DESC is planning to make the DC2 data products publicly available in the near future. 

\acknowledgments

This paper has undergone internal review by the LSST Dark Energy Science Collaboration. The internal reviewers were Chihway Chang, Richard Dubois, and Surhud More.

BA investigated variations in the sky model across the focal plane in imSim.
HA implemented the dithers and extracted the visit lists (and sensors for Run1) for the simulations.
YNB worked on the development of the Parsl pipeline and Parsl capabilities and extensions to enable the imSim workflows at NERSC and ALCF.
FEB contributed to the development and testing of the AGN model, as well as to the text of the paper.
GB managed the European computational grid work for Run~1.2i and Run~2.1i.
RB interfaced time domain science implementations with existing middleware software, compiled scientific desiderata for SN in DC2, coded the SN population including assignments to host galaxies, contributed text on SN, provided input to the cadence selection and planning and requirements for strong lensing injection.
JRB is responsible for creation of and access to the PostgreSQL database.
DB contributed to the the image processing pipeline configuration, deployment and tuning at CC-IN2P3 and to the validation of the various data products.
KC worked on the design and implementation of the imSim workflow and coordinated extensions to Parsl to meet the performance and scalability needs of the imSim workflow.
JC worked on imSim development, image validation, image processing development and debugging, and calibration product generation.
JCT was responsible for the definition, implementation, and deployment of the SRS pipeline at CC-IN2P3.
AJC led the development of the LSST simulation tools and contributed to the initial definition of the DESC data challenges and to the text of the paper.
ADW developed the LSST DESC exposure checker and organized the DC2 visual inspection effort.
RD assisted in organization, planning and obtaining computing resources.
SFD developed and ran the middleware connecting the cosmological simulations with the image simulator, including the injection of variable and transient sources according to models provided by RB and JBK.
SWD edited the paper text.
EG contributed to the field location and dither design and edited the paper text.
TG adapted PhoSim to HPC at NERSC; carried out extensive PhoSim testing; designed, implemented and ran workflows which produced the DC2 Run~1.2p data, worked on the production of certain calibration products, and assisted with management of DESC NERSC resources.
SH is the HACC team lead; he contributed to the assessment of image generation computational requirements, co-led the management of DESC NERSC resources, and contributed to the text of the paper.
APH led the development of the model for the extragalactic catalog.
KH was responsible for the overall organization of the project, was deeply involved in many aspects of the extragalactic catalog production, and contributed to the text of the paper.
FH implemented the mechanism for making the LSST Science Pipelines available online and usable both at CC-IN2P3 and at NERSC, managed the CC-IN2P3 data processing infrastructure used by the image processing pipeline and was responsible for the prompt data transfer between CC-IN2P3 and NERSC.
RH worked on the coordination and testing of simulated SN inside DC2, draft reading and editing.
JH was a core member of the extragalactic catalog production team, and created workflow figures for the paper.
MI read the draft and made comments.
MJ contributed to the imSim development effort for DC2, including a number of important updates to GalSim to improve both its speed and accuracy as used by imSim; MJ also contributed to the overall planning and design of DC2, especially with respect to its utility for weak lensing studies.
JBK was the main developer of the SL Sprinkler that inserted strongly lensed AGN into the instance catalogs and contributed the text of the paper relating to the SL Sprinkler.
HMK co-chaired the DM-DC2 task force which coordinated the DESC DM processing of the DC2 image simulation data, produced required calibration products, and managed the DESC software and data resources at NERSC.
EK contributed extensively to the production and validation of the extragalactic catalog, worked on the interface of the catalog with the image simulations and contributed to the text of the paper.
DK led the development of the model underlying the extragalactic catalog.
KSK contributed to the conceptual design of the simulated survey including determining which electronic effects to simulate and by association which master calibration products to include.
SWJ contributed to the validation of SN and the related text.
FL contributed the model for the knots component included in galaxy light profiles, and the implementation of said model in CatSim and imSim.
PL contributed significantly to the production of the extragalactic catalog.
CSL helped develop physical models of the CCD detectors, which allowed physically real simulations of tree rings and the brighter-fatter effect.
NL contributed to the generation of strongly lensed host galaxies of multiply lensed AGN and SNIa in the strong lensing systems sprinkled in the DDF.
EPL made contributions to the sky model in imSim. 
RHL contributed to the validation of the final data catalogs and provided support in using the LSST Science Pipelines.
RM organized analysis teams and synthesized input that factored into the overall DC2 design and validation, was engaged in the validation efforts, and contributed to the text of the paper. 
YYM contributed to the generation and validation of various DC2 data products, co-led the effort for internal data access and DESC Science Platform, and contributed to the text of the paper.
PJM helped design the survey regions and cadences, provided high-level scientific oversight, and contributed to the strong lensing requirements section of the paper.
JEM implemented the atmospheric model in imSim and helped validate the composite PSF model.
JWP contributed to the validation of AGN variability and SNe, implemented the \text{Sprinkler} modules for rendering postage-stamp images of strongly lensed SNe and AGN and generating the lensing truth tables, and contributed to the text of the paper.
JP contributed to the validation of various DC2 data products, and managed the Apache Spark tools at NERSC.
DJP implemented a model for LSST optical effects in imSim, assisted in the development of internal data access tools, contributed to the visual validation of DC2 images, and contributed to the text of the paper.
JP implemented a system for running imSim on the UK computational grid and used it to perform parts of Run~1.2i and Run~2.1i in Europe.
SP contributed to the validation of various DC2 data products, and managed the Apache Spark tools at NERSC.
AP contributed to many aspects of the underlying extragalactic catalog, significantly improved the multi-thread scaling performance of PhoSim, generated PhoSim calibration products, and performed initial studies of using imSim in containers.
ESR performed validation on the extragalactic catalog for clusters and red galaxies and contributed QA on galaxy photomety, especially red galaxies in clusters.
FJS carried out early planning for DC2, performed validation and verification, and contributed to the text of paper and figures.
SJS wrote the text for the photometric redshifts section.
DS contributed to modeling/strategy for injection of fake SN light-curves.
TDU made initial profiling runs of Phosim to characterize performance on KNLs, and managed execution and postprocessing of imSim for Run~1  on thousands of nodes of Theta and supporting clusters at Argonne.
ASV was responsible for early generation of instance catalogs, implementing the Parsl workflow for imSim on NERSC and ALCF resources, helping in initial validation of these outputs, and contributing to the text of the paper.
CWW carried out early planning for DC2, worked on development, testing and management of the imSim image simulation program, and contributed to the text of the paper.
MPW implemented the code to add lensed host galaxies to the lensed AGN and lensed SNe in the DC2 code.
MWV co-led the Data Access Task Force to provide DC2 data products to the DESC community and consulted on the LSST Science Pipelines processing of DC2.

DA, EA, RB, CFC, CC, EG, PG, ZI, SMK, LLG, MM, CBM, AN, POC, HYP, KR, CWS, AJR, AR, JAT and JZ are Rubin Observatory Builders (and at the same time DESC Full Members) and/or DESC Builders.

LSST DESC acknowledges ongoing support from the Institut National de Physique Nucl\'eaire et de Physique des Particules in France; the Science \& Technology Facilities Council in the United Kingdom; and the Department of Energy, the National Science Foundation, and the LSST Corporation in the United States. LSST DESC uses the resources of the IN2P3 / CNRS Computing Center (CC-IN2P3--Lyon/Villeurbanne - France) funded by the Centre National de la Recherche Scientifique; the Univ. Savoie Mont Blanc - CNRS/IN2P3 MUST computing center; the National Energy Research Scientific Computing Center, a DOE Office of Science User Facility supported by the Office of Science of the U.S.\ Department of Energy under contract No.\ DE-AC02-05CH11231; STFC DiRAC HPC Facilities, funded by UK BIS National E-infrastructure capital grants; and the UK particle physics grid, supported by the GridPP Collaboration.  This work was performed in part under DOE contract DE-AC02-76SF00515.

This research used resources of the Argonne Leadership Computing Facility, which is a DOE Office of Science User Facility supported under Contract DE-AC02-06CH11357.


The work of SH, APH, KH, JH, EK, PL, AP, TDU, and ASV at Argonne National Laboratory was supported under the U.S. DOE contract DE-AC02-06CH11357.
Support for YYM was provided by the Pittsburgh Particle Physics, Astrophysics and Cosmology Center through the Samuel P.\ Langley PITT PACC Postdoctoral Fellowship, and by NASA through the NASA Hubble Fellowship grant no.\ HST-HF2-51441.001 awarded by the Space Telescope Science Institute, which is operated by the Association of Universities for Research in Astronomy, Incorporated, under NASA contract NAS5-26555. 
EG and HA were supported by
the US Department of Energy Cosmic Frontier program, grants DE-SC0011636 and DE-SC0010008.
EPL, DS, and CWW were supported by
the US Department of Energy program, DOE grant DE-SC0010007.
HA also acknowledges support by the Rutgers Discovery Informatics Institute (RDI$^2$) Fellowship of Excellence in Computational and Data Science, Rutgers University \& Bevier Dissertation Completion Fellowship, Leinweber Postdoctoral Research Fellowship, and DOE grant DE-SC009193. 
CSL was supported by DOE grant DE-SC0009999 and Heising-Simons Foundation grant 2015-106.
RM was supported by the US Department of Energy Cosmic Frontier program, grant DE-SC0010118.
AJC and JBK were supported by DOE grant DE-SC0011635.
FEB acknowledges support from ANID-Chile grants Basal AFB-170002, FONDECYT Regular 1200495 and 1190818,
and Millennium Science Initiative – ICN12\_009.
SJS acknowledges support from DOE grant DE-SC 0009999 and NSF/AURA grant N56981C.
MI was supported by the US Department of Energy Cosmic Frontier program grant DE-SC0019206.
RB was supported by the research environment grant "Gravitational Radiation and Electromagnetic Astrophysical Transients (GREAT)" funded by the Swedish Research Council (VR) under Dnr 2016-06012, and by the research project grant "Understanding the Dynamic Universe" funded by the Knut and Alice Wallenberg Foundation under Dnr KAW 2018.0067.
We thank Lindsey Bleem for many discussions and contributions relevant to galaxy clusters. We also would like to thank Nicole Crumpler for carefully reading the manuscript and providing very helpful comments. We are grateful to the AGN SC for important discussions on the AGN modeling approach.
We are grateful to Bo Xin of the Rubin Observatory for both his calculated Rubin optical sensitivity matrix and the outputs of the Rubin Observatory simulations of the active optics system.

\bibliographystyle{yahapj}
\bibliography{ref,imsim,software}

\appendix

\section{Calibration Products}
\label{sec:calib}

The LSST Science Pipelines require a set of calibration data products in order to perform the image processing of LSST data. These data products characterize the photometric sensitivity of the LSST system and any detector and electronic readout features or anomalies that would be accounted for in the instrument signature removal (ISR) steps.  Since the DC2 image simulations do not include all of the instrumental features that are present in real data, only a subset of the standard calibration data products have been generated for DC2.  Specifically, the imSim code adds crosstalk, readout gain, bias offset levels, read noise, dark current, simulated cosmic rays, and bleed trails from saturated pixels to each CCD exposure.   Furthermore, as noted above, the silicon sensor model includes electrostatic effects that produce features in the simulated images, such as tree rings and the brighter-fatter PSF.   To account for these features, the DC2 calibration data products include master biases, master darks, master flats, master gain values, saturation levels, intra-CCD crosstalk matrices, and brighter-fatter correction kernels.  Instrumental features that are not present in the DC2 image simulations, but which would occur in real data, include vignetting, roll-off at CCD edges from electrostatic effects, non-linear detector response, fringing, optical ghosts and glints arising from light scattering off the telescope support structure, and pixel-level defects such as hot or dead pixels and charge traps.

The DC2 calibration products were generated using the software available in the \code{cp\_pipe}\footnote{\https{github.com/lsst/cp\_pipe}} package.   The various tasks in this package take as input the same format of raw image files that are produced for sky exposures.   As with the sky exposure processing, the raw image files undergo an appropriate level of ISR (e.g., overscan subtraction, cosmic-ray repair, crosstalk correction, pixel masking, saturation detection/interpolation, etc.) before they are combined to produce the corresponding master files. We generate sufficient numbers of input raw files in order to satisfy the requirements stipulated in the DM documentation\footnote{\https{dmtn-101.lsst.io}}, e.g., bias or dark corrections using the master frames should not increase the effective read noise of a single exposure by more than 2.5\%.  System gains, saturation levels, and crosstalk coefficients are set at nominal values that are based on laboratory measurements made by the Camera team on actual LSST sensors and electronics; and these quantities stored in configuration files used both by the imSim code and the image processing tasks. Since none of the effects captured by the calibration products are simulated to vary in time, one set of calibration products is used for all epochs of the 5-year DC2 dataset.

Generation of the brighter-fatter correction kernels for DC2 proceeded in a much more streamlined fashion than would be undertaken for real data.  Since the charge redistribution in the CCDs owing to electrostatic effects is modeled exactly the same way for all sensors, we need only to generate the input flats and the resulting brighter-fatter kernel for a single sensor, and we apply that kernel to all 189 science sensors.  For real CCDs, because of differences in intrinsic properties of the individual devices as well as in their voltages or other readout settings, the brighter-fatter effect would manifest itself somewhat differently in each CCD, thereby requiring full sets of flats and corresponding brighter-fatter kernels for each of the science sensors in the focal plane.

In order to account for sky background features on angular scales larger than a CCD, we have included the \code{skyCorrection.py} task in the image processing pipeline.   This task includes two components for modeling the sky in a given visit: an empirical background model that extends over the entire focal plane, and a scaled ``sky frame'', which is the mean response of the instrument to the sky.  The latter component requires the creation of a set of sky frame calibration products for each band.  For real observations, these will be generated from sky exposures that sample a range of observing conditions, in order to average out visit-specific structure in the sky background.  For the DC2 simulations, the instrument response to the sky is exactly uniform across the focal plane, so rather than constructing sky frames from our simulated observations, we instead have created perfectly flat sky frames and ingested those into our calibration data products repository.

\section{Glossary}
\label{sec:glossary}

Here we provide a summary of the acronyms and names of tools and special software packages used throughout the paper.

\begin{itemize}
    \item AGN: Active Galactic Nuclei
    \item ALCF: Argonne Leadership Computing Facility
    \item Brighter-fatter: Observed property of thick CCDs that, as the intensity of the light source increases (becomes brighter), the PSF becomes wider (fatter)~\citep{2006SPIE.6276E..09D, 2014JInst...9C3048A,  2015A&A...575A..41G,  2014SPIE.9150E..17R}
    \item  CatSim: LSST catalog simulation framework described in \cite{2010SPIE.7738E..1OC, 2014SPIE.9150E..14C}
    \item CC-IN2P3: Centre De Calcul -- Institut National de Physique Nucl\'eaire et de Physique des Particules du CNRS
    \item CMB: Cosmic microwave background
    \item CFHT: Canada-France-Hawaii Telescope
    \item cosmoDC2: Extragalactic catalog underlying the data challenge, DC2; cosmoDC2 is described in detail in~\cite{korytov} and is publicly available at this URL: \url{portal.nersc.gov/project/lsst/cosmoDC2/_README.html}
    \item COSMOS: Cosmic Evolution Survey, for more details see this URL: \url{cosmos.astro.caltech.edu/}
    \item DAGs: Directed Acyclic Graphs
    \item DCs: Data Challenges carried out in LSST DESC, DC1 is described in~\cite{dc1}, DC2 is described in this paper
    \item DC2: Data Challenge 2, full end-to-end data challenge
    \item DCR: Differential chromatic refraction 
    \item DDF: Deep Drilling Field
    \item DESCQA: Validation framework developed for DESC \citep{2018ApJS..234...36M}
    \item DESI: Dark Energy Spectroscopic Instrument \citep{Aghamousa:2016zmz}
    \item DIA: Difference Image Analysis
    \item DM: Rubin's LSST Data Management
    \item DPDD: Data Products Definition Document, for more information see: \https{ls.st/dpdd}
    \item FP: Fundamental Plane
    \item FWHM: Full width at half maximum
    \item GalSim: Software library for astronomical object rendering~\citep{2015A&C....10..121R} 
    \item GCR: Generic Catalog Reader, a custom Python package, originally developed as part of the DESCQA framework \citep{2018ApJS..234...36M}, available here: \https{github.com/LSSTDESC/gcr-catalogs},
serves well as a catalog registry and provides a unified access interface to data products that are stored in different formats
    \item GREAT3: Gravitational Lensing Accuracy Testing challenge \citep{2014ApJS..212....5M}
    \item HACC: Hardware/Hybrid Accelerated Cosmology Code, N-body code used for the gravity-only simulation, described in~\cite{2016NewA...42...49H}
    \item HEALPix: Hierarchical Equal Area iso-Latitude Pixelization
    \item HPC: High-Performance Computing
    \item HSC: Hyper Suprime-Cam
    \item HST/ASC: Hubble Space Telescope/Advanced Camera for Surveys 
    \item imSim: Software package to generate pixel data for simulated observations of the Rubin Observatory (publication in preparation), the code is available at this URL: \https{github.com/LSSTDESC/imSim} 
    \item KDE: Kernel density estimation 
    \item KNL:  Intel Knights Landing
    \item LSS: Large-scale structure
    \item LSST: Legacy Survey of Space and Time \citep{2009arXiv0912.0201L,2019ApJ...873..111I}
    \item LSSTCam: Rubin Observatory LSST Camera
    \item LSST DESC: LSST Dark Energy Science Collaboration \citep{Abate:2012za}
    \item LSST Science Pipelines: Rubin's LSST Data Management
Science Pipelines software stack, available at  \https{pipelines.lsst.io}
    \item MAF: LSST Metric Analysis Framework, described in \cite{MAF}
    \item minion\_1016: Early baseline cadence output described in \https{docushare.lsst.org/docushare/dsweb/View/Collection-4604} 
    \item NERSC: National Energy Research Scientific Computing Center
    \item OM10: Oguri and Marshall 2010 catalog, described in \cite{OM10} 
    \item OpSim: LSST Operations Simulator, which simulates 10 years of LSST operations and accounts for various factors such as the scheduling of observations, slew and downtime, and site conditions \citep{2016SPIE.9911E..25R, 2016SPIE.9910E..13D}
    \item PhoSim: Photon Monte Carlo codes to simulate astronomical images, described in~\cite{2015ApJS..218...14P} and publicly available at this URL \url{bitbucket.org/phosim/phosim_release/wiki/Home}
    \item protoDC2: Small extragalactic catalog covering$\approx$ 25 deg$^2$ out to $z=1$, used for Run 1
    \item PSF: Point-spread function
    \item Run 1: Engineering DC2 runs carried out with both imSim and PhoSim covering a limited area of 25 deg$^2$ out to redshift $z=1$
    \item Run 2: Main DC2 run of the Wide-Fast-Deep area, covering 300 deg$^2$ out to redshift $z=3$
    \item Run 3: Additional DC2 run focused on the Deep Drilling Field area and containing additional objects (AGN, strongly lensed galaxies and supernova and supernova at a higher rate)
    \item SALT2: Spectral Adaptive Lightcurve Template 2, described in \cite{2007A&A...466...11G}
    \item SAM: Semi-analytic model
    \item SDSS: Sloan Digital Sky Survey
    \item SED: Spectral energy distribution
    \item SL: Strong lensing
    \item SLSprinkler: Software package to enable addition of strongly lensed objects, available here: \https{github.com/lsstdesc/slsprinkler}.
    \item SNe: Supernovae
    \item SRS workflow: Workflow engine, originally developed at SLAC to handle the processing of data from the Fermi Gamma-ray Space Telescope \citep{2009ASPC..411..193F}
    \item WFD: Wide-Fast-Deep
\end{itemize}

\end{document}